\documentclass[useAMS, usenatbib, usegraphicx]{mn2e}
\usepackage{times}
\usepackage{amssymb}
\usepackage{caption}
\usepackage{lscape}
\usepackage[fleqn]{amsmath}
\usepackage{url}
\usepackage{graphicx}
\usepackage{subfigure}
\usepackage{float}
\usepackage{xcolor}
\usepackage{bm}
\usepackage{enumitem}
\usepackage[figuresright]{rotating}

\newcommand{\msunyr}{\ensuremath{\mathit{M}_{\odot}{\rm \ yr}^{-1}}}   
\newcommand{\msun}{\ensuremath{\mathit{M}_{\odot}}}                  
\newcommand{\lsun}{\ensuremath{\mathit{L}_{\odot}}}                  
\newcommand{\rsun}{\ensuremath{\mathit{R}_{\odot}}}                  

\newcommand{\lstar}{\ensuremath{\mathit{L}_{\star}}}                 
\newcommand{\mdot}{\ensuremath{\dot{M}}}                             
\newcommand{\mdota}{\ensuremath{\dot{M}_{\eta_{\mathrm{A}}}}}        
\newcommand{\mstar}{\ensuremath{\mathit{M}_{\star}}}                 
\newcommand{\rstar}{\ensuremath{\mathit{R}_{\star}}}                 
\newcommand{\tstar}{\ensuremath{\mathit{T}_{\star}}}                 

\def\ec{$\eta$~Car}
\def\eca{$\eta_{\mathrm{A}}$}
\def\ecb{$\eta_{\mathrm{B}}$}

\newcommand{\altion}[2]{\textup{#1}\,\textsc{#2}}

\defcitealias{madura12}{M12}
\defcitealias{hillier01}{H01}
\defcitealias{hillier06}{H06}
\defcitealias{okazaki08}{O08}
\defcitealias{parkin09}{P09}
\defcitealias{parkin11}{P11}
\defcitealias{groh12a}{G12a}
\defcitealias{groh12b}{G12b}
\defcitealias{pittard02}{PC02}
\defcitealias{damineli08a}{D08a}
\defcitealias{corcoran01}{C01}
\defcitealias{corcoran10}{C10}
\defcitealias{hamaguchi07}{H07}

\title[Constraints on decreases in \ec's mass loss]{Constraints on decreases in $\boldeta$ Carinae's mass loss from 3D hydrodynamic simulations of its binary colliding winds}

\author[Madura et al.]{T.~I. Madura$^{1}$\thanks{NASA Postdoctoral Program Fellow, Email: thomas.i.madura@nasa.gov},
T.~R. Gull$^{1}$,  A.~T. Okazaki$^{2}$, C.~M.~P. Russell$^{2}$, S.~P. Owocki$^{3}$,
\newauthor J.~H. Groh$^{4}$, M.~F. Corcoran$^{5,6}$, K. Hamaguchi$^{5,7}$ and M. Teodoro$^{1,8}$\\\\
$^{1}$ Astrophysics Science Division, Code 667, NASA Goddard Space Flight Center, Greenbelt, MD 20771, USA\\
$^{2}$ Hokkai-Gakuen University, Toyohira-ku, Sapporo 062-8605, Japan\\
$^{3}$ Department of Physics and Astronomy, University of Delaware, Newark, DE 19716, USA\\
$^{4}$ Geneva Observatory, Geneva University, Chemin des Maillettes 51, CH-1290, Sauverny, Switzerland\\
$^{5}$ CRESST and X-ray Astrophysics Laboratory, NASA Goddard Space Flight Center, Greenbelt, MD 20771, USA\\
$^{6}$ Universities Space Research Association, 10211 Wisconsin Circle, Suite 500, Columbia, MD 21044, USA\\
$^{7}$ Department of Physics, University of Maryland, Baltimore County, 1000 Hilltop Circle, Baltimore, MD 21250\\
$^{8}$ CNPq/Science without Borders Fellow}

\begin{document}

\pagerange{\pageref{firstpage}--\pageref{lastpage}} \pubyear{2013}

\maketitle

\label{firstpage}
\begin{abstract}
Recent work suggests that the mass-loss rate of the primary star \eca\ in the massive colliding wind binary $\eta$~Carinae dropped by a factor of $2 - 3$ between 1999 and 2010. We present results from large- ($\pm1545$~au) and small- ($\pm155$~au) domain, 3D smoothed particle hydrodynamics (SPH) simulations of \ec's colliding winds for three \eca\ mass-loss rates (\mdota = $2.4$, $4.8$, and $8.5 \times 10^{-4} \ M_{\odot}$~yr$^{-1}$), investigating the effects on the dynamics of the binary wind-wind collision (WWC). These simulations include orbital motion, optically thin radiative cooling, and radiative forces. We find that \mdota\ greatly affects the time-dependent hydrodynamics at all spatial scales investigated. The simulations also show that the post-shock wind of the companion star \ecb\ switches from the adiabatic to the radiative-cooling regime during periastron passage ($\phi \approx 0.985 - 1.02$). This switchover starts later and ends earlier the lower the value of \mdota\ and is caused by the encroachment of the wind of \eca\ into the acceleration zone of \ecb's wind, plus radiative inhibition of \ecb's wind by \eca. The SPH simulations together with 1D radiative transfer models of \eca's spectra reveal that a factor of two or more drop in \mdota\ should lead to substantial changes in numerous multiwavelength observables. Recent observations are not fully consistent with the model predictions, indicating that any drop in \mdota\ was likely by a factor $\lesssim 2$ and occurred after 2004. We speculate that most of the recent observed changes in \ec\ are due to a small increase in the WWC opening angle that produces significant effects because our line-of-sight to the system lies close to the dense walls of the WWC zone. A modest decrease in \mdota\ may be responsible, but changes in the wind/stellar parameters of \ecb, while less likely, cannot yet be fully ruled out. We suggest observations during \ec's next periastron in 2014 to further test for decreases in \mdota. If \mdota\ is declining and continues to do so, the 2014 X-ray minimum should be even shorter than that of 2009.
\end{abstract}

\begin{keywords}
hydrodynamics -- binaries: close -- stars: individual: Eta Carinae -- stars: mass-loss -- stars: winds, outflows
\end{keywords}

\section{Introduction} \label{intro}

Deep within Eta Carinae's spectacular bipolar ``Homunculus" nebula lies an extremely luminous ($L_{\mathrm{Total}} \gtrsim 5 \times 10^{6}$~L$_{\odot}$) colliding wind binary with a highly eccentric ($e \sim 0.9$), 5.54-year orbit \citep{damineli97, hillier01, damineli08a, damineli08b, corcoran10}. The primary of the system, \eca, is our closest \citep[$2.3 \pm 0.1 \ \mathrm{kpc}$,][]{smith06} example of a very massive star. A Luminous Blue Variable (LBV) with a current mass of $\sim 100$~\msun\ and a powerful stellar wind with a mass-loss rate \mdot~$\approx 8.5 \times 10^{-4}$~\msunyr\ and terminal speed $v_{\infty} \approx 420 \ \mathrm{km \ s}^{-1}$ (\citealt{hillier01, hillier06}; \citealt{groh12a}, hereafter H01, H06, and G12a, respectively), \eca\ is in a relatively short and poorly understood stage of stellar evolution. Such very massive stars can lose substantial amounts of mass in giant, violent outbursts as they rapidly transition from hydrogen burning to helium core burning \citep{conti84, humphreys94}. The most prominent illustration of this is the mid-1840s ``Great Eruption'' of Eta Carinae (\ec), wherein $\sim 10 - 20$~\msun\ was ejected over a period of $\sim 5 - 15$ years, forming the Homunculus nebula \citep{davidson97, smith06}. Evidence indicates that such strong mass loss dominates the evolution of the most massive ($\gtrsim 40$~\msun) stars \citep[see e.g.,][]{smithowocki06}. The physical mechanisms leading to these eruptions are not well understood, in either single stars \citep{smithowocki06, smithtownsend07, quataert12} or binaries \citep{ofek07, smith11}. Determining their cause is of fundamental importance for understanding how very massive stars evolve into supernovae (SNe), gamma ray bursts (GRBs), and black holes.

Ground- and space-based multiwavelength observations over the last two decades show that \eca\ has a less-luminous binary companion, \ecb\ \citep[][hereafter C10]{damineli97, ishibashi99, whitelock04, smith04b, damineli08a, damineli08b, gull09, eduardo10, corcoran10}. Some of the strongest evidence for binarity comes from extended X-ray monitoring of the system (\citealt{ishibashi99, corcoran05, hamaguchi07}; \citetalias{corcoran10}). \ec's \emph{Rossi X-ray Timing Explorer} (\emph{RXTE}) light curve is characteristic of a highly eccentric colliding wind binary, the variable X-ray emission arising in a wind-wind collision (WWC) shock zone formed between the stars (\citealt{pittard02, parkin09, parkin11}; \citetalias{corcoran10}). The hard X-ray spectrum ($kT \approx 4 - 5$~keV) requires that \ecb\ have an unusually high wind terminal speed of $\sim 3000$~km~s$^{-1}$, while modeling of the X-ray spectrum suggests an enormous mass-loss rate of $\sim 10^{-5}$~\msun~yr$^{-1}$ \citep{pittard02,parkin09}. Due to \eca's immense brightness, \ecb\ has never been directly observed. Nonetheless, studies of the photoionization effects of \ecb\ on the circumstellar ejecta known as the ``Weigelt blobs'' \citep{weigelt86} have helped constrain \ecb's temperature ($T_{\mathrm{eff}} \sim 37,000 - 43,000 \ \mathrm{K}$) and luminosity \citep[$\log L/\lsun \sim 5 - 6$,][]{verner05, mehner10}.

Monitoring of \ec\ with \emph{RXTE} and \emph{Swift} shows that, broadly speaking, the X-ray flux repeats itself from orbit-to-orbit with more consistency than is shown in any other waveband (\citetalias{corcoran10}; \citealt{corcoran11}). Surprisingly, the observed X-ray minimum in 2009 was $\sim 50\%$ shorter than the minima of 1998 and 2003.5 \citepalias[figure~3 of][]{corcoran10}. The sudden recovery of the X-ray flux in 2009 probably represents a change in the cooling of the WWC shock due to currently unknown reasons \citep[][hereafter P09 and P11]{parkin09, parkin11}. Simple calculations suggest a large reduction in \eca's \mdot, of a factor of $\sim 2 - 4$, is needed to match the reduced 2009 minimum \citepalias{corcoran10}.

Spectroscopic changes observed in stellar-wind emission features such as H$\alpha$ are also interpreted as being due to a secular decrease of \eca's mass-loss rate (by a factor of $2 - 3$ between 1999 and 2010, \citealt{mehner10, mehner11, mehner12}). However, monitoring of Balmer H$\delta$ wind lines shows no significant changes in the inner wind of \eca\ \citep{teodoro12}. Moreover, while a secular change is visible in various wind profiles for a direct view of the central source, little to no change is seen in wind profiles at high stellar latitudes or reflected off of the Weigelt blobs \citep{gull09, mehner12}. Model stellar spectra also predict that a factor of two or more change in the \mdot\ of \eca\ should lead to very significant changes in the observed spectrum \citepalias{hillier06}. He~I wind lines should become very bright, while lines of Fe~II should nearly disappear \citepalias[figure 11 of][]{hillier06}. Such changes have not been observed, and so the puzzle of what recently occurred in \ec\ remains.

Here we present results from large- ($\pm1545$~au) and small- ($\pm155$~au) domain, 3D smoothed particle hydrodynamics (SPH) simulations of \ec's massive colliding winds for three different mass-loss rates of \eca\ ($2.4$, $4.8$, and $8.5 \times 10^{-4} \ M_{\odot}$~yr$^{-1}$). The goal is to investigate how decreases in \eca's \mdot\ affect the 3D geometry and dynamics of \eca's optically-thick wind and spatially-extended WWC regions, both of which are known sources of observed X-ray, optical, UV, and near-infrared (IR) emission and absorption (\citealt{pittard02}; \citealt{vanboekel03}; \citealt{martin06a}; \citealt{nielsen07}; \citealt{weigelt07}; \citetalias{parkin09}; \citetalias{parkin11}; \citealt{gull09}, 2011; C10; \citealt{groh10a, groh10b}; \citealt{madura12}; \citealt{maduragroh12}; G12a; \citealt{teodoro12}). We use several computational domain sizes in order to better understand how the various observables that form at different length scales may be influenced by declines in \eca's \mdot. This is the first such parameter study of \ec\ in 3D, which is essential since orbital motion greatly affects the shape and dynamics of the WWC zones during periastron passage (\citealt{okazaki08}; \citetalias{parkin09}; \citetalias{parkin11}; \citealt{madura12}; \citealt{maduragroh12}). The simulations in this paper can help constrain \ec's recent mass-loss history and possible future state.

The following section describes the 3D SPH simulations, while Section~3 presents the results. A qualitative discussion of the observational implications is in Section~4. Section~5 summarizes our conclusions and outlines the direction of future work.

\section{The 3D SPH Simulations} \label{sims}

The hydrodynamical simulations in this paper were performed with an improved version of the 3D SPH code used in \citet{okazaki08} and \citet[][hereafter M12]{madura12}, which is based on a version originally developed by \citet{benz90} and \citet{bate95}, to which we refer the reader for further details. The linear and non-linear SPH artificial viscosity parameters are $\alpha_{\mathrm{SPH}} = 1$ and $\beta_{\mathrm{SPH}} = 2$, respectively \citep{okazaki08}. Optically thin radiative cooling is now implemented using the Exact Integration scheme of \citet{townsend09}. The radiative cooling function $\Lambda(T)$ is calculated with \textsc{Cloudy} 90.01 \citep{ferland98} for an optically thin plasma with solar abundances. The pre-shock stellar winds and rapidly-cooling dense gas in the WWC region are assumed to be maintained at a floor temperature $= 10^{4}~\mathrm{K}$ due to photoionization heating by the stars \citepalias{parkin11}. The same initial temperature is assumed for both winds for simplicity. The effect of the initial wind temperature ($T_{\mathrm{wind}}$) on the flow dynamics is negligible \citep{okazaki08}.

Radiative forces are incorporated in the SPH code via an ``anti-gravity" formalism developed by two of us (A.~T.~O. \& S.~P.~O.), the details of which can be found in Appendix~\ref{appenda} and \citet{russell13}. We parameterize the individual stellar winds using the standard ``beta-velocity law" $v(r) = v_{\infty} (1 - R_{\star}/r)^{\beta}$, where $v_{\infty}$ is the wind terminal velocity, $R_{\star}$ the stellar radius, and $\beta$ a free parameter describing the steepness of the velocity law. Effects due to ``radiative braking,'' in which one of the stellar winds experiences a sudden deceleration before reaching the WWC zone (\citealt{gayley97}; \citetalias{parkin11}), are not included. Such braking effects are not expected to play a prominent role in \ec\ (\citetalias{parkin09,parkin11}; \citealt{russell13}). We include the more important velocity-altering effects of ``radiative inhibition'', in which one star's radiation field reduces the net rate of acceleration of the opposing star's wind (\citealt{stevens94}; \citetalias{parkin09}; \citetalias{parkin11}). However, because we fix the stellar mass-loss rates in our anti-gravity approach, possible changes to the mass-loss due to radiative inhibition are not included. These mass-loss changes are not expected to be significant in \ec\ and should not greatly affect our results or conclusions (see Appendix~\ref{appa3} for details).

Using a standard $xyz$ Cartesian coordinate system, the binary orbit is set in the $xy$ plane, with the origin at the system center-of-mass (COM) and the major axis along the $x$-axis. The two stars orbit counter-clockwise when viewed from the $+z$ axis. By convention, $t = 0$ ($\phi = t/2024 = 0$) is defined as periastron. Simulations are started with the stars at apastron and run for multiple consecutive orbits. Orbits are numbered such that $\phi = 1.5$, $2.5$ and $3.5$ correspond to apastron at the end of the second, third, and fourth full orbits, respectively.

\begin{table}
\caption{Stellar, Wind, and Orbital Parameters of the 3D SPH Simulations}
\label{tab1}
\begin{center}
\begin{tabular}{l c c c}\hline
  Parameter & $\eta_{\mathrm{A}}$ & $\eta_{\mathrm{B}}$ & Reference \\ \hline
  $M_{\star}$ ($M_{\odot}$) & 90 & 30 & \protect\citetalias{hillier01,okazaki08} \\
  $R_{\star}$ ($R_{\odot}$) & 60 & 30 & \protect\citetalias{hillier01, hillier06} \\
  $T_{\mathrm{wind}}$ ($10^{4}$ K) & 3.5 & 3.5 & \protect\citetalias{okazaki08}; this work \\
  $\dot{M}$ ($10^{-4} M_{\odot}$ yr$^{-1}$) & 8.5, 4.8, 2.4 & 0.14 & \protect\citetalias{groh12a,parkin09} \\
  $v_{\infty}$ (km s$^{-1}$) & 420 & 3000 & \protect\citetalias{groh12a}; \protect\citetalias{pittard02} \\
  $\beta$ & 1 & 1 & \protect\citetalias{hillier01, groh12a} \\
  $\eta$ & \multicolumn{2}{c}{0.12, 0.21, 0.42} & this work \\
  $q$ & \multicolumn{2}{c}{22.28, 15.22, 9.59} & \protect\citetalias{hillier01,parkin09}; this work \\
  $P_{\mathrm{orb}}$ (days) & \multicolumn{2}{c}{2024} & \protect\citetalias{damineli08a}\\
  $e$ & \multicolumn{2}{c}{0.9} & \protect\citetalias{corcoran01,parkin09} \\
  $a$ (au) & \multicolumn{2}{c}{15.45} & \protect\citetalias{corcoran01, okazaki08} \\\hline
\end{tabular}
\end{center}
\textbf{Notes:} $M_{\star}$ and $R_{\star}$ are the stellar mass and radius. $T_{\mathrm{wind}}$ is the initial wind temperature. $\dot{M}$, $v_{\infty}$, and $\beta$ are the stellar-wind mass-loss rate, terminal speed, and velocity-law index, respectively. $\eta \equiv (\dot{M} v_{\infty})_{\eta_{\mathrm{B}}}/(\dot{M} v_{\infty})_{\eta_{\mathrm{A}}}$ is the secondary/primary wind momentum ratio, $q$ is the primary/secondary stellar luminosity ratio assuming $L_{\star, \eta_{\mathrm{A}}} = 5 \times 10^{6}$~\lsun, $P_{\mathrm{orb}}$ is the period, $e$ is the eccentricity,  and $a$ is the length of the orbital semimajor axis.\\
\textbf{References:} C01~=~\citet{corcoran01}; H01~=~\citet{hillier01}; PC02~=~\citet{pittard02}; H06~=~\citet{hillier06}; D08a~=~\citet{damineli08a}; O08~=~\protect\citet{okazaki08}; P09~=~\citet{parkin09}; G12a~=~\citet{groh12a}.
\end{table}

The outer spherical simulation boundary is set at $r = 10a$ and $100a$ from the system COM for the small- and large-domain simulations, respectively, where $a = 15.45$~au is the length of the orbital semimajor axis. Particles crossing this boundary are removed from the simulations. The total number of SPH particles used in the small- and large-domain simulations is roughly the same (between $\sim 5 \times 10^{5}$ and $9 \times 10^{5}$, depending on the value of the primary \mdot). The small-domain simulations are better suited for studying the complex physics and geometry of the WWC zone very close to the central stars where most of the instantaneous changes observed in numerous spectral lines, the optical/near-IR/radio continuum fluxes, and X-rays across \ec's 5.54-year orbital cycle occur \citep[see e.g. ][]{damineli08b}.

The large-domain simulations are comparable in size to past and planned \emph{Hubble Space Telescope}/Space Telescope Imaging Spectrograph (\emph{HST}/STIS) mapping observations of the interacting stellar winds in \ec's central core \citep[$\sim \pm 0.67'' \approx \pm 1540 \ \mathrm{au}$,][M12, Teodoro et al. 2013]{gull11}. As demonstrated by \citet{gull11} and \citetalias{madura12}, 3D simulations at this scale are necessary for understanding and modeling the extended, time-variable forbidden line emission structures that are spatially and spectrally resolved by \emph{HST}/STIS. We note that the SPH formalism is ideally suited for such large-scale 3D simulations due to the significantly decreased computational cost, as compared to grid-based hydrodynamics codes \citep{price04, monaghan05}.

The adopted simulation parameters (Table~\ref{tab1}) are consistent with those derived from the available observations, although there has been some debate on the exact value of \eca's mass-loss rate (hereafter \mdota). For the current study, we investigate three values of \mdota, which we refer to as Cases A $-$ C. The largest adopted value, $8.5~\times~10^{-4}~M_{\odot}$~yr$^{-1}$ (Case~A), is based on the latest \texttt{CMFGEN} radiative transfer model fits to \emph{HST}/STIS spatially-resolved spectroscopic observations of \ec\ \citepalias{groh12a}. \citetalias{hillier01,hillier06}; \citet{davidson95}; and \citet{cox95} derived a mass-loss rate of $10^{-3} \ M_{\odot}$~yr$^{-1}$ and a wind terminal speed of $500$~km~s$^{-1}$ for \eca. The good match between the synthetic [\altion{Fe}{iii}] spectroimages of \citetalias{madura12} and \emph{HST}/STIS observations taken between 1998 and 2004 \citep{gull09} further suggests \mdota\ was $\sim 10^{-3} \ M_{\odot}$~yr$^{-1}$ during that time. \citetalias{groh12a} improved upon the original model fits of \citetalias{hillier01,hillier06} by using a slightly lower \mdota\ ($8.5 \times 10^{-4} \ M_{\odot}$~yr$^{-1}$) and $v_{\infty}$ ($420$~km~s$^{-1}$). The optical and UV spectra modeled by \citetalias{groh12a} were taken by \emph{HST} in March 2000 and April 2001, corresponding to a phase near apastron. We adopt the \citetalias{groh12a} value and assume it as the initial \mdota\ before any possible drop after 2001. We also use the same $v_{\infty}$ for \eca\ ($v_{\infty, \eta_{\mathrm{A}}} = 420$~km~s$^{-1}$) in each simulation.

Our second adopted \mdota\ ($4.8~\times~10^{-4}~M_{\odot}$~yr$^{-1}$, Case~B) is consistent with that derived by \citetalias{parkin09} and used in the 3D adaptive mesh refinement (AMR) hydrodynamical simulations of \citetalias{parkin11}. The lowest value of \mdota\ used in our simulations ($2.4~\times~10^{-4}~M_{\odot}$~yr$^{-1}$, Case~C) is nearly identical to the \mdota\ obtained by \citet{pittard02} in their attempts, using 2D hydrodynamical simulations of the colliding winds, to fit a \emph{Chandra} grating spectrum of \ec\ collected near apastron. This same \mdota\ was later used by \citet{okazaki08} in a 3D SPH simulation to model \ec's \emph{RXTE} X-ray light curve. We note that all of the X-ray modeling studies mentioned assume wind terminal speeds of 500~km~s$^{-1}$ and 3000~km~s$^{-1}$ for \eca\ and \ecb, respectively, and that we use the same \mdot\ and $v_{\infty}$ as \citetalias{parkin11} for \ecb\ (Table~\ref{tab1}).

\section{Results} \label{res}

\subsection{Small-domain Simulations}\label{R10}

\begin{figure*}
\includegraphics[width=15.25cm]{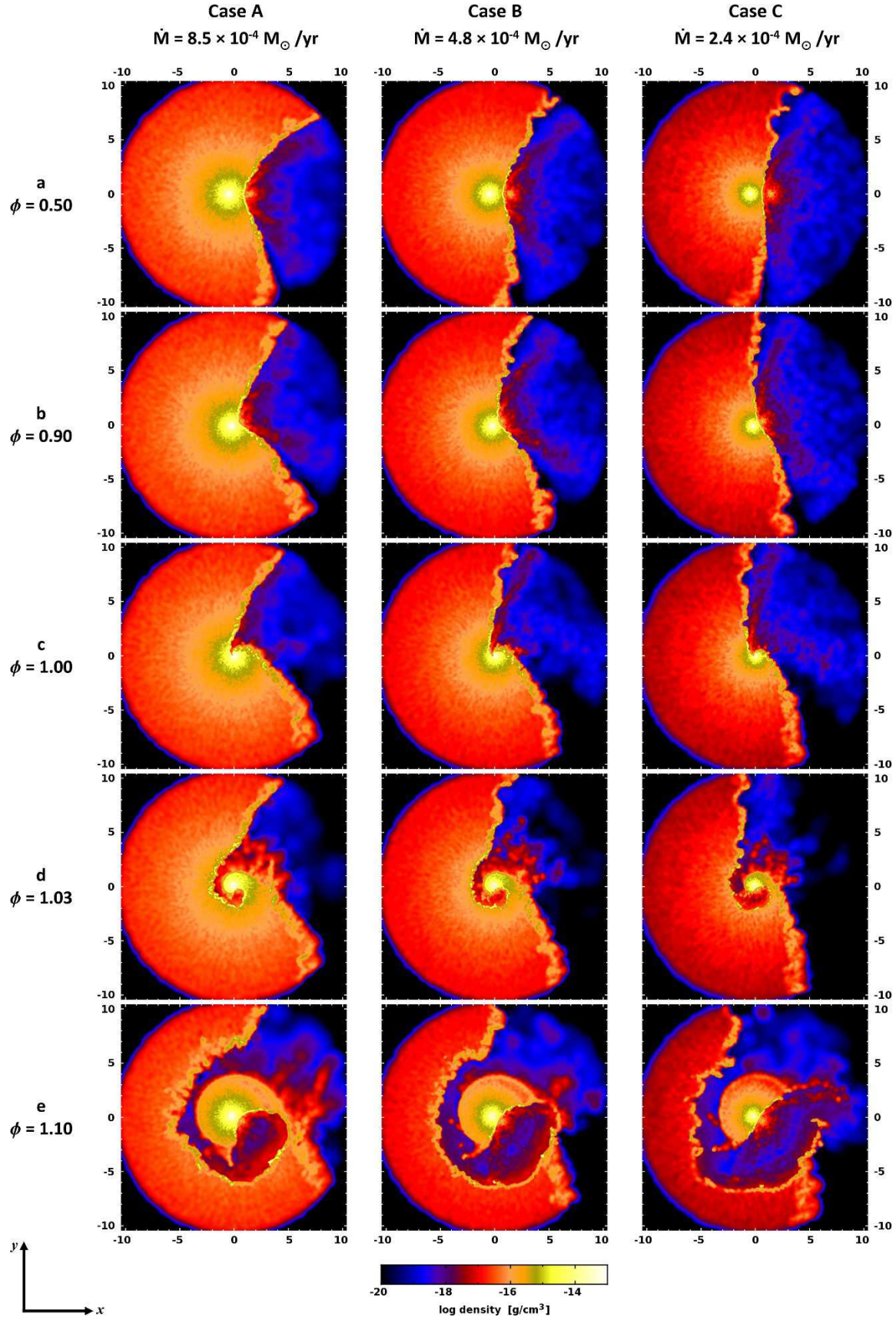}
\caption{Slices in the orbital plane from the small-domain 3D SPH simulations of \ec\ (Table~1) at orbital phases $\phi = 0.5$ (apastron), 0.9, 1.0 (periastron), 1.03, and 1.10 (rows, top to bottom). Columns correspond to the three assumed \eca\ mass-loss rates. Color shows log density in cgs units. The spherical computational domain size is $r = 10a \approx 155$~au $\approx 0.067''$ ($D = 2.3$~kpc). Axis tick marks correspond to an increment of $1a \approx 15.45$~au. The orbital motion of the stars is counterclockwise. \eca\ is to the left and \ecb\ is to the right at apastron.}
\label{fig1}
\end{figure*}

\begin{figure*}
\includegraphics[width=15.25cm]{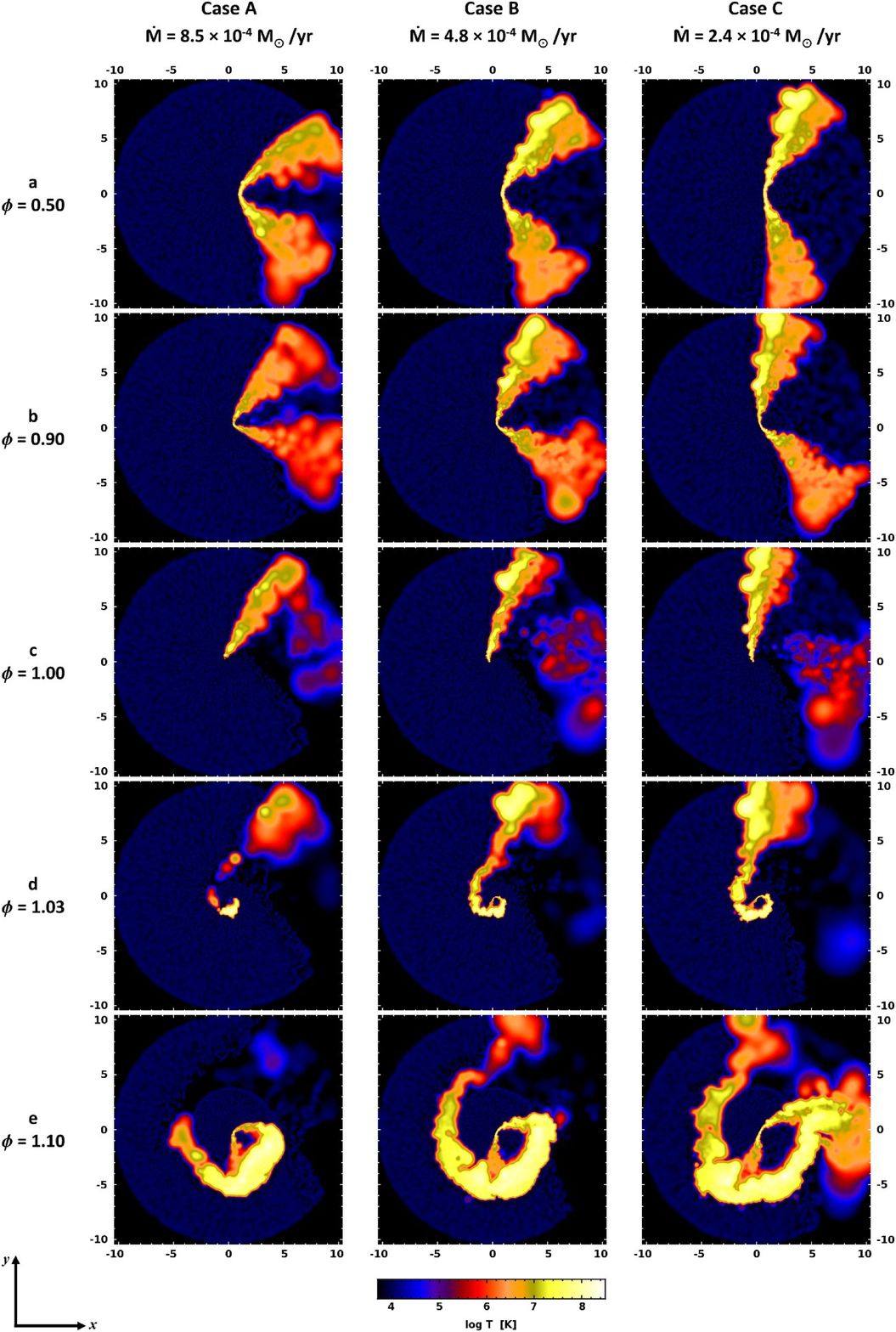}
\caption{Same as Figure~\protect\ref{fig1}, but with color showing log temperature.}
\label{fig2}
\end{figure*}

We begin with the results of our $r = 10a$ simulations, focusing on slices in the orbital plane. Results for slices in the $xz$ and $yz$ planes are in Appendix~\ref{appb1}. Movies of the simulations are also available in the online version of this paper (see Supporting Information).

Figures~\ref{fig1} and \ref{fig2} display the density and temperature at five select orbital phases, demonstrating the key features of the WWC and the effects of orbital motion. As in previous 3D simulations (\citealt{okazaki08}; \citetalias{parkin11}; \citetalias{madura12}; \citealt{maduragroh12}), the fast, low-density wind of \ecb\ carves a large cavity out of the slower, denser wind of \eca\ for most of the 5.54-year orbit. Around apastron (row~a of both figures) when orbital speeds are their lowest ($\sim 14$~km~s$^{-1}$ for \ecb\ with respect to the system COM), this cavity and the WWC zone maintain the expected axisymmetric conical shape. The measured half-opening angle of each shock cone ($55^{\circ}$, $65^{\circ}$, and $75^{\circ}$ for $\eta \approx 0.12$, $0.21$, and $0.42$, respectively) is consistent with that predicted by the analytic asymptotic formula (28) of \citet{canto96} ($53.5^{\circ}$, $62.4^{\circ}$, and $74.2^{\circ}$) and increases as the value of \mdota\ decreases. This increase is due to the changing wind momentum balance, which also moves the apex of the WWC zone closer to \eca. At apastron, the WWC apex is $\sim 22$~au ($18$~au) from \eca\ when $\eta \approx 0.12$ ($0.42$).

The WWC zone in each simulation consists of a distended shock containing hot ($T \gtrsim 10^{6} \ \mathrm{K}$), low-density \ecb\ wind material and a much thinner, colder ($T \approx 10^{4} \ \mathrm{K}$) region of dense post-shock \eca\ wind material, separated by a contact discontinuity (CD). The post-shock density of the \eca\ wind is roughly an order of magnitude larger than the pre-shock density. The post-shock density of the \ecb\ wind is approximately a factor of four higher than the pre-shock density, as expected for adiabatic shocks. The slower, denser \eca\ wind radiates more efficiently and has a characteristic cooling parameter $\chi \equiv t_{\mathrm{cool}}/t_{\mathrm{escape}}= v_{8}^4 d_{12} / \dot{M}_{-7} \ll 1$ (where $v_{8}$ is the pre-shock wind velocity in units of 1000~km~s$^{-1}$, $d_{12}$ is the distance to the CD in units of $10^{12}$~cm, and $\dot{M}_{-7}$ is the mass-loss rate in units of $10^{-7}$~\msun\ yr$^{-1}$, see \citealt{stevens92}). The lower-density, high-velocity \ecb\ wind has $\chi > 1$ and is adiabatic for most of the orbit. The maximum expected post-shock temperature in the \ecb\ wind \citep[$T = 3\bar{m}v_{w}^{2}/16 k$, where $\bar{m} = 10^{-24}$~g is the average mass per particle assuming solar abundances, $v_{w}$ is the pre-shock wind speed, and $k$ is Boltzmann's constant,][]{stevens92} at apastron is $\sim 1.2 \times 10^{8}$~K, which is very near what is observed in the simulations at the WWC apex ($\sim 10^{8}$~K, Figure~\ref{fig2}).

The overall fraction of \ecb's wind shocked to high temperatures increases with \mdota. The areas where the post-shock \ecb\ wind is hottest (in bright yellow in rows a and b of Figure~\ref{fig2}) increase in size as \mdota\ is decreased. This is because for a given pre-shock wind speed, oblique shocks (like those in Case~A) are less efficient at thermalizing the flow and produce lower post-shock temperatures \citep{pittard09}. In contrast, Cases~B and C have WWC zones that are more normal to the pre-shock flow, resulting in higher post-shock temperatures in the outer wings of the WWC region.

The post-shock \eca\ wind region appears to become somewhat thinner and less dense the lower the value of \mdota. Since the colliding winds have very different speeds there is a velocity shear across the CD that can excite Kelvin-Helmholtz (KH) instabilities (\citealt{stevens92}; \citetalias{parkin11}). The thin, dense and rapidly-cooling post-shock \eca\ wind is also subject to non-linear thin-shell instabilities (NTSIs, \citealt{vishniac94}; \citetalias{parkin11}). As \mdota\ is decreased, the WWC zone seems to become more unstable.  One possible reason for this is because the time scale for exponential growth of the KH instability is $\propto \sqrt{\rho_{1} \rho_{2}}/(\rho_{1} + \rho_{2})$, where $\rho_{1}$ and $\rho_{2}$ are the densities on the two sides of the CD \citep{stevens92}. Therefore, as \mdota\ is lowered, $\rho_{1}$ decreases and the growth timescale for the KH instability shortens. We note, however, that standard SPH schemes are notorious for under-resolving the KH instability \citep{agertz07, price08}, so these results should be interpreted with caution. In the case of the NTSI, the overall stability of a dense shell depends on the shell's thickness \citep{vishniac94, blondin98}. The thicker, denser compressed `shell' of primary wind in Case~A is thus probably less prone to the NTSI.

The aberration angle and degree of downstream curvature of the WWC zone are determined by the ratio of the orbital speed to the pre-shock wind speed \citep{parkin08, pittard09}. As the stars approach periastron, the orbital speed of \eca\ relative to \ecb\ ($\approx 360$~km~s$^{-1}$) increases to a value near its wind speed ($420$~km~s$^{-1}$), thus highly distorting the WWC zone (Figure~\ref{fig1}, rows b and c). The increasing orbital speeds furthermore cause the post-shock \ecb\ gas in the leading arm of the WWC zone to be heated to higher temperatures than the gas in the trailing arm (rows b and c of Figure~\ref{fig2}), a result also found by \citetalias{parkin11}.

Because of the high orbital eccentricity, at $\phi \sim 0.97 - 1.0$, \ecb\ and the WWC apex become deeply embedded in \eca's dense wind (row~c of Figures~\ref{fig1} and \ref{fig2}). The rapid decrease in stellar separation at these phases and the adopted $\beta = 1$ velocity laws cause the two winds to collide before \ecb's wind can reach its terminal speed. This, combined with radiative inhibition of \ecb's wind by \eca, leads to a significant decrease in the pre-shock velocity of \ecb's wind (compared to phases around apastron, see Appendix~\ref{appenda}). The associated drop in \ecb's wind momentum moves the WWC apex closer to \ecb\ and decreases the WWC-zone's opening angle. Because of these effects, the trailing wind of \ecb\ is unable to collide with \eca's downstream wind. As a result, the post-shock \ecb\ wind in the trailing arm of the WWC zone cools below $10^{6}$~K, making the hot shock noticeably asymmetric just before and at periastron (Figure~2, row~c). The higher the value of \mdota, the earlier this process starts, with the hot gas in the trailing arm vanishing at $\phi \approx 0.985$ (0.995) in Case~A (Case~C). There is also a ``collapse'' of the WWC zone around periastron (see Section~\ref{collapse} for details).

Rapid orbital motion during periastron passage distorts the WWC region, giving it a distinct spiral shape that is clearly visible at $\phi = 1.03$ (Figures~\ref{fig1} and \ref{fig2}, row d). The wind of \ecb\ also carves a narrow, low-density spiral cavity in the back side of \eca's wind. This cavity flows in the $-x$ and $-y$ directions following periastron (rows d and e of Figures~\ref{fig1} and \ref{fig2}). The width of the cavity increases with decreasing \mdota\ due to the change in the wind momentum balance. The high-velocity \ecb\ wind in the cavity also continuously ploughs into the denser, slower \eca\ wind that flows in the $-x$ and $-y$ directions. This makes the compressed wall or `shell' of \eca\ wind material grow in density and accelerate to speeds a few hundred km~s$^{-1}$ greater than $v_{\infty, \eta_{\mathrm{A}}}$. The lower the \mdota, the greater the outflow speed of the shell due to the more similar wind momenta. As the shell sweeps up more mass, it gradually slows to velocities comparable to $v_{\infty, \eta_{\mathrm{A}}}$. The thickness and density of the spiral shell of \eca\ wind also decrease with decreasing \mdota.

At $\phi \approx 1.03$, the dense layer of post-shock \eca\ wind located in the trailing arm of the WWC region is photo-ablated by the intense stellar radiation fields. \citetalias{parkin11} originally identified this photo-ablation in their 3D AMR simulations. Interestingly, only in Case~A does the photo-ablation cut-off or `snip' the spiral tail of the \ecb\ wind cavity (row~d of Figures~\ref{fig1} and \ref{fig2}). This snipping-off of the cavity prevents the \ecb\ wind from colliding with and heating the downstream \eca\ wind, causing a noticeable asymmetry in the temperature structure of the post-shock \ecb\ gas in the spiral cavity. In Case~A, only \ecb\ wind that lies in the cavity in the forward direction remains hot, while material in the trailing direction cools to $T \sim 10^{4}$~K. In Cases~B and C the entire \ecb\ wind cavity is hot, with $T \gtrsim 10^{6}$~K. The temperature asymmetry seen in Case~A remains as the gas flows outward and is also visible at $\phi = 1.1$ (Figure~\ref{fig2}, row e).

Recovery of the WWC zone following periastron takes longer the higher the value of \mdota. In Case~A, the hot \ecb-wind shock does not fully return until $\phi \approx 1.025$, whereas in Case~C, the hot shock returns by $\phi \approx 1.015$ (see Section~\ref{collapse} for details). After periastron, \ecb\ moves in front of \eca, its wind colliding with and heating the denser wind of \eca\ that flows unobstructed in the $+x,-y$ direction (row~e of Figures~\ref{fig1} and \ref{fig2}). At $\phi \approx 1.1$, the arms of the WWC zone are so distorted by orbital motion that the leading arm collides with the old trailing arm formed before periastron passage. The lower the value of \mdota, the sooner this collision takes place. The leading arm of the WWC zone, including the portion that collides with the old trailing arm, helps form another compressed shell of \eca\ wind that flows in the $+x$ and $-y$ directions after periastron (row~e). The higher the \mdota, the greater the mass in this shell and the slower it travels outward. Eventually, orbital speeds decrease and the WWC zone and \ecb\ wind cavity regain their axisymmetric conical shape as the system moves slowly back toward apastron.

\subsubsection{``Collapse'' of the WWC Zone During Periastron Passage}\label{collapse}

\citetalias{parkin09} and \citetalias{parkin11} showed that radiative inhibition (RI) likely reduces the pre-shock velocity of \ecb's wind during periastron passage, resulting in lower post-shock temperatures and rapid cooling of the post-shock \ecb\ gas. Such rapid cooling can catastrophically disrupt the WWC zone via NTSIs, leading to a ``collapse'', with dense fragments of post-shock \eca\ wind being driven into the acceleration zone of \ecb's wind and sometimes colliding with \ecb\ \citepalias{parkin11}. It has been postulated that such a WWC collapse could explain \ec's observed extended X-ray minima in 1998 and 2003, with the shorter 2009 minimum caused by the lack of a collapse for some unknown reason (\citetalias{parkin09, parkin11}; \citetalias{corcoran10}; \citealt{russell13}).

3D AMR simulations of \ec\ by \citetalias{parkin11} surprisingly did not show the expected collapse of the WWC zone around periastron. Rapid orbital motion, which increases the pre-shock wind velocity and post-shock pressure of \ecb's wind, is one key factor that helped prevent a collapse \citepalias{parkin11}. A second likely factor is the relatively weak coupling between \eca's radiation field and \ecb's wind adopted for the simulations. \citetalias{parkin11} suggest that different couplings could result in stronger inhibition that leads to a disruption and collapse of the WWC zone onto \ecb. Due to the high computational cost of 3D AMR simulations, \citetalias{parkin11} did not investigate other radiation-wind couplings. They were thus unable to demonstrate a WWC collapse around periastron in a 3D simulation that includes orbital motion.

\begin{figure*}
\includegraphics[width=16.5cm]{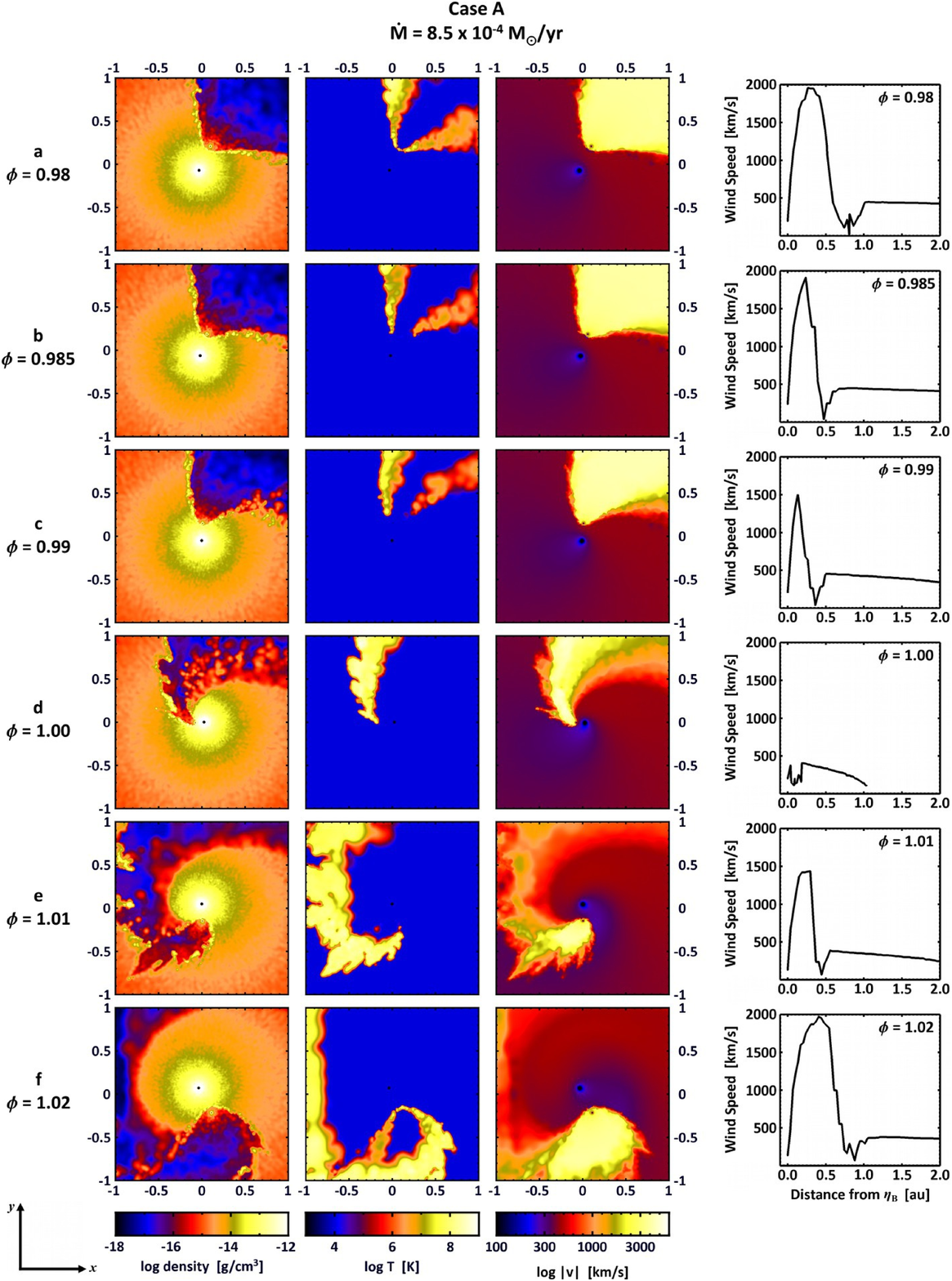}
\caption{\emph{First three columns}: Log density, temperature, and total wind speed (left to right, respectively, cgs units) in the orbital plane at phases $\phi = 0.98$, 0.985, 0.99, 1.00, 1.01, and 1.02 (rows) from the very-small-domain 3D SPH simulation of \ec\ assuming \mdota$= 8.5 \times 10^{-4} \ M_{\odot}$~yr$^{-1}$ (Case~A). All plots show the inner $\pm 1a$ region. Axis tick marks correspond to an increment of $0.1a \approx 1.54$~au. \emph{Rightmost column}: Line plots of the total wind speed ($\sqrt{v_{x}^{2}+v_{y}^{2}+v_{z}^{2}}$) along the line of centers between \ecb\ and \eca\ from the simulation at each phase. Distance is measured from the surface of \ecb\ in au. Note that all line plots extend only to $r = 2$~au from \ecb\ in order to emphasize the WWC and pre-shock wind speeds. The stellar radius of \ecb\ is R$_{\star, \eta_{\mathrm{B}}} \approx 0.14$~au.}
\label{fig3}
\end{figure*}

Since orbital motion, radiative cooling, and the velocity-altering effects of RI are included in our simulations, they provide a means to investigate further the possibility of a WWC collapse around periastron. Here we focus on the inner $\pm 1a \approx \pm 15.45$~au region in the orbital plane during periastron passage ($\phi = 0.97 - 1.03$). In order to ensure that we are adequately resolving the WWC shocks and any possible WWC collapse, we performed a series of very-small-domain 3D SPH simulations with a computational domain size of $r = 1.5a$ and eight times the number of particles used in the $r = 10a$ simulations, leading to a roughly factor of two improvement in the overall spatial resolution compared to the $r = 10a$ simulations. We begin with the \mdota~$= 8.5 \times 10^{-4} \ M_{\odot}$~yr$^{-1}$ simulation results (Figure~\ref{fig3}).

At $\phi = 0.98$ (Figure~\ref{fig3}, row a), we find that the pre-shock speed of \ecb's wind along the line of centers is significantly reduced to $\sim 1950$~km~s$^{-1}$. The stellar separation at $\phi = 0.98$ ($\approx 5$~au) is small enough for the two winds to collide while \ecb's wind is still accelerating. The expected pre-shock speed of \ecb's wind along the line of centers at $\phi = 0.98$ when assuming $\beta = 1$ wind-velocity laws and taking into account orbital motion is $\sim 2770$~km~s$^{-1}$. RI of \ecb's wind by \eca\ explains the difference. Using the analysis in Appendix~\ref{appa2}, the computed expected total \ecb\ wind speed along the line of centers ($v_{2, \mathrm{tot}}$) at the location of ram pressure balance $r_{b} \approx 1.03$~au (Equation~\ref{eqA16}) is $\approx 1953$~km~s$^{-1}$, in excellent agreement with the $1950$~km~s$^{-1}$ found in the SPH simulation. Even at apastron, we find that \ecb's pre-shock wind speed in the simulation is reduced to $\sim 2850$~km~s$^{-1}$, compared to the expected $2944$~km~s$^{-1}$ when using simple $\beta = 1$ velocity laws (Figure~\ref{figA2}).

\begin{figure}
\begin{center}
\includegraphics[width=8.4cm]{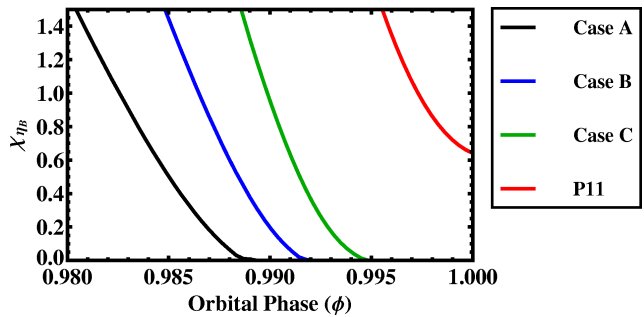}
\end{center}
\caption{Cooling parameter $\chi$ in \ecb's wind as a function of orbital phase $\phi$ assuming the simulation parameters in Table~\protect\ref{tab1} (Cases~A $-$ C) and \citetalias{parkin11}. The pre-shock wind speeds are computed using Equations~(\protect\ref{eqA15}) $-$ (\protect\ref{eqA16}).}
\label{fig4}
\end{figure}

Due to the lower wind speed and shorter distance to the CD at $\phi = 0.98$, the colliding \ecb\ wind heats to lower temperatures. The shorter distance to the CD also increases the pre- and post-shock \ecb\ wind densities. The cooling parameter $\chi$ in the post-shock \ecb\ wind near the WWC apex drops to $\sim 1.6$ (Figure~\ref{fig4}). The post-shock temperature in the \ecb\ wind at the WWC apex is reduced, which decreases the post-shock thermal pressure that helps support the WWC zone \citepalias{parkin11}. The lowered \ecb\ wind speed also alters the wind momentum ratio and decreases the WWC opening angle.

Between $\phi = 0.98$ and $0.99$, the stellar separation drops to $\sim 3$~au and the winds collide at progressively deeper locations within the acceleration zone of \ecb's wind. RI of \ecb's wind by \eca\ also becomes stronger (Figure~\ref{figA2}). During this time, the pre-shock \ecb\ wind speed along the line of centers in the simulation drops to $\sim 1500$~km~s$^{-1}$ (Figure~\ref{fig3}). The WWC opening angle decreases further, the WWC apex moves closer to \ecb, $\chi$ drops to well below unity (Figure~\ref{fig4}), and radiative cooling takes over. The post-shock \ecb\ wind near the WWC apex now cools quickly to $T \approx 10^{4}$~K. At this point we observe a ``collapse'' \citepalias{parkin09,parkin11} or ``discombobulation'' \citep{davidson02, martin06a} of the inner WWC region. Since a WWC zone is still present (although highly unstable due to the two radiative shocks), we believe it is more appropriate to say that the post-shock \ecb\ wind undergoes a ``cooling-transition'' phase. The entire apex and old trailing arm of the hot \ecb-shock vanish (Figure~\ref{fig3}, rows b and c) as the post-shock \ecb\ gas switches from the adiabatic to the radiative-cooling regime. With the thermal pressure from the hot \ecb\ wind gone, the cold, dense \eca\ wind can penetrate deep into \ecb's wind and come very close to the star, $\sim 3.5$~R$_{\star, \eta_{\mathrm{B}}}$ ($\sim 0.5$~au) from \ecb's surface at $\phi = 0.99$.

At periastron, there is almost a complete collapse of the apex of the WWC region in the simulation (row~d of Figure~\ref{fig3}). The dense wind of \eca\ directly between the stars comes within $\sim 1.5$~R$_{\star, \eta_{\mathrm{B}}}$ of \ecb's surface. \ecb\ can only effectively drive a wind in directions away from \eca. Thus, only regions downstream where \ecb's wind collides with the highly distorted spiral leading arm of the dense WWC zone are shock-heated to high $T$ (yellow areas in second column of Figure~\ref{fig3}, row~d). However, these regions have very low densities and volumes and are enshrouded by dense primary wind.

The equations in Appendix~\ref{appa2} predict that \ecb\ should not be able to drive a wind along the line of centers at periastron due to the strong RI by \eca. In our SPH simulations, however, we launch the \ecb\ wind at a value of twice the local sound speed (see Appendix~\ref{appa1}). This, combined with the large transverse velocity of the stars at periastron, produces enough of an \ecb\ wind to create a ram pressure balance at periastron in our simulations. In the absence of other forces (e.g. radiative braking), a full WWC collapse should occur along the line of centers at periastron.

The disappearance of the hot, inner WWC zone (i.e. the total cooling-transition phase) lasts $\sim 2.5$~months ($\phi \approx 0.983 - 1.022$) in the simulation of Figure~\ref{fig3}. While the \ecb\ wind speed at $\phi = 1.01$ is enough to restore a WWC at a few \rstar\ above \ecb's surface (row~e of Figure~\ref{fig3}), it is not until $\phi \approx 1.02$ that \ecb\ moves far enough from \eca\ that
the hot, inner conical shock can start to be fully restored (Figure~\ref{fig3}, row f). Therefore, the hottest plasma near the WWC apex responsible for the observed X-ray emission is absent from $\phi \approx 0.983 - 1.022$. The pre-shock speed of \eca's wind along the line of centers is also slightly reduced by $\sim 30 - 40$~km~s$^{-1}$ just after periastron ($\phi \sim 1.01 - 1.03$) due to the rapid orbital motion.

\begin{figure*}
\includegraphics[width=17.7cm]{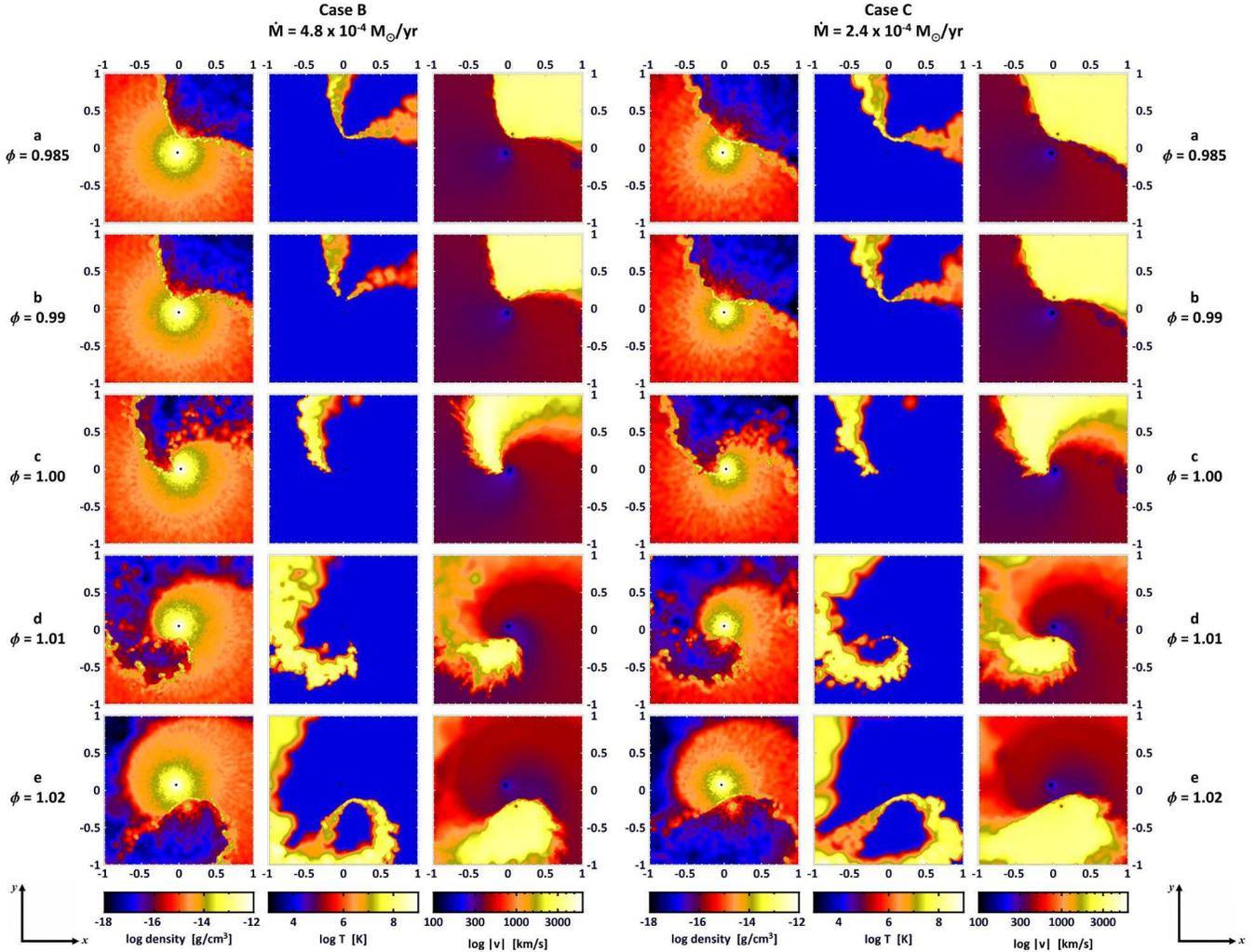}
\caption{Same as the first three columns of Figure~\protect\ref{fig3}, but for simulation Cases B (left three columns) and C (right three columns).}
\label{fig5}
\end{figure*}

\emph{Figures~\ref{fig4} and \ref{fig5} show that even with lowered \eca\ mass-loss rates, a cooling-transition phase occurs during periastron passage in the post-shock \ecb\ wind located near the WWC apex.} Although the details are somewhat different, the overall behavior is the same as described above. In each case, \eca's wind penetrates deep into the acceleration zone of \ecb's wind and comes very close to the star at periastron. \emph{The key point is that the lower the value of \mdota, the later the start of the cooling-transition phase and the earlier the recovery of the hottest X-ray emitting gas.} In the Case~B (Case~C) simulation the switch to the radiative-cooling regime occurs at $\phi \approx 0.987$ ($0.991$), with the hot \ecb-shock apex fully recovering by $\phi \approx 1.018$ ($1.014$). The cause of this dependence on \mdota\ is the decreased wind momentum of \eca\ relative to that of \ecb, which places the CD farther from \ecb\ and allows its wind to obtain larger pre-shock velocities and lower post-shock densities. This increases the value of $\chi$ since $\chi \propto v_{\mathrm{wind}}^4$ and decreases the importance of radiative cooling in \ecb's wind around periastron.

\begin{figure*}
\includegraphics[width=17.7cm]{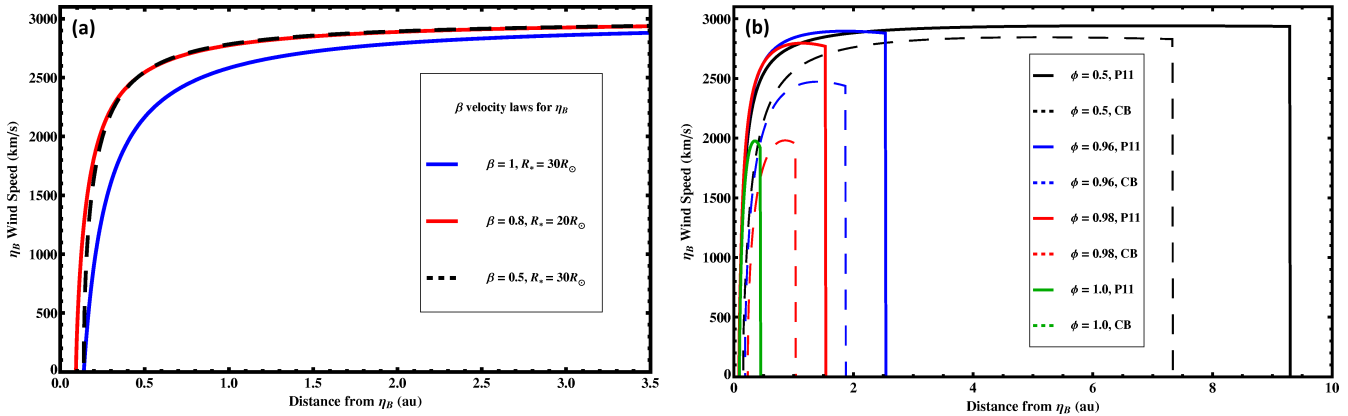}
\caption{\ecb\ wind speed along the line of centers as a function of distance from \ecb. \emph{Left panel}: Standard $\beta$-wind velocity laws computed for the combinations of $\beta$ and \rstar\ that correspond to the \ecb\ parameters of Table~1 (blue) and \protect\citetalias{parkin11} (red). Included is the velocity profile for $\beta = 1/2$ and \rstar\ = 30\rsun\ (dashed black line) that also roughly matches the velocity profile of \ecb\ used by \protect\citetalias{parkin11}. \emph{Right panel}: Total \ecb\ wind speed $v_{2,\mathrm{tot}}$ via Equations~(\protect\ref{eqA15}) $-$ (\protect\ref{eqA16}), using the parameters in Table~1 with \mdota $= 4.8 \times 10^{-4} \ M_{\odot}$~yr$^{-1}$, $q = 15.22$ (the Case~B simulation, dashed lines) and the simulation parameters of \protect\citetalias{parkin11} (solid lines), for binary separations $D$ corresponding to $\phi = 0.5$ (black), 0.96 (blue), 0.98 (red), and 1.00 (green). Each curve is truncated at the radius of ram pressure balance $r_{b}$ given by Equation~\ref{eqA16}. There is no stable balance at $\phi = 1$ using the Case~B parameters, implying a collapse of the apex of the WWC region.}
\label{fig6}
\end{figure*}

The results of our 3D simulations are quite different from those of \citetalias{parkin11}. The analysis in Appendix~\ref{appa2} can be used to help understand why. First, while the wind of \ecb\ in our simulations is parameterized using a $\beta = 1$ velocity law, \citetalias{parkin11} use standard CAK parameters for \ecb\ that are more consistent with $\beta \approx 0.8$. Our simulations also use a slightly larger \rstar\ for \ecb\ ($30$\rsun\ vs. $20$\rsun). As illustrated in Figure~\ref{fig6}a, even without RI, larger \rstar\ and $\beta$ lead to substantially lower \ecb\ wind speeds at distances $\lesssim 1.5$~au from \ecb. For our adopted \rstar, a $\beta =1/2$ would be required to roughly match the \ecb\ wind velocity profile used by \citetalias{parkin11}. However, simply reducing \rstar\ would also result in a wind velocity law that closely matches that of \citetalias{parkin11}. Thus, the assumed \rstar\ of \ecb\ can have as much of an influence on the overall wind velocity profile as the $\beta$ index.

Inclusion of the velocity-altering effects of RI makes our pre-shock \ecb\ wind speeds even lower compared to those of \citetalias{parkin11}. Setting $\beta = 0.8$ in Equation~(\ref{eqA4}) and substituting this into Equation~(\ref{eqA12}), we can numerically integrate Equation~(\ref{eqA12}) using the simulation parameters of \citetalias{parkin11}. We can then compare the results to those for the parameters in Table~\ref{tab1}. To allow for a more direct comparison with \citetalias{parkin11}, we investigate here parameters for our Case~B simulation with \mdota~$= 4.8 \times 10^{-4} \ M_{\odot}$~yr$^{-1}$ and $L_{\eta_{\mathrm{B}}} = 3.3 \times 10^{5}$\lsun.

Figure~\ref{fig6}b shows the results for several $D$ corresponding to $\phi = 0.5$, $0.96$, $0.98$, and $1.0$. At apastron, the difference between our expected total pre-shock \ecb\ wind speed and that using the parameters of \citetalias{parkin11} is relatively small, $\sim 100$~km~s$^{-1}$. This is true for most of the orbit. However, the difference between the two models becomes much more pronounced as periastron is approached. At $\phi = 0.96$, our pre-shock \ecb\ wind speed is $\sim 440$~km~s$^{-1}$ lower than that of \citetalias{parkin11}, while at $\phi = 0.98$, it is $\sim 815$~km~s$^{-1}$ lower. The largest, most important difference occurs at periastron, where for our model parameters \ecb\ is unable to drive a wind along the line of centers. \ecb\ can, however, drive a wind to a rather large pre-shock velocity of $\sim 1920$~km~s$^{-1}$ when using the parameters of \citetalias{parkin11}.

\citetalias{parkin11} speculated that an increase in the wind velocities when moving into periastron helped stabilize the WWC in their 3D simulations and prevented a collapse. While this may be true, we find that the \rstar\ and wind-velocity law assumed for \ecb\ also have a big influence on determining whether there will be a cooling-transition phase in \ecb's post-shock wind around periastron. The larger \rstar\ and $\beta$ used in our simulations lead to a significant reduction in the pre-shock \ecb\ wind speed, which results in strong, rapid cooling of the post-shock \ecb\ gas around periastron. Using the parameters of \citetalias{parkin11}, we compute a $\chi < 1$ in the post-shock \ecb\ wind only between $\phi \approx 0.997$ and $1.003$ (Figure~\ref{fig4}), and a minimum value of $\chi \approx 0.64$ at periastron, implying that the reduction in \ecb's wind speed by \eca\ is insufficient to cause \ecb's wind to switch strongly to the radiative-cooling regime. This helps explain the absence of a WWC collapse in the 3D simulations of \citetalias{parkin11}.

While radiative forces and inhibition are important factors that help determine whether a cooling-transition phase and WWC collapse occur in \ec\ at periastron, we agree with \citetalias{parkin11} that the key factor is radiative cooling. Earlier 3D SPH simulations of \ec\ that launched the stellar winds at their terminal speed and were isothermal \citep{okazaki08} or adiabatic (\citetalias{madura12}; \citealt{maduragroh12}) did not show any cooling-transition phase or WWC collapse. Investigating further, we performed two very-small-domain ($r = 1.5a$) 3D SPH simulations with \mdota~$= 8.5 \times 10^{-4} \ M_{\odot}$~yr$^{-1}$, focusing on phases around periastron. One simulation includes $\beta = 1$ velocity laws, RI effects, and assumes adiabatic cooling, while the second simulation includes radiative cooling, but no radiative forces or inhibition (i.e. the winds are launched at terminal speed).

\begin{figure}
\begin{center}
\includegraphics[width=8.25cm]{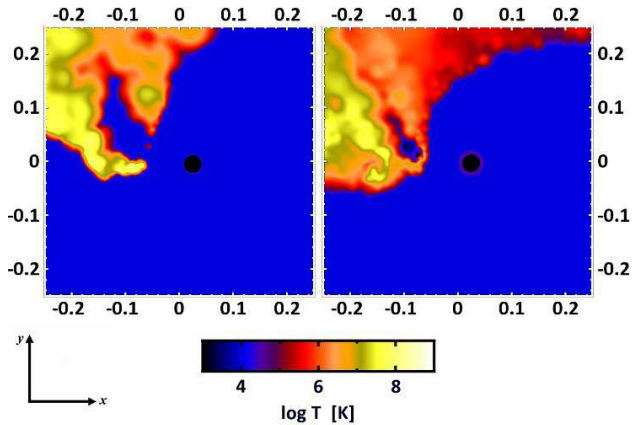}
\end{center}
\caption{Slices showing log temperature in the orbital plane at periastron from very-small-domain SPH simulations of \ec\ that use radiative cooling with the stellar winds launched at terminal speed (left panel) and adiabatic cooling with $\beta = 1$ velocity-law winds and RI effects (right panel). All plots show the inner $\pm 0.25a$ region.}
\label{fig7}
\end{figure}

We find that a cooling-transition phase \emph{only} occurs in the radiative cooling simulation (left panel of Figure~\ref{fig7}). However, because the winds are launched at terminal speed and there is no RI, the pre-shock \ecb\ wind speeds are much higher and the cooling-transition occurs over a much narrower range of phases ($\phi \approx 0.998 - 1.002$). In the adiabatic simulation, the hot apex of the \ecb\ shock survives throughout the entire simulation and there is no cooling transition phase (Figure~\ref{fig7}, right panel). Due to the twisted WWC zone and embedding of \ecb\ in \eca's dense wind at late phases ($\phi \approx 0.995$), the most distant parts of the trailing arm of the WWC region still cool to $T \lesssim 10^{6}$~K, producing a temperature asymmetry between the leading and trailing arms in the outermost post-shock \ecb\ wind. This, however, is not a collapse or radiative cooling/inhibition effect, but a consequence of rapid orbital motion in a highly eccentric orbit. \emph{Our results thus suggest that radiative cooling of the post-shock \ecb\ wind is primarily responsible for any disappearance of hot X-ray emitting gas and/or WWC collapse in \ec\ during periastron passage.} Radiative forces (e.g. $\beta = 1$ wind-velocity laws and inhibition) assist the cooling transition by helping slow \ecb's wind prior to periastron, allowing the transition to start earlier and last longer compared to a situation in which the winds collide at near their terminal speeds, since $\chi \propto v_{\mathrm{wind}}^4$.

\subsubsection{The Influence of Radiative Braking}\label{radbraking}

Our anti-gravity approach does not currently permit the implementation of radiative braking in the SPH simulations \citep[for details see][]{gayley97, russell13}. \citetalias{parkin09} demonstrated that \eca\ cannot brake \ecb's wind, irrespective of what CAK parameters they used. \citet{kashi09} claim there should be a normal ram balance between \ec's colliding winds at periastron and no radiative braking. However, \citetalias{parkin09} found that \ecb\ can radiatively brake \eca's wind if it comes extremely close to the stellar surface, e.g. within $\sim 80$\rsun\ ($\sim 0.37$~au) of \ecb. While under normal circumstances this means radiative braking would not occur before the two winds collide, it implies that radiative braking may be able to prevent the collision of dense \eca\ wind material with the surface of \ecb\ at periastron.

The two conditions required for radiative braking can be summarized as (1) $\hat{d} \equiv d/d_{rb} > 1$ and (2) $\hat{P} \equiv P_{12}/P_{rb} > \hat{d}^{2}$, where $d \equiv D/R_{2}$ is the binary separation scaled by the secondary's stellar radius, $P_{12} \equiv (\dot{M}_{1} v_{1})/(\dot{M}_{2} v_{2})$ is the primary/secondary wind momentum ratio, $P_{rb} \equiv 4\beta^{\beta} d_{rb}^{2} / (2 + \beta)^{2+\beta}$ for a velocity law with index $\beta$, and $d_{rb}$ is found by solving the transcendental expression $d_{rb} = 1 + (d_{rb}/\zeta)^{(1-\alpha)/(1+\alpha)}$ for standard CAK power index $\alpha$ and constant $\zeta$ given by

\begin{equation}
\zeta = \frac{(L_{2}/L_{1})^{1/(1-\alpha)}}{(1+\alpha) \alpha^{\alpha/(1-\alpha)}} \frac{2 G M_{1}}{R_{2} v_{\infty,1}^{2}} \ \ \ \mathrm{(Gayley 1995)}. \label{eq1}
\end{equation}\\
Condition (1) implies that a photospheric collision of star 1's wind with the surface of star 2 is prevented, while condition (2) implies that there is no ``normal'' ram pressure balance.

Using the values in Table~\ref{tab1} with \mdota $= 8.5 \times 10^{-4} \ M_{\odot}$~yr$^{-1}$, $q = 22.28$, and $\alpha = 2/3$, we see that condition (1) is satisfied throughout \ec's 5.54-year orbit, which means that \ecb's radiation is capable of preventing \eca's wind from impacting \ecb, even at periastron. For most of the orbit, condition (2) is \emph{not} satisfied and there is no radiative braking. However, for $0.99 \lesssim \phi \lesssim 1.01$, Equations~(\ref{eqA15}) through (\ref{eqA16}) indicate that there is no normal ram pressure balance along the line of centers because of the RI of \ecb's wind by \eca. If there is indeed no ram balance between $\phi = 0.99$ and $1.01$, then both (1) and (2) are satisfied and there \emph{will} be radiative braking of \eca's wind by \ecb. This is consistent with the findings of \citetalias{parkin09}. The reason \citet{kashi09} determined that there would be no radiative braking at periastron is because they neglected the effects of RI and orbital motion and assumed that $P_{12}$ is constant throughout the orbital cycle. As shown in Appendix~\ref{appa2}, \ec's wind momentum ratio \emph{is not constant} and depends strongly on the binary separation. While our analysis shows that radiative braking may occur between $\phi = 0.99$ and $1.01$ in \ec, this does not significantly affect our results or conclusions since the main effect of such braking would be the prevention of the collision of dense primary wind with \ecb's surface at periastron if/when the WWC zone `collapses'. Radiative braking, however, has important implications for models in which \ecb\ is assumed to accrete wind material from \eca\ at periastron, such as that proposed by \citet{kashi09}, since the braking could likely prevent such accretion.

\subsection{Large-domain Simulations}\label{R100}

\begin{figure*}
\includegraphics[width=15.25cm]{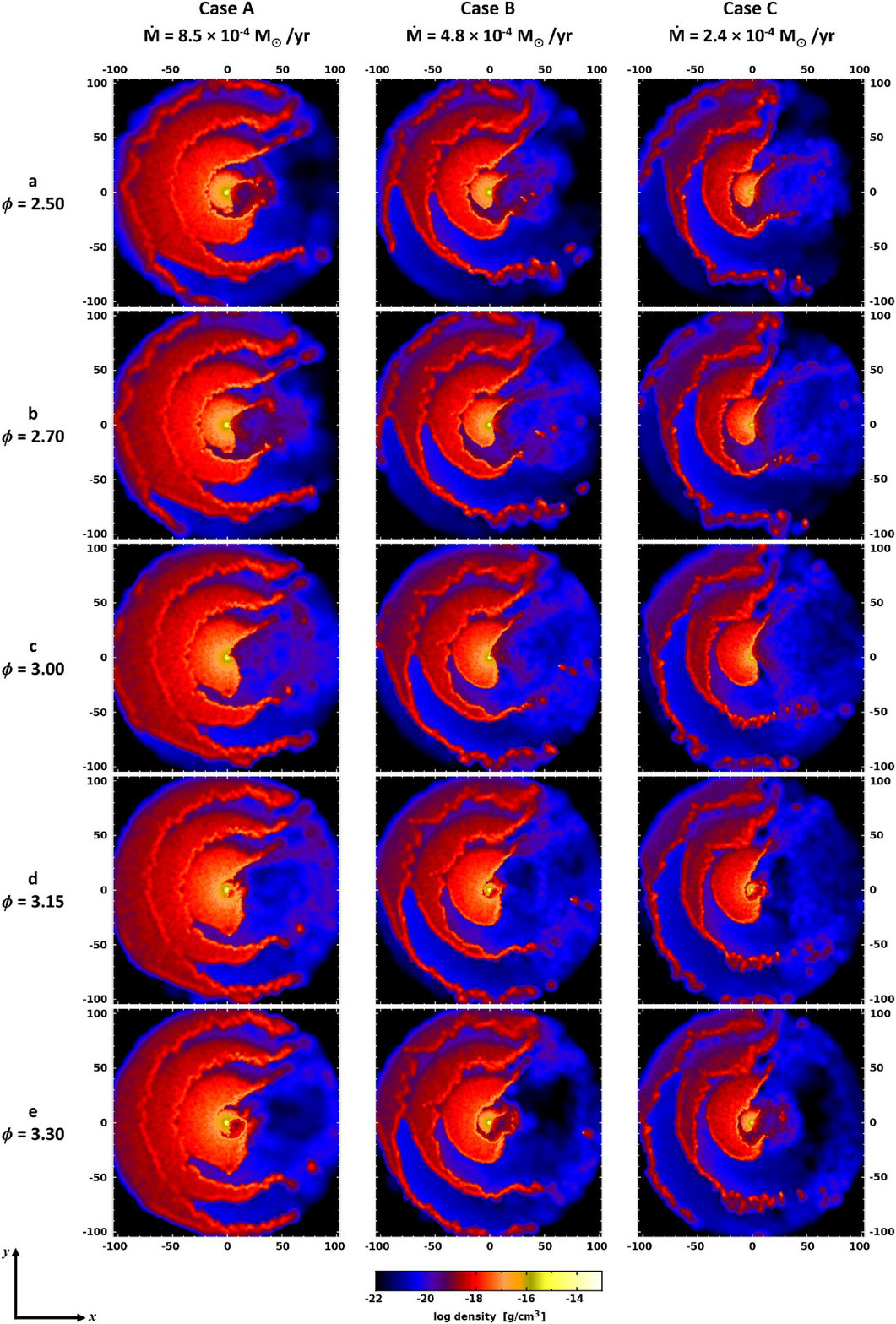}
\caption{Slices in the orbital plane from the large-domain 3D SPH simulations of \ec\ (Table~1) at orbital phases $\phi = 2.5$ (apastron), 2.7, 3.0 (periastron), 3.15, and 3.30 (rows, top to bottom). Columns correspond to the three assumed \eca\ mass-loss rates. Color shows log density in cgs units. The spherical computational domain size is $r = 100a \approx 1545$~au $\approx 0.67''$ ($D = 2.3$~kpc). Axis tick marks correspond to an increment of $10a \approx 155$~au $\approx 0.067''$. The orbital motion of the stars is counterclockwise. \eca\ is to the left and \ecb\ is to the right at apastron.}
\label{fig8}
\end{figure*}

\begin{figure*}
\includegraphics[width=15.25cm]{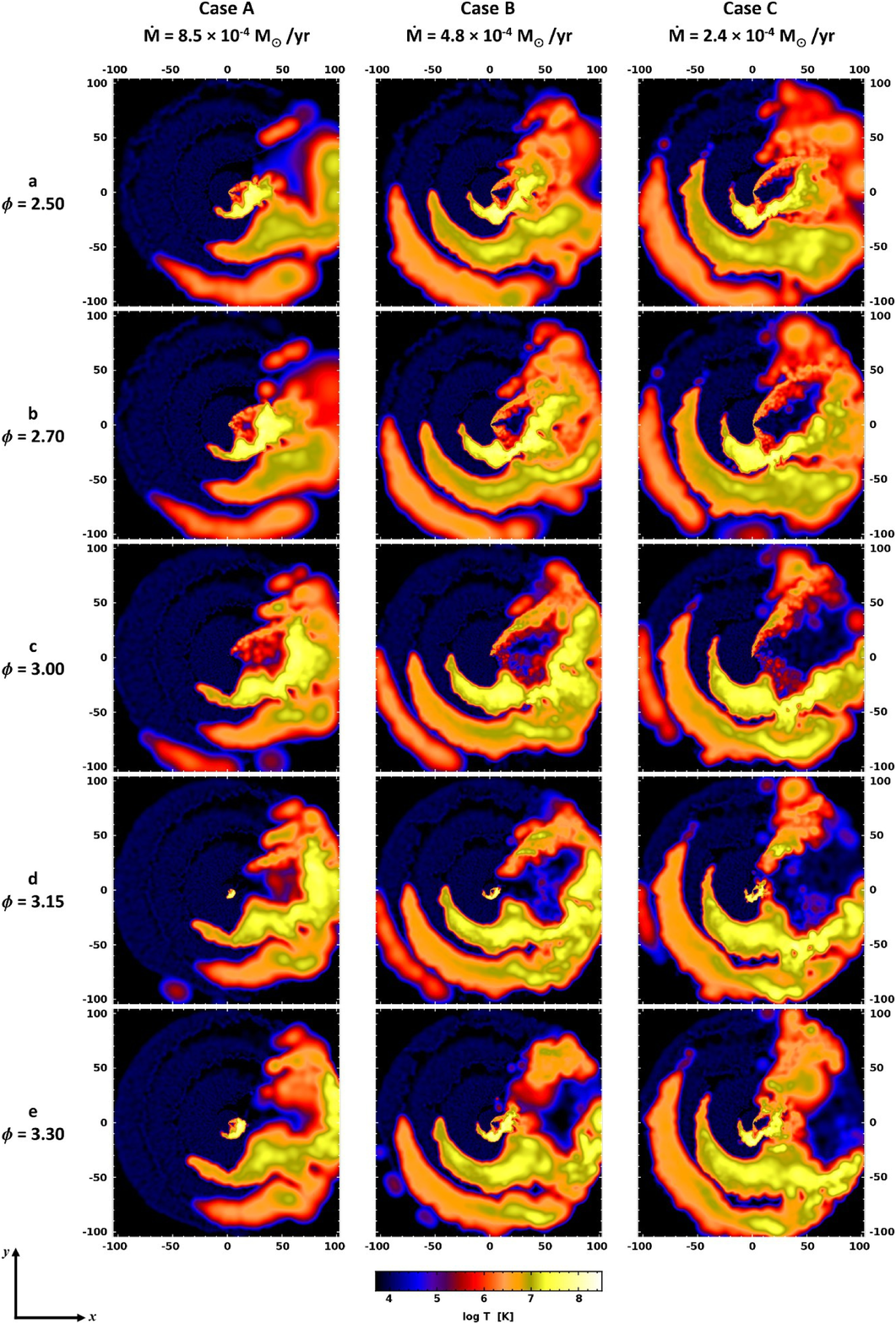}
\caption{Same as Figure~\protect\ref{fig8}, but with color showing log temperature.}
\label{fig9}
\end{figure*}

Figures \ref{fig8} and \ref{fig9} show, respectively, the density and temperature in the orbital plane at five select phases from the $r = 100a$ simulations. Results for slices in the $xz$ and $yz$ planes are in Appendix~\ref{appb2}, while movies of the simulations are available in the online version of this paper (see Supporting Information).

The dense arcs and shells of \eca\ wind visible in the outer ($> 20a$) regions of Figure~\ref{fig8} indicate that the binary has already undergone multiple orbits. Narrow cavities carved by \ecb\ in \eca's wind during each periastron passage also exist on the back (periastron) side of the system. The width of these cavities increases with decreasing \mdota\ due to the change in the wind momentum balance. The larger cavities in Cases~B and C also contain hotter \ecb\ wind material and extend farther in the $+y$ direction (Figure~\ref{fig9}).

Bordering the wind cavities on the periastron side of the system are compressed, high-density shells of \eca\ wind that form as a result of the rapid WWC. These shells flow outward at $v \gtrsim v_{\infty, \eta_{\mathrm{A}}}$, accelerated by the collision with \ecb's higher-velocity wind. The speed of the shells increases with decreasing \mdota, while the density and thickness of the shells decrease with \mdota. The innermost shell closest to the stars has the highest outward velocity due to the more recent collision with \ecb's wind. The outer shells move slower, at approximately $v_{\infty, \eta_{\mathrm{A}}}$ due to the gradual increase in the amount of swept-up \eca-wind mass in each shell.

The counter-clockwise direction of orbital motion in the $xy$ plane also produces a noticeable difference in the thickness of the shells and width of the \ecb\ wind cavities located in the $+y$ and $-y$ directions (Figure~\ref{fig8}). The shells of \eca\ wind that lie in the $-y$ direction are thinner and more compressed, while the \ecb\ wind cavities are wider. This size difference becomes more apparent as the value of \mdota\ is lowered, with Case~C having the thinnest shells and widest cavities in the $-y$ direction.

Following periastron, \ecb\ returns to the apastron side of the system, its wind colliding with the \eca\ wind that flows in the $+x,-y$ direction. At $\phi \approx 3.1$ the arms of the WWC region become extremely distorted by orbital motion and the leading arm collides with the old trailing arm from before periastron, forming another dense, compressed shell of \eca\ wind that flows in the $+x$ and $-y$ directions (rows d and e of Figure~\ref{fig8}). The overall stability of this expanding shell depends on the shock thickness (\citealt{vishniac94}; \citealt{wunsch10}; \citetalias{parkin11}). Portions of the shell moving in the $-y$ direction appear to be the most stable due to the large amount of \eca\ wind that borders it in this direction. The shells are also more stable and remain intact longer the higher the value of \mdota. In Case~A, the upper part of the shell expanding in the $+x, +y$ direction does not start to fragment until $\phi \approx 3.34$. In Case~C, the shell is completely disrupted in the $+x, +y$ direction by $\phi \approx 3.1$, the wind of \ecb\ having ploughed through it, causing it to separate from the leading arm of the WWC zone (see rows d and e of Figure~\ref{fig8}, and row e of Figures~\ref{fig1} and \ref{figB1}). This separation produces a pair of dense `arcs' of \eca\ wind on the apastron side of the system. Multiple pairs of arcs, formed during previous orbital cycles, are visible in Figure~\ref{fig8}. They are also quite spatially extended; by the time the system is back at periastron the arcs formed during the previous periastron are up to $80a \approx 1235$~au from the central stars.

As the dense arcs flow outward, the wind of \ecb\ collides with and propels them at a speed slightly greater than $v_{\infty, \eta_{\mathrm{A}}}$ into the low-density cavity created during the previous orbital cycle (Figure~\ref{fig8}). This produces another shock located $\sim 20a$ from the central stars (along the $+x$ axis at $\phi = 0.5$) that heats the outer \ecb\ wind to $T > 10^{7}$~K (Figure~\ref{fig9}). The arcs also expand as they move outward. Since they are bordered on both sides by low-density, high-speed \ecb\ wind, the arcs eventually become unstable and fragment, mixing with the surrounding \ecb\ wind material. The lower the value of \mdota, the thinner and less dense the expanding arcs, and the sooner they fragment and mix with the \ecb\ wind. An example of this fragmentation and mixing can be seen in the outermost arc/shell located at the bottom of the panels for Case~C in Figure~\ref{fig8}.

\section{Implications for Various Observational Diagnostics} \label{disc}

The SPH simulations show that for the range of explored values, a factor of two or more decrease in \mdota\ alters significantly the time-dependent 3D density, temperature, and velocity structure of \ec's spatially-extended primary wind and WWC zones. Therefore, any such drop in \mdota\ should result in detectable differences in numerous observational diagnostics (compared to earlier cycles). Guided by the SPH results, we qualitatively discuss below some of the expected changes and suggest observations to perform in order to further test whether \mdota\ is decreasing. We focus here on changes that are likely to be most important. Detailed quantitative modeling of the vast array of multiwavelength observations of \ec\ using the 3D simulations is beyond the scope of this paper and is left for future work.

\subsection{Effects on the Amount of Material in Line-of-Sight}

\begin{figure*}
\includegraphics[width=17.55cm]{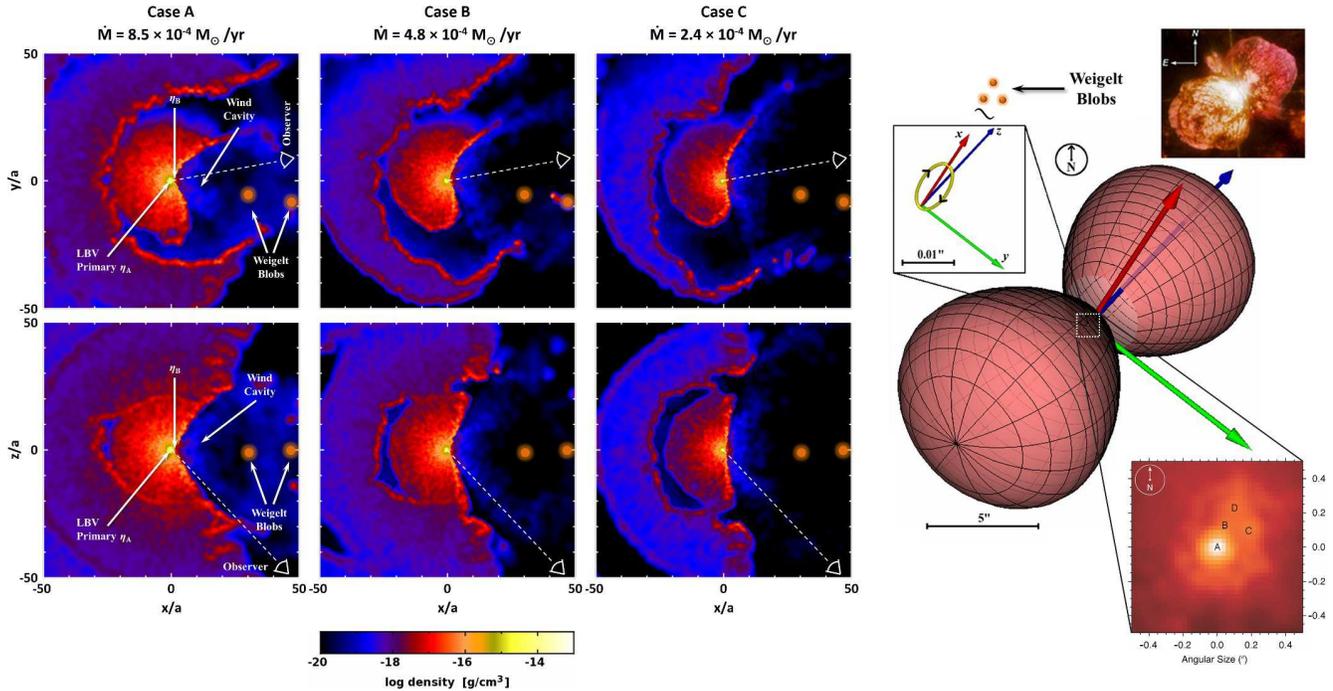}
\caption{Geometry and location of the colliding stellar winds relative to the observer and the central stars, and incorporating the apparent position of two Weigelt blobs, at an orbital phase near apastron. \emph{Top row of three panels}: Slices showing log density in the orbital plane from the 3D SPH simulations for the three different values of \mdota. The locations of the stars, \ecb\ wind cavity, and Weigelt blobs are indicated by arrows. The dashed line and eye indicate the projected line-of-sight to the observer. Axis tick marks correspond to an increment of $10a \approx 155$~au $\approx 0.067''$. \emph{Bottom row of three panels}: Same as top row, but for slices in the $xz$ plane. \emph{Far right panel}: Illustration of \ec's binary orbit (upper left inset, yellow) on the sky relative to the Homunculus nebula assuming the binary orientation derived by \protect\citetalias{madura12} ($i = 138^{\circ}$, $\omega = 263^{\circ}$, and $\mathrm{PA}_{z} = 317^{\circ}$). The $+z$ orbital axis (blue) is closely aligned with the Homunculus polar axis in 3D. \ecb\ orbits clockwise on the sky relative to \eca\ (black arrows in inset), and apastron is on the observer's side of the system. The semimajor ($+x$) and semiminor ($+y$) axes are indicated by the red and green arrows, respectively, and North is up. Included for reference are insets showing a \emph{HST} WFPC2 image of the Homunculus (top right, Credit: NASA, ESA and the Hubble SM4 ERO Team) and a $1'' \times 1''$ \emph{HST}/ACS F550M image (lower right) showing the location of the central stars (A) and the bright Weigelt blobs (B, C, D).}
\label{fig10}
\end{figure*}

\citetalias{madura12} tightly constrained, for the first time, the 3D orientation of \ec's binary orbit using a 3D dynamical model for the broad, spatially-extended [\altion{Fe}{iii}] emission observed by the \emph{HST}/STIS \citep{gull09}. \citetalias{madura12} find that the observer's line-of-sight (LOS) has an argument of periapsis $\omega \approx 240^{\circ}$ to $285^{\circ}$, with the binary orbital axis closely aligned with the Homunculus polar axis at an inclination $i \approx 130^{\circ}$ to $145^{\circ}$ and position angle on the sky $\mathrm{PA}_{z} \approx 302^{\circ}$ to $327^{\circ}$, implying that apastron is on the observer's side of the system and that \ecb\ orbits clockwise on the sky (see rightmost panel of Figure~\ref{fig10}). This orientation and clockwise motion of \ecb\ are consistent with all known observations of \ec\ to date (see the discussion in \citetalias{madura12} and \citealt{maduragroh12}).

Assuming the orbital orientation in Figure~\ref{fig10}, a factor of two or more drop in \mdota, for the range of explored values, would change significantly the time-variable amount of material in LOS to the observer. For example, if $\mdota = 8.5 \times 10^{-4} \ M_{\odot}$~yr$^{-1}$, then the half-opening angle of the WWC zone ($\approx 55^{\circ}$) is quite close in value to the inclination angle of the binary away from the observer ($\approx 48^{\circ}$). Therefore, the observer's LOS lies very near the edge of the WWC cavity (see bottom left panel of Figure~\ref{fig10}), with portions of the dense, outer interacting winds passing through our LOS. The column density is thus expected to be greater for sightlines to the observer than to most other parts of the system, especially toward the Weigelt blobs, which lie in/near the orbital plane \citep{davidsonetal97, smith04a} surrounded by the low-density wind from \ecb. As discussed in \citetalias{groh12a}, this may resolve the long standing puzzle of why our direct view of the central star suffers more extinction than the Weigelt blobs located only $0.3''$ away (\citealt{davidson95}; H01). Furthermore, as suggested by \citet{smith10} and \citetalias{groh12a}, dust formation in the densest parts of the outer interacting winds might explain the peculiar ``coronograph" described by \citetalias{hillier06}.

A decrease of less than a factor of two in \mdota\ is enough to increase the opening angle of the WWC cavity and cause the observer's LOS to no longer intersect portions of the WWC region (bottom row of Figure~\ref{fig10}). For $\mdota = 8.5 \times 10^{-4} \ M_{\odot}$~yr$^{-1}$, the difference in angle between the observer's LOS and the dense wall of the WWC zone is $\sim 5^{\circ} - 10^{\circ}$, whereas for $4.8 \times 10^{-4} \ M_{\odot}$~yr$^{-1}$, the difference is $\sim 15^{\circ} - 20^{\circ}$. Figure~\ref{fig10} shows that the lower the value of \mdota, the more the observer's LOS samples only low-density \ecb\ wind. The column density in LOS to most other directions on the apastron side of the system, e.g. toward the Weigelt blobs, is not nearly as strongly affected by decreases in \mdota. Interestingly, since 1998, the Weigelt blobs have not brightened significantly, even though the central source has brightened by one stellar magnitude in the optical \citep{davidson05, gull09}. A change in $\eta$ caused by a slight decrease in \mdota\ might explain this since the larger opening angle would decrease the amount of extinction in LOS to the observer, causing an apparent increase in the brightness of the central source while leaving the Weigelt blobs unaffected due to their location in the orbital plane. Further observations are needed in order to determine that such a change in opening angle occurred and is due solely to a decrease in \mdota\ and not, say, an increase in the \mdot\ and/or $v_{\infty}$ of \ecb.

\subsection{Effects on X-rays}

\subsubsection{Column density and the X-ray minimum}

Changes to the wind momentum ratio will change the WWC opening angle and alter the distribution of the shock-heated gas, which should have observable consequences in the X-ray band. An increase in the WWC opening angle of $\sim 10^{\circ} - 20^{\circ}$ should cause a delay of $\sim 20 - 40$~days in the observed sharp rise in column density ($N_{\mathrm{H}}$, \citealt{hamaguchi07}) that occurs prior to periastron. The decrease in $N_{\mathrm{H}}$ seen after periastron should also begin $\sim 20 - 40$~days earlier. When combined with the reduced duration of the cooling-transition phase of \ecb's post-shock wind, the net result should be a shorter X-ray minimum during periastron passage for lower values of \mdota. This simple analysis reveals that for the range of explored \mdota, the total X-ray minimum should be $\sim 40 - 80$~days shorter than that of a previous orbital cycle\footnote{For detailed quantitative modeling of \ec's X-ray observations that support this result, see \protect\citet{russell13}.}.

One mystery regarding \ec's X-ray light curve is why no significant change was observed in 2009 just prior to the X-ray minimum. Based on the simulations, the start of the 2009 minimum should have also been delayed if \mdota\ dropped by a factor of two or more. Yet, only the timing of the X-ray recovery changed. One might argue that this is evidence of a sudden significant (e.g. $\gtrsim$ factor of 2) change in \mdota\ during the 2009 event, with the apparent observed decrease in X-ray hardness between 2004 and 2009 due to a smaller, gradual decrease in \mdota\ that was not enough to change the timing of the start of the 2009 X-ray minimum. Since the recovery from the 2009 event, the $3 - 10$~keV X-ray flux has been $\sim 20\%$ below the level seen in the previous two cycles \citep{corcoran13}. A decrease in flux outside of the minimum is expected if \mdota\ decreased. However, a drop in \mdota\ by a factor of two should result in a flux decrease of about a factor of two \citep{russell13}. The implication is that, if \mdota\ decreased around the 2009 event, either (1) the change was by a factor $\lesssim 2$ or (2) the decrease occurred suddenly and by a factor $\gtrsim 2$, but immediately after the 2009 event, \mdota\ quickly returned to its `normal' value, resulting in an overall X-ray flux decrease of $\sim 20\%$ at times away from periastron.

Furthermore, the 2003 X-ray minimum did not start earlier, which seems to rule out any large change in \mdota\ prior to mid-2003. The first significant sign of a possible recent change in \ec\ was the early recovery of the 2009 X-ray minimum. Up until that point, most differences in \ec\ were viewed as typical cycle-to-cycle variations and/or caused by the continuous decrease in the amount of circumstellar extinction in LOS \citep{martin06b, smith10}. Detailed analysis by \citetalias{corcoran10} of \ec's X-rays over the last three cycles revealed later that the X-ray hardness and flux had decreased slightly between 2004 and 2009. The X-ray spectrum was also softer 100~days after the 2009 minimum than it was 100~days after the 2003.5 minimum, with the $E > 3$~keV flux lower as well. Moreover, it now appears that a high \mdota\ of $\sim 10^{-3} \ M_{\odot}$~yr$^{-1}$ is necessary in 3D hydrodynamical simulations to reproduce the extended \emph{RXTE} X-ray minima in 1998 and 2003.5, and the X-ray flux level observed between events (\citealt{russell13}; Russell et al., in preparation). Therefore, the X-ray data suggest that if there was a recent large decrease in \mdota, it occurred sometime around/after 2004. The lack of any change in the \altion{He}{ii}~$\lambda 4686$ emission before the 2009 event supports this (see Section~\ref{Heii}).

If the extended X-ray minima of 1998 and 2003.5 were the result of a long cooling-transition phase, perhaps the shorter minimum in 2009 was the result of some change in the effectiveness of radiative cooling in \ecb's post-shock wind. A modest decrease in \mdota\ of $\sim 20\%$ (or a bit more) might have moved the WWC apex far enough from \ecb\ to decrease the effectiveness of radiative cooling, but a factor of $2 - 4$ drop does not appear to be fully consistent with the current X-ray data. On the other hand, if \mdota\ is indeed decreasing and continues to do so, the 2014 X-ray minimum should be even shorter than that of 2009. Continued decrease of \mdota\ would eventually lead to an even shorter X-ray minimum.

\subsubsection{The CCE component}

During the deep minimum the X-ray emission is dominated by the ``Central Constant Emission'' (CCE) component \citep[][hereafter H07]{hamaguchi07}. \emph{Chandra} data indicate the CCE is pointlike ($\lesssim 2000$~au for $D = 2.3$~kpc). The X-ray luminosity of and column density to the CCE component suggest that it is a stable (over the course of the deep minimum), luminous X-ray source that exists beyond the primary wind in front of the WWC apex at periastron \citepalias{hamaguchi07}. The large-domain SPH simulations (Figures~\ref{fig8} $-$ \ref{fig9} and \ref{figB5} $-$ \ref{figB8}) show that, at periastron, the collision of \ecb's wind with the arcs of compressed \eca\ wind material located on the apastron side of the system could produce significant amounts of hot ($\gtrsim 10^{7}$~K) gas in front of the primary at distances $< 2000$~au in size. The simulations further show that this emission varies slowly on timescales of months, and so this faint, hot gas is a logical candidate for the source of the CCE. Any decrease of \mdota\ will propagate into the outer wind structures of \ec, affecting the thermalization of the CCE component and/or the absorption to the CCE in LOS. Measurements of the X-ray emission measure, column density in LOS, and ionization state of the CCE component during \ec's next periastron in 2014 would thus be useful. When compared to the same data from the 2009 event, the 2014 observations could help confirm or rule out a recent factor of $2 - 4$ drop in \mdota.

\subsubsection{X-ray emission line profiles}

X-ray emission line profiles provide important information on the temperature and density distribution of the hot, post-shock \ecb\ gas flowing along the boundary of the WWC zone \citep{henley08}. They can also be used to constrain the opening angle of the WWC cavity and the orientation of the system relative to our LOS (\citealt{henley08}; \citetalias{parkin11}). \emph{Chandra} high energy grating spectra show that strong, narrow emission lines are relatively unshifted during most of \ec's orbit. Just prior to the X-ray minimum ($\phi \approx 0.98 - 0.99$) the strongest lines become increasingly blueshifted and broader, reaching velocities $\approx -2000$~km~s$^{-1}$ as the flow of shocked \ecb\ wind along the trailing arm of the WWC zone swings rapidly past the observer's LOS \citep{henley08,corcoran11}.

The small-domain simulations show that an increase in the WWC opening angle causes the hot trailing arm of the WWC zone to cross the observer's LOS (with $\omega \approx 260^{\circ}$) at later orbital phases as the system approaches periastron. The observer will thus not see hot, high-velocity post-shock \ecb\ gas until much closer to periastron. This should result in a detectable phase delay in when the X-ray lines become blueshifted and broader. The magnitude of this delay will be proportional to the increase in WWC opening angle.

\subsubsection{X-ray flaring}

\ec's \emph{RXTE} light curve also exhibits short-term rapid variations in brightness or ``flares'' (\citealt{corcoran97}; \citealt{moffat09}; \citetalias{corcoran10}). These flares become more frequent and shorter in duration as periastron is approached. \citet{moffat09} presented a series of models for explaining the flares, their preferred model consisting of large stochastic wind clumps from \eca\ that enter and compress the hard X-ray-emitting WWC zone. As an alternative, they considered that the flares are the result of instabilities intrinsic to the WWC zone. \citetalias{parkin11} indicate that such instabilities, especially if combined with a clumpy LBV wind, may be a viable explanation for the X-ray flares. The SPH simulations suggest that a decrease in the value of \mdota\ leads to a more unstable, clumpy WWC region. A decrease in \mdota\ would therefore result in a higher flare frequency, especially as periastron is approached. The relative duration and amplitude of the flares may also change, although detailed numerical modeling is required to test this.

\subsubsection{Changes to \ecb's wind parameters?}

An increase in the opening angle of the WWC zone combined with a shortened cooling-transition phase during periastron passage might explain the shortened duration of the 2009 X-ray minimum. However, these effects may not necessarily be a result of a drop in \mdota. A comparable increase in the \mdot\ and/or $v_{\infty}$ of \ecb's wind would also increase the WWC opening angle. One argument against a recent increase in the $v_{\infty}$ of \ecb's wind is that such a change should increase the X-ray hardness at all phases. However, the X-ray hardness appears to have declined somewhat between 2004 and 2009 \citepalias{corcoran10}. A significant increase in the \mdot\ of \ecb\ is harder to quantify since the densities in the post-shock \ecb\ wind would increase, affecting the X-ray emission measure. Recent flux measures are lower compared to previous cycles, suggesting a decrease in the emission measure of the shocked \ecb\ wind. One argument against an increase in \ecb's \mdot\ is that such an increase should strengthen the effectiveness of radiative cooling in the post-shock wind at phases close to periastron. One would then expect a longer cooling-transition phase that leads to a longer minimum, as opposed to the observed shorter minimum. Finally, a decrease in stellar radius or change in the velocity-law profile (e.g. $\beta$) of \ecb\ could also allow its wind to reach higher pre-shock velocities without drastically altering $v_{\infty}$, moving the WWC zone closer to \eca\ and again decreasing the effectiveness of radiative cooling during periastron passage. Such changes could in principle produce a shortened X-ray minimum compared to previous cycles, although the physical reasons for a sudden alteration of \ecb's stellar parameters seem inexplicable. Numerical simulations investigating \ecb's stellar/wind parameters are needed in order to determine if such changes can explain the X-ray observations.

\subsection{The \altion{He}{II} $\mathbf{\lambda 4686}$ Emission}\label{Heii}

Somewhat related to the X-ray emission is the peculiar behavior of the \altion{He}{ii} $\lambda 4686$ emission around periastron \citep{steiner04, martin06a, mehner11, teodoro12}. The \altion{He}{ii} $\lambda 4686$ line intensity increases suddenly by a factor of ten just before phase $1.0$ and then drops sharply to zero, after which it recovers to a second peak in intensity before declining back to zero again. A highly luminous source of He$^{+}$-ionizing photons with $h \nu > 54.4$~eV is required for the formation of this line. The necessary soft X-rays are thought to come from the shocked \eca\ wind, while the \altion{He}{ii}~$\lambda 4686$ emission is hypothesized to originate in the \eca\ wind just before the WWC and/or in the cold, dense post-shock \eca\ wind \citep{martin06a, abraham07, mehner11, teodoro12}. Earlier alternative explanations in which the \altion{He}{ii} $\lambda 4686$ arises in the wind of \ecb\ can be found in \citet{steiner04} and \citet{soker06}.

\citet{mehner11} argue that the \altion{He}{ii} emission behaved differently during the 2009 spectroscopic event and included a second intensity peak just after periastron that was not present during the 2003 event. However, \citet{teodoro12} claim that a second \altion{He}{ii} peak was present during the 2003 event, and that the two events were not drastically different. In any case, both \citet{mehner11} and \citet{teodoro12} support a scenario in which there is a WWC `collapse' during periastron that helps explain the behavior of the \altion{He}{ii} $\lambda 4686$ line. A WWC disruption seems necessary to explain the fact that the \altion{He}{ii} behavior is remarkably similar for both reflected spectra from the polar region of \eca\ and a direct view of the central stars \citep{stahl05, mehner11}.

\subsubsection{Formation of the \altion{\emph{He}}{\emph{II}} $\lambda$\emph{4686} emission in \ec}

Before elaborating on how the \altion{He}{ii} emission should change if there is a large drop in \mdota, it is important to clarify how/where the \altion{He}{ii} emission forms in \ec, and why the equivalent width (EW) behaves the way it does around periastron (see e.g. figure~4 of \citealt{mehner11} or figure~3 of \citealt{teodoro12}). Some of the ideas below have been discussed in \citet{martin06a, mehner11}; and \citet{teodoro12}, to which we refer the reader for details. However, we present several new concepts that are crucial for understanding the formation of the observed \altion{He}{ii} $\lambda 4686$ emission and its dependence on \mdota.

As discussed by \citet{martin06a, mehner11}; and \citet{teodoro12}, the low-density pre-shock wind of \ecb, the acceleration region of the wind of \ecb, and the post-shock wind of \ecb\ can all easily be excluded as the source of the \altion{He}{ii}~$\lambda 4686$ emission, based simply on energy and density requirements. While it is often assumed that the \altion{He}{ii} emission arises in the dense wind of \eca, two key remaining questions are (1) why is strong \altion{He}{ii} emission only observed between $\phi \sim 0.98$ to $1.03$ and (2) how/where do the required He$^{+}$ ions form?

According to figure~9 of \citetalias{hillier01}, for $r >10 R_{\star, \eta_{\mathrm{A}}}$, helium is \emph{neutral} in the dense wind of \eca\ (assuming $\mdota=10^{-3} \ M_{\odot}$~yr$^{-1}$). Therefore, under normal circumstances, there can be no \altion{He}{ii} emission for most of \ec's cycle since there is no He$^{+}$ near the WWC zone to be ionized. Moreover, for most of the orbit, soft X-rays formed in the WWC zone cannot penetrate the dense primary wind that lies between \ecb\ and the small He$^{+}$ zone at $r<10 R_{\star, \eta_{\mathrm{A}}}$. There is no other known source of photons capable of ionizing the inner \eca\ He$^{+}$ zone. One might try to argue that radiation from the WWC shocks ionizes the neutral helium in \eca's wind located near the WWC region. However, the WWC shocks are unsuitable for this since the $3000$~km~s$^{-1}$ shocked wind of \ecb\ produces mainly hard X-rays with energies above $500$~eV, while the $420$~km~s$^{-1}$ wind of \eca\ produces photons with energies on the order of a few hundred eV. Both are highly inefficient for ionizing neutral helium, and only the shocked wind of \eca\ is suitable for providing photons to ionize He$^{+}$.

The only other possible source of neutral-helium ionizing photons at times away from periastron is \ecb. As noted by \citet{nielsen07} and \citet{humphreys08}, during most of the binary orbit, \ecb\ \emph{may} photoionize neutral helium in a small region of the pre-shock \eca\ wind located just beyond the WWC zone (similar to figure~12 of \citealt{humphreys08})\footnote{We say `may' because it is possible that the dense post-shock \eca\ wind will prevent \ecb\ from ionizing neutral helium in the pre-shock \eca\ wind, in which case there would be no \altion{He}{ii}~$\lambda 4686$ emission. Detailed 3D radiative transfer simulations are needed to test this though.}. Neutral helium in portions of the colder, denser post-shock \eca\ wind would also be ionized by \ecb. The He$^{+}$ in the dense post-shock \eca\ wind will very quickly recombine, producing \altion{He}{i} emission. Some of the He$^{+}$ in the lower-density, pre-shock \eca\ wind would also recombine and create \altion{He}{i} emission \citep{humphreys08}. However, we suggest that \emph{if} \ecb\ can photoionize neutral helium in the pre-shock wind of \eca, some of the resulting He$^{+}$ will be additionally ionized by soft X-rays created in the colliding \eca\ wind, producing a very small amount of \altion{He}{ii}~$\lambda 4686$ emission at times away from periastron. The amount of \altion{He}{ii} emission is likely to be small because of the lower densities in the pre-shock \eca\ wind near the WWC region, and because of the likely very small volume of He$^{+}$ created by \ecb's radiation. This is consistent with the observations and explains why little to no \altion{He}{ii}~$\lambda 4686$ emission is detected during most of \ec's 5.54-year cycle \citep{steiner04,stahl05,martin06a,mehner11,teodoro12}.

\begin{figure}
\begin{center}
\includegraphics[width=8.35cm]{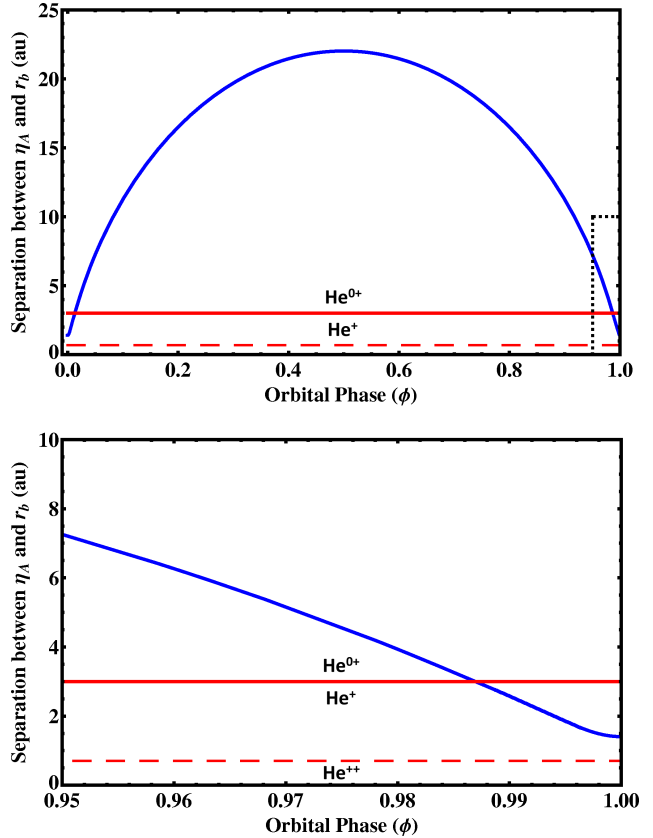}
\end{center}
\caption{Plot of the separation between \eca\ and the apex of the WWC zone as a function of orbital phase $\phi$, assuming the parameters in Table~1 with \mdota $= 8.5 \times 10^{-4} \ M_{\odot}$~yr$^{-1}$, $q = 22.28$, and using Equations~(\ref{eqA15}) and (\ref{eqA16}) to determine the location of ram pressure balance ($r_{b}$). Apastron occurs at $\phi = 0.5$ and periastron at $\phi = 0$, 1. The solid-red horizontal line indicates the radius in the wind of \eca\ where, based on figure~9 of \protect\citetalias{hillier01}, helium transitions from being neutral to singly ionized. The dashed-red line marks the radius in \eca's wind where helium transitions from being singly to doubly ionized ($r \sim 0.7$~au). The lower plot is a zoom of the dashed-box region in the upper plot ($0.95 \leq \phi \leq 1.0$).}
\label{fig11}
\end{figure}

Earlier proposed scenarios for the \altion{He}{ii}~$\lambda 4686$ emission also have difficulty explaining the emission peaks observed just prior to and after periastron. Here we present a new, simple explanation for these peaks. Figure~9 of \citetalias{hillier01} shows that He$^{+}$ exists in \eca's wind only between $\sim 2.4 R_{\star, \eta_{\mathrm{A}}}$ and $10 R_{\star, \eta_{\mathrm{A}}}$ ($0.7~\mathrm{au} - 3$~au). The densities in this region ($\sim 10^{11} - 10^{13}$~cm$^{-3}$) are capable of explaining the observed \altion{He}{ii} emission if the He$^{+}$ can be ionized by soft X-rays. If we assume that the $54 \ \mathrm{eV} - 500$~eV photons needed to ionize the He$^{+}$ come from the post-shock \eca\ wind, we find that for most of \ec's orbit, the WWC zone is too far from \eca's He$^{+}$ zone to produce any \altion{He}{ii} emission (Figure~\ref{fig11}).

Plotting the separation between \eca\ and the WWC apex as a function of phase assuming the parameters in Table~1, we see that only for $0.986 \lesssim \phi \lesssim 1.014$ does the WWC apex penetrate into the He$^{+}$ zone of \eca\ (bottom panel of Figure~\ref{fig11}). Therefore, the soft X-rays from the WWC region can ionize the He$^{+}$ only during these times. This is almost \emph{exactly} the same range of spectroscopic phases that strong \altion{He}{ii}~$\lambda 4686$ emission is detected. Furthermore, this is roughly the same range of phases that the post-shock \ecb\ wind is in the radiative-cooling regime. During this time, the post-shock \ecb\ will also produce photons capable of ionizing He$^{+}$.

\citet{martin06a} and \citet{mehner11} speculated that high densities in the post-shock \ecb\ gas around periastron could lead to a switchover to the radiative-cooling regime, which would make the WWC zone unstable and collapse, and lead to the increased generation of the soft X-rays required to form \altion{He}{ii}. According to our SPH simulations, a complete WWC collapse is not required for the additional generation of soft X-rays to ionize He$^{+}$. A mass-ejection event, accretion by \ecb, and/or changes to a latitudinally-dependent primary wind \citep{martin06a, soker06, mehner11} are also not required. \emph{Instead, we suggest that the strong \altion{\emph{He}}{\emph{II}}~$\lambda 4686$ emission arises in the inner He$^{+}$ zone of \eca's dense wind and that the first peak in EW observed just before the spectroscopic event ($\phi \approx 0.98 - 0.995$) is due to a combination of two important effects; (1) penetration of the WWC region into the He$^{+}$ zone of \eca\ that allows soft X-rays generated in the WWC zone to ionize the He$^{+}$ and (2) a switchover of the post-shock wind of \ecb\ to the radiative-cooling regime that results in additional soft X-rays that can ionize the He$^{+}$ and produce extra \altion{\emph{He}}{\emph{II}} emission.} The fact that the timing of both matches almost perfectly with the observed first \altion{He}{ii} peak provides strong support for our scenario. The sharp drop in \altion{He}{ii} EW at $\phi \approx 1.0$ may be due, at least in part, to a physical collapse of the WWC zone at periastron.

We further suggest that the second, smaller emission peak in \altion{He}{ii}~$\lambda 4686$ that occurs $\sim+20$~days after $\phi = 1.0$ represents the period just after perisatron when \ecb\ returns to our side of the system and begins to carve a new cavity within \eca's wind. During this time, the WWC zone may still be highly unstable, but two important facts will remain. Namely, the WWC region will be within the He$^{+}$ zone of \eca\ and the post-shock wind of \ecb\ will be in the radiative-cooling regime and supplying additional soft X-rays to produce \altion{He}{ii}. The decreased height of the second \altion{He}{ii} peak could be due to a combination of instability in the WWC zone and increased optical depth in LOS caused by the dense \eca\ wind that flows in the apastron direction during periastron passage. The \altion{He}{ii} behavior should look relatively similar for both a direct view of the central source and reflected spectra corresponding to high stellar latitudes, with some possible changes in the line profiles.

\subsubsection{Behavior of the \altion{\emph{He}}{\emph{II}} $\lambda$\emph{4686} with decreasing \mdota}

If \mdota\ is lowered to $\sim 2.5 \times 10^{-4} \ M_{\odot}$~yr$^{-1}$, keeping the other stellar and wind parameters constant, 1D \texttt{CMFGEN} models show that \eca's He$^{+}$ zone increases greatly and extends to $r \sim 120$~au. As a result, the WWC zone is within the He$^{+}$ region for the entire 5.54-year cycle. Soft X-rays generated in the post-shock \eca\ wind should ionize the He$^{+}$ and produce significant, detectable \altion{He}{ii} emission at all phases, with the exception of a possible short, sharp drop in \altion{He}{ii} EW at periastron. This emission, though, is not observed.

If \mdota\ is lowered to $\sim 5.0 \times 10^{-4} \ M_{\odot}$~yr$^{-1}$ while keeping the other parameters constant, \texttt{CMFGEN} models show that \eca's He$^{+}$ zone extends to $r \sim 7.5$~au, more than a factor of two larger than in the $10^{-3} \ M_{\odot}$~yr$^{-1}$ model. Combined with the movement of the WWC apex closer to \eca, this should lead to a significant change in the timing of the \altion{He}{ii} emission peaks. The WWC apex is predicted to penetrate the He$^{+}$ zone at $\phi \sim 0.94$, or $\sim 85$~days earlier. The first \altion{He}{ii} peak should therefore also appear $\sim 85$~days earlier. While this first peak may appear earlier, its initial amplitude may not be as large. This is because a lower \mdota\ produces a shorter cooling-transition phase in \ecb's post-shock wind, resulting in the generation of less soft X-rays to ionize the He$^{+}$. Therefore, an initial, broad \altion{He}{ii} peak could begin at $\phi \sim 0.94$, created when the WWC apex penetrates the He$^{+}$ zone and the soft X-rays formed in the post-shock \eca\ wind produce \altion{He}{ii} emission. Another sharper peak could then occur near $\phi \sim 0.99$ when \ecb's wind switches to the radiative-cooling regime and produces an additional burst of soft X-rays to further ionize the He$^{+}$ and create additional \altion{He}{ii} emission. Since the WWC zone will be more stable in a lower \mdota\ situation, the sharp drop in \altion{He}{ii} EW near $\phi = 1.0$ may be delayed, while the recovery to a second \altion{He}{ii} peak after periastron may be sped up. Because the WWC zone will remain in the He$^{+}$ zone longer after periastron, the height and duration of the second \altion{He}{ii} peak should both increase. The time taken for the \altion{He}{ii} to decline back to near zero EW should also increase.

Based on these results, it is clear that for the explored parameter space, any change in \mdota\ of a factor $\gtrsim 2$ should lead to an equally significant change in the behavior of the \altion{He}{ii}~$\lambda4686$ EW. Interestingly, as shown by both \citet{mehner11} and \citet{teodoro12}, the first \altion{He}{ii} peak during the 2009 event occurred almost exactly 2023~days after the 2003.5 event. This and the fact that the sharp drop in X-rays also occurred nearly 2023~days after the 2003 periaston very strongly suggest that there was no large decrease of \mdota\ before 2009. As the available observations make it unclear whether a second \altion{He}{ii} peak occurred during the 2003 event, there is a possibility \mdota\ dropped during the 2009 event. However, such a drop would likely have to be by a factor $\lesssim 2$ (in contrast to \citetalias{corcoran10}), otherwise, there would be a detectable increase in the amount of \altion{He}{ii}~$\lambda4686$ emission at phases far from periastron. Such emission is currently not seen. If \mdota\ has decreased, or continues to do so, changes in the behavior of \altion{He}{ii}~$\lambda4686$ during the next spectroscopic event in 2014 should be clearly visible.

\subsection{Effect on Line Profiles of Broad Wind-Emission Features}

\subsubsection{Expected changes}

Using 2D radiative transfer models, \citetalias{groh12a} showed that the WWC cavity significantly affects the formation of H$\alpha$, H$\beta$, and \altion{Fe}{ii} lines in \eca's dense extended wind, even at apastron, by reducing the amount of P-Cygni absorption in LOS. Ionization of portions of \eca's extended wind by \ecb\ also affects the shape and strength of these stellar wind lines (G12a; \citealt{groh12b}, hereafter G12b). The WWC cavity furthermore induces a latitudinal and azimuthal dependence in these line profiles at orbital phases around apastron, explaining the stronger P-Cygni absorption observed in spectra scattered off of the poles of the Homunculus (G12a,b), which was originally interpreted as being caused by rapid stellar rotation \citep{smith03, mehner12}. Observed latitudinal changes in the line profiles during periastron, specifically the appearance of strong P-Cygni absorption in LOS, is explained by the flow of dense primary wind toward the observer for a brief period when \ecb\ is behind \eca\ at periastron (G12a,b; \citealt{maduragroh12}, see also rows c and d of Figures~\ref{fig2} and \ref{figB1}).

Since the broad wind-emission features of H$\alpha$, H$\beta$, and \altion{Fe}{ii} are so greatly affected by the WWC cavity, a change to \mdota\ that alters the WWC opening angle should have significant observable consequences for these and similar lines. We can make some predictions using the SPH simulations and the previous spectral models of \citetalias{hillier01,hillier06} and G12a,b. Figure~11 of \citetalias{hillier06} shows that, keeping the other parameters (e.g. \rstar, \lstar, etc.) fixed, as \mdota\ is lowered, the Balmer P-Cygni emission and absorption in LOS weaken, with the absorption eventually disappearing. The \altion{Fe}{ii} emission lines weaken as well, but emission lines of \altion{He}{i} strengthen. Eventually, for low enough values of \mdota\ ($\sim 2.5 \times 10^{-4} \ M_{\odot}$~yr$^{-1}$), lines of \altion{N}{iii} and \altion{He}{ii}~$\lambda 4686$ come into emission \citepalias{hillier06}.

It is important to note that the results of \citetalias{hillier01,hillier06} were derived from 1D spherically-symmetric radiative transfer models and do not take into account the presence of the WWC cavity or \ecb. Inclusion of the WWC cavity will further reduce the amount of Balmer P-Cygni absorption in LOS \citepalias{groh12a}. The effect of the cavity on the absorption for a direct view of the central source may be difficult or impossible to detect though since for low values of \mdota\ ($\sim 2.5 \times 10^{-4} \ M_{\odot}$~yr$^{-1}$), the P-Cygni absorption is sometimes completely absent in the 1D models, depending on the assumed \rstar. There is no P-Cygni absorption in H$\alpha$ in the models only if $R_{\star, \eta_{\mathrm{A}}}$ is small ($\sim 60$\rsun). Assuming \lstar$= 5 \times 10^{6}$\lsun, a model with $60$\rsun\ has a temperature at the hydrostatic core of \tstar$\approx 35,000$~K, while the model with $300$\rsun\ has \tstar$\approx 15,000$~K. This results in a big difference in the ionizing fluxes between models, which ultimately determines the amount of H$\alpha$ absorption.

However, the increase in cavity opening angle \emph{will} alter the latitudinal and azimuthal variation of the P-Cygni absorption \citepalias{groh12a}. The photoionization cavity created by \ecb\ will also increase in size. Based on Figure~\ref{figB1}, a factor of two or more drop in \mdota\ should reduce the \eca\ wind density, and thus the amount of P-Cygni emission and absorption in H$\alpha$, over the stellar poles \citepalias[see also][]{hillier06}. A factor of four drop in \mdota\ might result in almost no P-Cygni absorption in H$\alpha$ over the poles of \eca\ at phases far from periastron (assuming that the dense, post-shock \eca\ wind is ionized). The models predict that any larger drop in \mdota\ will increase the opening angle of the WWC cavity in such a way that it starts to bend toward \eca, likely resulting in little to no P-Cygni absorption in H$\alpha$ at high stellar latitudes during most of \ec's cycle.

The SPH simulations also suggest that a drop in \mdota\ should result in the detection of less P-Cygni absorption in LOS in H$\alpha$ and similar lines during periastron passage. This is because, as shown in Figures~\ref{fig1} and \ref{figB1}, as \mdota\ is decreased, the volume of \eca\ wind that flows toward the observer during periastron also decreases, producing a thinner, less-dense wall of \eca\ wind between the star and the observer after periastron. If \mdota\ is low enough ($\sim 2.5 \times 10^{-4} \ M_{\odot}$~yr$^{-1}$), for \rstar $=60$\rsun, there may be \emph{no} increase in P-Cygni absorption in LOS during periastron since \eca\ will be able to fully ionize hydrogen in this outflowing primary-wind region \citepalias{hillier06}. The decreased thickness and density of the shell will cause it to break apart quicker after periastron as well. The observed duration during which any increased absorption is detected in LOS around periastron should thus also decrease with \mdota.

\subsubsection{Comparison to recent observed changes}

\begin{figure*}
\includegraphics[width=17.5cm]{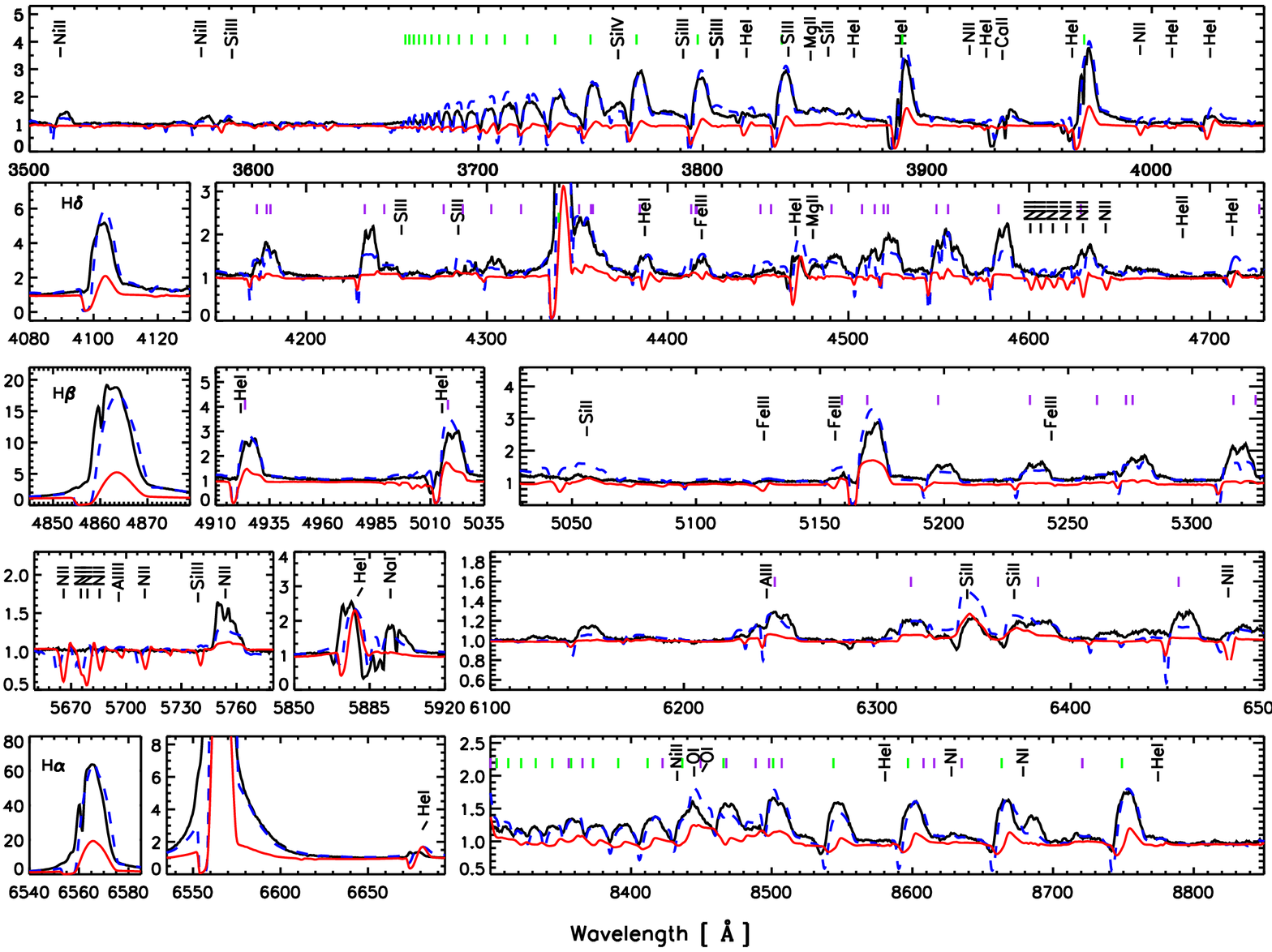}
\caption{Comparison of the spectrum of \ec\ observed with \emph{HST}/STIS in 2001 Apr 17 ($\phi = 10.603$; black solid line) with 1D \texttt{CMFGEN} models assuming $v_{\infty} = 420$~km~s$^{-1}$ with \mdota\  $= 8.5 \times 10^{-4} \ M_{\odot}$~yr$^{-1}$ (blue dashed line) and $2.5 \times 10^{-4} \ M_{\odot}$~yr$^{-1}$ (red solid line). The fluxes are continuum-normalized. The main spectral lines are identified, with green and purple ticks marking the laboratory wavelength of H and \altion{Fe}{ii} transitions, respectively. Details on the modeling approach can be found in \protect\citetalias{groh12a}. We note that a stellar core radius of $300$\rsun\ was used for the $2.5 \times 10^{-4} \ M_{\odot}$~yr$^{-1}$ model (versus $60$\rsun\ when \mdota\ $= 8.5 \times 10^{-4} \ M_{\odot}$~yr$^{-1}$) in order to keep the \altion{He}{i} $\lambda 6678$ line contribution from \eca\ within the observed levels.}
\label{fig12}
\end{figure*}

\citet{mehner11,mehner12} show that between $\sim 1998$ and 2011, the strengths of various broad stellar-wind emission lines (e.g. H$\alpha$, H$\delta$, and blends of \altion{Fe}{ii}, [\altion{Fe}{ii}], \altion{Cr}{ii}, [\altion{Cr}{ii}]) for our direct view of the central source of \ec\ decreased by factors of $1.5 - 3$ relative to the continuum, while at the same time the line strengths for reflected spectra corresponding to an \eca\ stellar latitude of $\sim 75^{\circ}$ showed no major variations (figures~6 and 7 of \citealt{mehner12}). The terminal velocity was found to be similar at all stellar latitudes and relatively unchanged as well \citep{mehner11,mehner12}. The lack of any significant changes in \ec's reflected polar spectra poses a serious problem for the hypothesis that there has been a recent large global modification of \mdota. Spherically-symmetric 1D radiative transfer models of \ec\ should be representative of the stellar spectrum at high latitudes, which are mostly unaffected by the WWC cavity and \ecb\ radiation field \citepalias{groh12b}. To illustrate expected changes, we computed a 1D \texttt{CMFGEN} model similar to those described in \citetalias{hillier01, hillier06, groh12a}, but with \mdota\ decreased to $2.5 \times 10^{-4}$~\msun~yr$^{-1}$ and $R_{\star, \eta_{\mathrm{A}}}$ increased to 300~\rsun. This model shows that a factor of $3 - 4$ decrease in \mdota\ should produce obvious changes in the reflected polar spectra (Figure~\ref{fig12}). The H$\delta$ emission should decrease by a factor of $\sim 2.5$, while H$\alpha$ and H$\beta$ should show a decrease in emission of a factor of $3 - 4$. The broad \altion{Fe}{ii}/\altion{Cr}{ii} blend between $4570$ {\AA} and $4600$ {\AA} and other \altion{Fe}{ii} lines should practically vanish. The emission and absorption of numerous other lines, like \altion{He}{i}, would also be affected. Based on these results and the observed lack of any changes over the stellar poles, any recent decrease in \mdota\ was probably by a factor $\lesssim 2$. It is beyond the scope of this paper to provide a detailed fit to the post-2009 observations of \eca, since other stellar parameters could have changed in addition to \mdota. Such detailed analysis will be reported elsewhere (Groh et al., in preparation). Additional evidence for the lack of a recent large drop in \mdota\ comes from \emph{HST}/STIS measurements of the H$\alpha$ EW at Weigelt blobs C and D, which show a long-term decrease in emission strength of only $\sim 10\% - 20\%$ \citep{gull09, mehner12}. The continuum flux at Weigelt D also has not greatly changed, even though in LOS to \eca\ it has increased by a factor of ten since 1998 \citep{gull09, mehner12}.

\subsection{Effects in the UV}

\citetalias{groh12a} showed that the WWC cavity in \ec\ also has a big influence on the observed UV spectrum, which is strongly dominated by bound-bound transitions of \altion{Fe}{ii}. Spherically-symmetric radiative transfer models have difficulties fitting the observed UV spectrum, even at apastron \citepalias{hillier06,groh12a}. However, the presence of the WWC cavity, introduced in the context of a 2D model, reproduces well both the observed UV and optical spectra of \ec\ by reducing the optical depth of \altion{Fe}{ii} transitions in LOS to \eca\ \citepalias{groh12a}.

Given the extremely large UV-emitting region of \eca\ \citepalias{hillier06}, any large change in the half-opening angle of the WWC cavity (of $\gtrsim20^{\circ}$, G12a) will affect the observed UV spectrum. This is in addition to any other changes in the UV caused by a decrease in \mdota. Figure~11 of \citetalias{hillier06} shows that a factor of four drop in \mdota\ drastically reduces both the emission and absorption of lines between $1250$~{\AA} and $1450$~{\AA}. Note also the dramatic increase in emission strength (nearly a factor of four) of the lines between $1390$~{\AA} and $1410$~{\AA}. Detailed 2D models of \ec's UV spectrum for different \mdota\ and taking into account the WWC cavity are needed to better determine what changes are expected and can be observed.

\citet{mehner11} note that the depth of the minimum in the UV during the 2009 photometric event, as observed in the F250W and F350W filters on \emph{HST}, was approximately twice as deep as the minimum of the 2003.5 event (their figure~1), speculating that the increased depth of the 2009 event may be due to a sudden increase in \eca's mass-loss during periastron. However, it is unclear what would cause such enhanced mass-loss. One would normally expect a decrease in \mdota\ to reduce the size of the UV photosphere and make it hotter \citep{mehner11}. The hypothesized increase in mass-loss around periastron may instead be due to an increase in density in LOS caused by the wind of \eca\ that flows toward the observer during periastron passage. The primary wind between \eca\ and the observer at periastron may now be blocking a larger relative amount of flux in LOS, leading to an increase in the depth of the UV minimum. The increase in WWC opening angle will, however, also alter greatly the amount of UV flux that can escape from the hotter, inner regions of \eca's wind \citep{maduragroh12}. It is unclear how the UV flux in LOS would be affected by such a change, especially given the unusual circumstellar extinction. Therefore, it is difficult to determine whether a decrease in \mdota\ is consistent with the observed increased minimum-depth in the UV photometry. Nevertheless, the results of \citet{maduragroh12} show that any large change in WWC opening angle should affect the time-dependent directional UV illumination of circumbinary material in the system \citep[see also][]{smith04b}.

\subsection{The Spatially-Extended Forbidden Line Emission}

\citet{gull09} presented an analysis of \emph{HST}/STIS observations taken between 1998 and 2004, identifying spatially extended (up to $0.8''$), velocity-resolved forbidden emission lines from low- and high-ionization\footnote{Low- and high-ionization refer here to atomic species with ionization potentials (IPs) below and above the IP of hydrogen, 13.6 eV.} species. Using 3D SPH simulations, \citetalias{madura12} developed a 3D dynamical model for the broad [\altion{Fe}{iii}] emission described in \citet{gull09} and showed that the broad high-ionization forbidden emission arises in portions of the dense, extended WWC structures photoionized by \ecb. The first detailed spatial maps showing the time evolution of the broad [\altion{Fe}{iii}] emission \citep{gull11} provide strong support for the results of \citetalias{madura12}.

\begin{figure*}
\includegraphics[width=17.5cm]{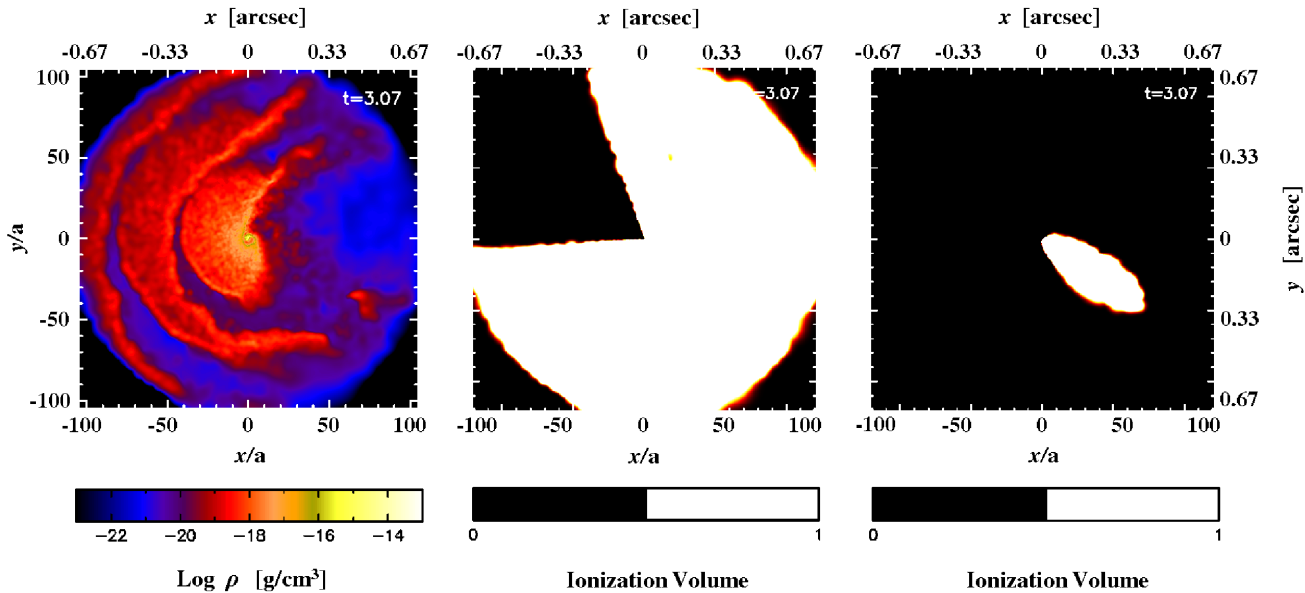}
\caption{Slices in the orbital plane from a \mdota~$= 2.5 \times 10^{-4}~M_{\odot}$~$\mathrm{yr}^{-1}$ 3D SPH simulation of \ec\ showing the density structure (left panel) and photoionization region created by \ecb\ (right two panels) at $\phi = 3.07$, assuming H$^{0}$ ionizing fluxes of $10^{49.48}$~photons~s$^{-1}$ (middle panel) and $10^{48.2}$~photons~s$^{-1}$ (right panel). The method for computing the photoionization region is the same as that in \protect\citetalias{madura12}, to which we refer the reader for details.}
\label{fig13}
\end{figure*}

\citetalias{madura12} found that the observed broad forbidden line emission depends strongly on \mdota\ and the ionizing flux from \ecb. If the flux from \ecb\ remains constant, but \mdota\ drops by a factor of two or more, the photoionization region created by \ecb\ should grow considerably (Figure~\ref{fig13}). The spatial extent and flux of the observed high-ionization forbidden lines should thus change. The emissivity should decrease with \mdota\ since, for densities much less than the line's critical density, the emissivity scales with the density squared \citepalias{madura12}. The phase dependence should differ as well, with the high-ionization emission vanishing at later phases (compared to earlier orbital cycles) when going into periastron, and reappearing at earlier phases afterward. Changes in the WWC opening angle will result in the broad forbidden line emitting regions exhibiting different Doppler velocities, leading to observed variations with slit position angle and spatial location on the sky. New components at different spatial locations and Doppler velocities may appear, while components identified in earlier observations might vanish.

The ring-like [\altion{Fe}{iii}] emission structures identified by \citet{gull09} and modeled by \citetalias{madura12} should disappear in future observations if \mdota\ drops by a factor of two or more. This is because the dense arcs of \eca\ wind where the broad high-ionization forbidden lines form disperse sooner after periastron the lower the value of \mdota\ (see Section~\ref{R100}). The flux, location, and velocity of broad, red-shifted, lower-ionization lines of [\altion{Fe}{ii}] that form in the periastron side of \eca's wind (\citealt{gull09}; Teodoro et al. 2013) should also change if \mdota\ drops. The flux should decrease due to the lower wind density, while the spatial distance from \eca\ and overall outward velocity of the emitting material should increase due to the increased momentum of \ecb's wind relative to that of \eca.

During periastron passage, the high-ionization forbidden line emission vanishes, both the broad features in the interacting stellar winds and the narrow features in the Weigelt blobs \citep{verner05, gull09, mehner10}. \citet{gull09} found that the high-ionization forbidden emission was absent during the 2003.5 event between $\phi = 1.0$ and $1.068$, but reappeared strongly by $\phi = 1.122$\footnote{These are spectroscopic phases, which are not necessarily equal to the orbital phase. The two are not expected to differ by more than $4 - 6$ weeks.}. Thus, there is an effective shutting-off of the UV flux from \ecb\ as it becomes temporarily enshrouded in the dense, extended wind of \eca\ around periastron. Therefore, during the 2003.5 event, \mdota\ was large enough to significantly reduce the size of the photoionization region created by \ecb, from a few thousand au across to just several au wide. An \mdota~$\sim 10^{-3} \ M_{\odot}$~yr$^{-1}$ is consistent with this result, assuming \ecb\ is an O5 giant with $T_{\mathrm{eff}} \approx 40,000$~K and an H$^{0}$ ionizing flux of $10^{49.48}$~photons~s$^{-1}$ (\citealt{verner05,martins05,mehner10,madura10}; \citetalias{madura12}). A factor of four drop in \mdota\ would create a much larger photoionization region during periastron passage (e.g. middle panel of Figure~\ref{fig13}), resulting in large amounts of high-ionization forbidden line emission.

In order to avoid a large photoionization region at $\phi \approx 1.068$ with such a low \mdota, one must drastically reduce the number of ionizing photons from \ecb, by a factor of $\sim 20$ (right panel of Figure~\ref{fig13}). This has important implications for the spectral type of \ecb. According to \citet{martins05}, there are no O-type giants or supergiants with hydrogen ionizing fluxes near $10^{48.2}$~photons~s$^{-1}$. Rather, an O8.5 $ - $ O9 main sequence star would be required, with a $T_{\mathrm{eff}} \approx 32,000$~K $-~34,000$~K and $\log L/L_{\odot} \approx 4.75 - 4.85$ \citep{martins05}. However, such a star lies outside the range of acceptable candidates for \ecb\ from both the works of \citet{verner05} and \citet{mehner10} (see their figure~12). Thus, a large drop in \mdota\ should reveal itself in the broad forbidden line emission during \ec's next periastron passage in 2014. If not, the implication is either (1) \mdota\ has not dropped significantly since 2004 or (2) our current understanding of \ecb\ is incorrect and the star has a lower effective temperature and hydrogen ionizing flux.

Another major issue regarding the forbidden line emission concerns the population of available Fe$^{+}$ to be photoionized to Fe$^{2+}$ by \ecb\ so that [Fe~{\scriptsize III}] emission can form. It is generally accepted that the energy required to ionize Fe$^{+}$ to Fe$^{2+}$ comes from photons provided by \ecb\ \citep{verner05, gull09, mehner10}. Thus, the iron in any [Fe~{\scriptsize III}] emitting region needs to initially be in the Fe$^{+}$ state. \citetalias{hillier01, hillier06} show that the ionization structure of \eca's wind and the amount of Fe$^{+}$ present depend strongly on charge exchange processes between Fe$^{2+}$ and neutral H via the reaction Fe$^{2+}$ + H $\leftrightarrow$ Fe$^{+}$ + H$^{+}$. According to figure~9 of \citetalias{hillier01}, which assumes \mdota~$= 10^{-3}~M_{\odot}$~$\mathrm{yr}^{-1}$, Fe$^{+}$ is the dominant ionization state of iron for radii $r \gtrsim 50$~au in \eca's wind. Below this, higher ionization states dominate. However, in \texttt{CMFGEN} models of \eca\ that assume \mdota~$= 2.5 \times 10^{-4}~M_{\odot}$~$\mathrm{yr}^{-1}$ (with the other parameters held fixed), H recombination never occurs and the iron is primarily Fe$^{2+}$ in the outer wind. \citetalias{hillier01, hillier06} discuss this issue and use it as a reason for favoring a higher \mdota, as it provides a better fit to the observed spectrum.

Thus, if \mdota\ were to drop to $\sim 2.5 \times 10^{-4}~M_{\odot}$~$\mathrm{yr}^{-1}$, Fe$^{2+}$ would be the dominant ionization state of iron in the outer \eca\ wind. \ecb\ would then be unnecessary for [Fe~{\scriptsize III}] emission to form, provided the \eca\ wind is in the right density and temperature regimes and has not recombined to Fe$^{+}$ or been excited to higher states. This has a number of major ramifications. First, while a hot \ecb\ would not be necessary for [Fe~{\scriptsize III}] emission, \ecb's UV flux would be necessary for forming the other observed higher ionization lines of [Ne~{\scriptsize III}] and [Ar~{\scriptsize III}] \citep{verner05,gull09}. A large drop in \mdota\ would hence lead to huge differences between the wind structures observed in lines like [Ar~{\scriptsize III}] and those of [Fe~{\scriptsize III}].

Secondly, if Fe$^{2+}$ became the dominant state of iron, then broad, \emph{red-shifted} [Fe~{\scriptsize III}] emission would become observable in most directions for most of the binary orbit, arising in the periastron side of \eca's wind and portions of the WWC region that are moving away from the observer. More importantly, the broad [Fe~{\scriptsize III}] emission would not completely vanish during periastron passage. The sudden appearance of large amounts of [Fe~{\scriptsize II}] emission during periastron passage, as previously observed \citep{gull09}, would also not occur. If the photoionization region extended far enough, the narrow high-ionization lines that arise in the Weigelt blobs and which also vanish during periastron would remain visible as well.

It is conceivable that in such a low \mdota\ situation, the Fe$^{2+}$ in \eca\ wind that gets compressed in the WWC regions will recombine to Fe$^{+}$. Such material on the \ecb\ side of the system could then be photoionized to reform Fe$^{2+}$ and allow [Fe~{\scriptsize III}] emission. Iron in the dense shells of compressed primary wind on the periastron side of \eca\ created during periastron passage could also recombine to Fe$^{+}$. As these shells would be shielded from \ecb's radiation, the gas in them would remain in the Fe$^{+}$ state and produce broad red-shifted [Fe~{\scriptsize II}] emission. However, the uncompressed \eca\ wind would still consist of Fe$^{2+}$ and exhibit significant amounts of broad red-shifted [Fe~{\scriptsize III}] emission. This emission would appear at nearly all phases in directions to the southeast on the sky at Doppler velocities up to $v_{\infty, \eta_{\mathrm{A}}}$. The latest \emph{HST} observations show no such emission \citep{gull11}. However, future \emph{HST} observations of the forbidden lines could help constrain any new changes to \mdota.  A major benefit of studying the extended, broad forbidden line structures is that they provide an excellent record of \ec's mass-loss over the past $2 - 3$ orbital cycles (\citealt{gull09,gull11}; \citetalias{madura12}; \citealt{teodoro13}).

\subsection{Other Expected Changes}

There are other observable changes expected from a large decrease in \mdota\ that are worth briefly mentioning. First, a drop in \mdota\ could affect any periodic dust formation during periastron passage, as mentioned in \citet{smith10}. A large decrease in \mdota\ would decrease the density in the compressed WWC zone where dust is expected to form, leading to a decrease in the near-IR flux and/or changes to the spectral energy distribution. An overall decrease in primary wind density and WWC-region density could also allow the intense radiation from the stars to more easily destroy any dust that forms. Older dust that lies in extended regions farther from the stars could also be destroyed or heated if intense radiation flowing through the WWC cavity can reach it. This may help explain the recent decrease in extinction and increase in brightness of the central source in LOS, e.g. an increase in WWC opening angle that allows more radiation to escape in LOS and destroy dust between us and the stars, thereby altering the extinction in LOS.

The wavelength-dependent apparent size of \eca\ is also controlled by the wavelength- and radially-dependent opacity and emissivity, which depend on \mdota\ \citepalias{hillier01,hillier06,groh12a}. As \mdota\ drops, the continuum extension of \eca\ in the optical and near-IR should drop as well. In other words, \eca\ should appear smaller than it was when observed between 2003 and 2004 \citep{vanboekel03, weigelt07, groh10a}. Continuum flux changes at a given wavelength that occur when the apex of the WWC cavity penetrates the surface of optical depth $\tau = 2/3$ \citep[e.g. the `bore-hole' effect, see][]{maduraowocki10, madura10} would also be affected, changing the behavior of the optical and near-IR photometric eclipse-like events observed to occur every 5.54-years around periastron \citep{whitelock04, eduardo10, madura10}. A reduced \mdota\ should cause the eclipse-like events to start later due to the longer time it takes for the WWC cavity to reach optical depth $\tau = 2/3$ in the wind photosphere of \eca. Similarly, the events should end earlier as the apex recedes from the location of optical depth $\tau = 2/3$. For low enough \mdota\ ($\lesssim 10^{-4}~M_{\odot}$~$\mathrm{yr}^{-1}$) the WWC cavity would never penetrate \eca's optically-thick wind photosphere and the photometric eclipse-like events would very likely cease. As of the writing of this paper, no significant decrease in the apparent continuum extension of \eca\ in the near-IR has been observed. No major differences are seen between the 2003.5 and 2009 events in the ground-based optical photometry either \citep{eduardo10}.

\section{Summary and Conclusions} \label{summ}

Using 3D SPH simulations, we showed that \mdota\ has profound effects on the 3D time-dependent hydrodynamics of \ec's binary colliding winds on a wide range of spatial scales. These simulations lay the foundation for future work generating synthetic data that can be compared directly to past and future observations of \ec. Such modeling provides information that is important for helping constrain \ec's recent mass-loss history and possible future state.

The 3D SPH simulations and 1D \texttt{CMFGEN} models of \eca's spectra show that for the range of explored values, a factor of two or more drop in \mdota\ results in substantial changes to numerous observables across a wide range of wavelengths. Below we summarize the most important changes expected.

\begin{enumerate}[leftmargin=*, label=\arabic*.]
  \item For \mdota~$\sim 10^{-3}~M_{\odot}$~$\mathrm{yr}^{-1}$, our LOS intersects portions of the dense, outer WWC region. For lower \mdota\ (by factors of two or more), our LOS passes primarily through the wind cavity carved by \ecb. An increase in WWC opening angle may thus explain the observed brightening of the system and changes to various wind-emission line profiles seen in LOS, with little to no changes at high stellar latitudes or at the Weigelt blobs.
  \item Increases in WWC opening angle cause lines-of-sight to the WWC apex to move into \eca's dense wind at later $\phi$ as the system goes into periastron, and to move out of \eca's wind at earlier $\phi$ as the system leaves periastron. Combined with the decreased cooling-transition phase of \ecb's post-shock wind, the net result is a shorter X-ray minimum (by $\sim 40 - 80$~days) during periastron passage for simulation Cases~B and C, versus Case~A.
  \item A decrease in \mdota\ leads to a corresponding drop in X-ray flux at phases far from periastron. Decreases in \mdota\ also propagate into \ec's extended interacting winds and may change the thermalization and/or absorption of the CCE component.
  \item Increases in WWC opening angle cause the trailing arm of hot post-shock \ecb\ gas to cross the observer's LOS at later $\phi$ as the system approaches periastron, delaying when the strongest observed X-ray lines become increasingly blueshifted and broader.
  \item As \mdota\ is lowered, the spatial extent of the He$^{+}$ zone in \eca's dense inner wind increases considerably, altering the timing of the emission peaks observed in \altion{He}{ii} $\lambda 4686$ around periastron. Assuming an initial $\mdota = 10^{-3}~M_{\odot}$~$\mathrm{yr}^{-1}$, a factor of four drop should lead to the detection of significant \altion{He}{ii} emission at all orbital phases. A factor of two drop should change the timing of the \altion{He}{ii} event, with the first peak occurring $\sim 3$~months earlier and the second peak lasting $\sim 2 - 3$~months longer.
  \item An increase in WWC opening angle will greatly affect broad wind-emission features of H$\alpha$, H$\beta$, and \altion{Fe}{ii} lines. The amount of Balmer P-Cygni absorption in LOS will be reduced and possibly entirely absent. The latitudinal and azimuthal variation of the Balmer P-Cygni absorption will also change. For $\mdota \lesssim 2.5 \times 10^{-4}~M_{\odot}$~$\mathrm{yr}^{-1}$, there could be little to no absorption in reflected spectra over the stellar poles of \eca.
  \item The observed increase in Balmer P-Cygni absorption in LOS during \ec's spectroscopic events should decrease for decreased \mdota. The density and amount of primary wind that flows toward the observer during periastron passage will also decrease. For $\mdota \lesssim 2.5 \times 10^{-4}~M_{\odot}$~$\mathrm{yr}^{-1}$, the absorption in LOS during the events should vanish since \eca\ will be able to easily ionize hydrogen throughout its extended wind.
  \item A decrease in \mdota\ by a factor of $2 - 4$ should decrease the emission strength of H$\alpha$, H$\beta$, and H$\delta$ by a factor of $2 - 3$. The broad Fe II/Cr II blend between 4570{\AA} and 4600{\AA} and other Fe II lines should practically vanish. Such changes should be visible not only in LOS, but also at high stellar latitudes and in other spatial directions, such as at the Weigelt blobs.
  \item Given the extremely large UV-emitting region of \eca, large changes in WWC half-opening angle of $\sim 20^{\circ}$ will alter the observed UV spectrum. Detailed 2D radiative transfer models of the resulting spectrum are needed, but current 1D models show that a factor of four drop in \mdota\ reduces both the emission and absorption of lines between 1250{\AA} and 1450{\AA}, and increases the emission strength of lines between 1390{\AA} and 1410{\AA}.
  \item The observed broad forbidden line emission depends strongly on \mdota\ and the ionizing flux from \ecb. As \mdota\ drops, the size of the photoionization region where the highest ionization forbidden lines form grows considerably. The size, location, flux, and time variability of the various forbidden lines will thus change. Ring-like emission structures that form in the dense arcs of primary wind on the apastron side of the system will be less structured and could vanish altogether since the arcs disintegrate sooner after periastron for lower \mdota.
  \item A decreased \eca\ wind density due to a large drop in \mdota\ may lead to the inability of \eca's wind to trap the ionizing radiation from \ecb\ at periastron. This would cause a drastic change in the behavior of the low- and high-ionization forbidden emission lines during periastron passage. The high-ionization lines would no longer vanish at periastron, while the low-ionization lines would fail to appear. This behavior could occur in both the extended wind structures of \ec\ and the Weigelt blobs.
  \item For $\mdota \lesssim 2.5 \times 10^{-4}~M_{\odot}$~$\mathrm{yr}^{-1}$, Fe$^{2+}$ becomes the dominant ionization state of iron at large radii in \eca's wind, resulting in large amounts of broad, red-shifted [\altion{Fe}{iii}] emission in most directions over most of the orbital cycle.
  \item A large decrease in \mdota\ could affect periodic dust formation during periastron passage since it would decrease the density in the compressed WWC zone and increase the likelihood of heating and/or destruction by the intense stellar radiation fields.
  \item The wavelength-dependent apparent size of \eca\ in the optical and near-IR continuum decreases with \mdota. Continuum flux changes that occur when the WWC apex penetrates the surface of high optical depth in \eca's wind photosphere would be affected, making the periodic photometric eclipse-like events shorter due to the smaller continuum size of \eca. For $\mdota \lesssim 10^{-4}~M_{\odot}$~$\mathrm{yr}^{-1}$, the eclipse-like events would likely cease.
\end{enumerate}

Given all of the multi-wavelength changes expected, a recent factor of two or more decrease in \mdota\ should be discernable with the right observations. However, as discussed in the previous section, there are a large number of inconsistencies between what is observed and what numerous models predict should occur if \mdota\ were to drop by a factor of $2 - 4$ from its historically accepted value of $\sim 10^{-3}~M_{\odot}$~$\mathrm{yr}^{-1}$. One of the most problematic is the lack of any observed changes in the broad wind-emission lines at high stellar latitudes or different spatial locations \citep{mehner12}. Also puzzling is the fact that the timing of the start of the 2009 spectroscopic event in X-rays, \altion{He}{ii} $\lambda 4686$, and the ground-based optical photometry was not delayed by any noticeable amount. It furthermore seems odd that \mdota\ would change with no corresponding change in wind terminal speed. Taken together with the other unobserved but expected changes, it appears that contrary to earlier claims, there has not been a recent significant drop in \mdota. Current evidence suggests that any decrease in \mdota\ was likely by a factor $\lesssim 2$ and occurred sometime after 2004, quite possibly during the 2009 event itself.

We speculate that most of the observed recent changes in \ec\ are due to a slight ($\lesssim10^{\circ}$) increase in the half-opening angle of the WWC zone, which altered the amount of extinction in LOS to the central source. This would explain why the continuum flux and broad wind-emission features in LOS have changed, while features at high latitudes and in other directions (e.g. at the Weigelt blobs) remain relatively constant. It would also explain why the start of the 2009 event at multiple wavelengths did not change. A modest decrease in \mdota\ by a factor $\lesssim 2$ sometime after 2004 could be responsible, and would hardly be surprising given \ec's nature. However, changes in \ecb's wind/stellar parameters, while less likely, cannot be fully ruled out at this time. Other less likely alternatives to a drop in \mdota\ are discussed in \citet{mehner10,mehner12}.

Finally, we note that the results of Sections~\ref{collapse} and \ref{radbraking} depend on rather complicated shock and radiative-wind-driving physics. In addition to the known effects of radiative inhibition and braking, there is the rather new phenomenon of so-called self-regulated shocks (SRSs), in which ionizing X-rays from the WWC shocks inhibit the wind acceleration of one or both stars, leading to lower pre-shock velocities and lower shocked plasma temperatures \citep{parkin13}. For most of \ec's orbit, the SRS effect is likely to be negligible due to the large binary separation. However, during periastron passage, SRSs may further inhibit the acceleration of \ecb's wind, helping trigger a cooling-transition phase of \ecb's post-shock wind and a possible WWC `collapse'. SRSs may cause the cooling-transition phase to occur earlier and/or last longer compared to what our simulations predict. The potential for a WWC collapse onto \ecb\ in the context of SRSs is difficult to determine without a detailed time-dependent model since radiative braking (not included in the analysis of \citealt{parkin13}) may prevail over the SRS effect and prevent a wind-photosphere collision \citep{parkin13}. Inclusion of SRSs in future models of \ec\ is thus worthwhile and may further elucidate the peculiar behavior of X-rays, \altion{He}{ii}~$\lambda 4686$, and other lines around periastron.

Many recent works suggest that very massive stars experience extreme \ec-like mass ejections of $\sim 1 - 10$~\msun\ in the year to decade before they explode as Type~II SNe \citep{kotak06, smith07, ofek07, ofek10, ofek13, quataert12, chevalier12}. The narrow emission lines indicative of Type~IIn SNe are interpreted as signatures of the SN ejecta interacting with this freshly expelled circumstellar material \citep{smith07, ofek07, ofek13}. In fact, the most likely detection of a progenitor before its explosion as a Type~IIn involved an LBV \citep{galyam09, ofek13}, thus raising the question of \ec's near-term fate. \ec's numerous massive eruptions point strongly to repeated phases of instability and may be an indication of its imminent demise as a powerful Type~IIn SN \citep{smith07, stritzinger12}. Constraining \ec's recent mass-loss history is therefore crucial for determining its possible near- and long-term states. The 3D simulations and analysis in this paper provide detailed information that is important for helping achieve this goal. Extensive monitoring of \ec's next periastron passage in 2014 will undoubtedly play a pivotal role as well.


\section*{Acknowledgements}
T.~I.~M. is supported by an appointment to the NASA Postdoctoral Program at the Goddard Space Flight Center, administered by Oak Ridge Associated Universities through a contract with NASA. M.~T. is supported by CNPq/MCT-Brazil through grant 201978/2012-1.

\clearpage

\appendix

\section{An anti-gravity approach for implementing radiative forces in SPH simulations of colliding stellar winds}\label{appenda}

\subsection{Basic approach}\label{appa1}

The SPH method used for the simulations in this paper follows the Lagrangian motion of the modeled gas particles, removing the need for a complex spatial grid. In the SPH scheme, spatial derivatives are calculated via analytical differentiation of interpolation formulae \citep{monaghan92}, as opposed to Eulerian grid-based methods where derivatives are computed using finite-difference or finite-volume techniques. While the SPH approach allows us to perform large-scale 3D simulations with significantly less computational cost compared to a grid-based code, the lack of a spatial grid complicates the calculation of the projected velocity gradient required for computing the CAK \citep{castor75} line force that drives the individual stellar winds. We therefore employ an alternative approach, described below, that produces a stellar wind that follows the desired beta-velocity law and is consistent with the results of standard CAK theory.

We start by considering the radial equation of motion for a steady, spherically symmetric stellar wind driven by a radiation force $g_{\mathrm{rad}} = \kappa \lstar / (4\pi c r^{2})$ and competing against stellar gravity $G \mstar / r^{2}$~:

\begin{equation}
v \frac{dv}{dr} = g_{\mathrm{rad}} - \frac{G \mstar}{r^{2}} \ . \label{eqA1}
\end{equation}\\
Here, $v$ is the wind speed, $r$ is the local radius, \mstar\ and \lstar\ are the stellar mass and luminosity, $\kappa$ is the wind opacity, and $c$ is the speed of light. We neglect standard sound speed terms that are, in practice, negligible. Since the key to driving a stellar wind is overcoming gravity, it is convenient to define a dimensionless equation of motion that is scaled by the gravity,

\begin{equation}
\frac{dw}{dx} = \Gamma - 1 \ , \label{eqA2}
\end{equation}\\
where the dependent variable $w \equiv v^{2}/v_{\mathrm{esc}}^{2}$ is the ratio of the radial kinetic energy to the surface escape energy $v_{\mathrm{esc}}^{2}~=~2G\mstar/\rstar$, and the independent variable is the inverse radius coordinate $x~\equiv~1 - \rstar/r$, measured from the stellar surface radius \rstar. $\Gamma \equiv r^{2} g_{\mathrm{rad}} / (G \mstar)$ is the ratio of the local radiative force to gravity, sometimes called the ``Eddington parameter''.

A simple model to consider is one in which we assume that, at some \rstar, $g_{\mathrm{rad}}$ suddenly jumps from below gravity to some fixed factor $\Gamma > 1$ above gravity. In such a simple ``anti-gravity'' model, static material at the surface \rstar\ effectively ``falls'' away from the star with a local net outward acceleration at radius $r$ of $(\Gamma~-~1)G \mstar/r^{2}$. In terms of Equation~(\ref{eqA2}), integration yields a solution $w(x) = (\Gamma - 1)x$, which in dimensional form gives $v(r) = v_{\infty}(1-\rstar/r)^{1/2}$, where $v_{\infty} = v_{\mathrm{esc}} \sqrt{\Gamma - 1}$. This is just a specific $\beta = 1/2$ form of the beta-velocity law $v(r)~=~v_{\infty}(1-\rstar/r)^{\beta}$.

While we considered only a simple example above, this anti-gravity approach can readily give a beta-velocity law with any index $\beta$. To demonstrate this, we rewrite Equation~(\ref{eqA2}) in terms of $v(x)$ and require the Eddington parameter to have a spatial variation $\Gamma(x)$. Then,

\begin{equation}
\frac{2~v(x)}{v_{\mathrm{esc}}^2} \frac{d}{dx} v(x) = \Gamma(x) - 1 \ . \label{eqA3}
\end{equation}\\
With $v(x) = v_{\infty}x^{\beta}$ and defining $w_{\infty} \equiv v_{\infty}^{2}/v_{\mathrm{esc}}^2$, the above gives

\begin{equation}
\Gamma(x) = 2 \beta w_{\infty}~x^{2\beta - 1} + 1 \ . \label{eqA4}
\end{equation}\\
Obtaining the desired velocity laws simply requires specifying the correct opacity for each wind. Using Equation~(\ref{eqA4}) together with $\Gamma(r) = r^{2} g_{\mathrm{rad}}/(G \mstar) = \kappa \lstar/(4 \pi G \mstar c)$ yields,

\begin{equation}
\kappa_{i}(r_{i}) = \frac{4 \pi c}{L_{i}} \left[v_{\infty ,i}^{2} R_{i} \beta_{i} (1 - R_{i}/r_{i})^{2 \beta_{i} - 1} + G M_{i} \right] \ , \label{eqA5}
\end{equation}\\
where $i = 1$, $2$ correspond to the individual stars and $r_{i}$ is measured with respect to star $i$. Achieving the different wind terminal speeds requires $\kappa_{1} \neq \kappa_{2}$. However, regions where the two winds mix requires weighting the local opacity by the relative mass contribution from each star,

\begin{equation}
\bar{\kappa} = \kappa_{1} \left(\frac{\rho_{1}}{\rho} \right) +  \kappa_{2} \left(\frac{\rho_{2}}{\rho} \right) = \kappa_{1} + (\kappa_{2} - \kappa_{1}) \frac{\rho_{2}}{\rho} \ , \label{eqA6}
\end{equation}\\
where $\rho = \rho_{1} + \rho_{2}$ is the total density.

In order to ensure that the stellar mass-loss rates correspond to what is expected from CAK theory, e.g.

\begin{equation}
\mdot_{\mathrm{CAK}} = \frac{\lstar}{c^{2}} \frac{\alpha}{1 - \alpha} \left[\frac{\bar{Q}\Gamma}{1 - \Gamma} \right]^{(1 - \alpha)/\alpha} \ \ \mathrm{(Gayley 1995)}, \label{eqA7}
\end{equation}\\
we require that the ratio of the two stellar luminosities satisfy

\begin{equation}
\frac{L_{2}}{L_{1}} = \left(\frac{M_{2}}{M_{1}} \right)^{1- \alpha} \left(\frac{\mdot_{2}}{\mdot_{1}} \right)^{\alpha} \ , \label{eqA8}
\end{equation}\\
where $\alpha$, typically $= 2/3$, is the CAK power index. Given an observed primary star luminosity $L_{1}$, the above equation can be used to find that of the secondary, $L_{2}$. Therefore, with $L_{1}$ and $M_{i}$, $R_{i}$, $\mdot_{i}$, $\beta_{i}$, and $v_{\infty, i}$ specified for each of the stars, the total force per mass $\vec{g}_{\mathrm{tot}}$ exerted on the wind particles can be computed as corrections to the vector gravities $\vec{g}_{i}$ of the stars,

\begin{equation}
\vec{g}_{\mathrm{tot}} = \left(1 - \frac{\bar{\kappa} L_{1}}{4 \pi G M_{1} c} \right)\vec{g}_{1} +  \left(1 - \frac{\bar{\kappa} L_{2}}{4 \pi G M_{2}~c} \right)\vec{g}_{2} \ . \label{eqA9}
\end{equation}\\

With the above formalism in place, we incorporate the effect of radiative forces into the 3D SPH simulations by:

\begin{enumerate}[leftmargin=*, label=\arabic*.]
  \item Specifying the input parameters $L_{1} = 5 \times 10^{6}$~\lsun\ and $M_{i}$, $R_{i}$, $\mdot_{i}$, $\beta_{i}$, and $v_{\infty, i}$ for each star (Table~\ref{tab1}).
  \item For the given $L_{1}$ and assumed stellar mass and mass-loss rate ratios, using Equation~(\ref{eqA8}) with $\alpha = 2/3$ to obtain $L_{2}$.
  \item Using Equation~(\ref{eqA5}) to compute $\kappa_{1}$ and $\kappa_{2}$. Particles from each star are distinguished via their specified distinct masses.
  \item Computing $\bar{\kappa}$ at each location in the wind using Equation~(\ref{eqA6}).
  \item Computing the associated $\vec{g}_{\mathrm{tot}}$ at each point with Equation~(\ref{eqA9}).
\end{enumerate}

\begin{figure}
\begin{center}
\includegraphics[width=8.4cm]{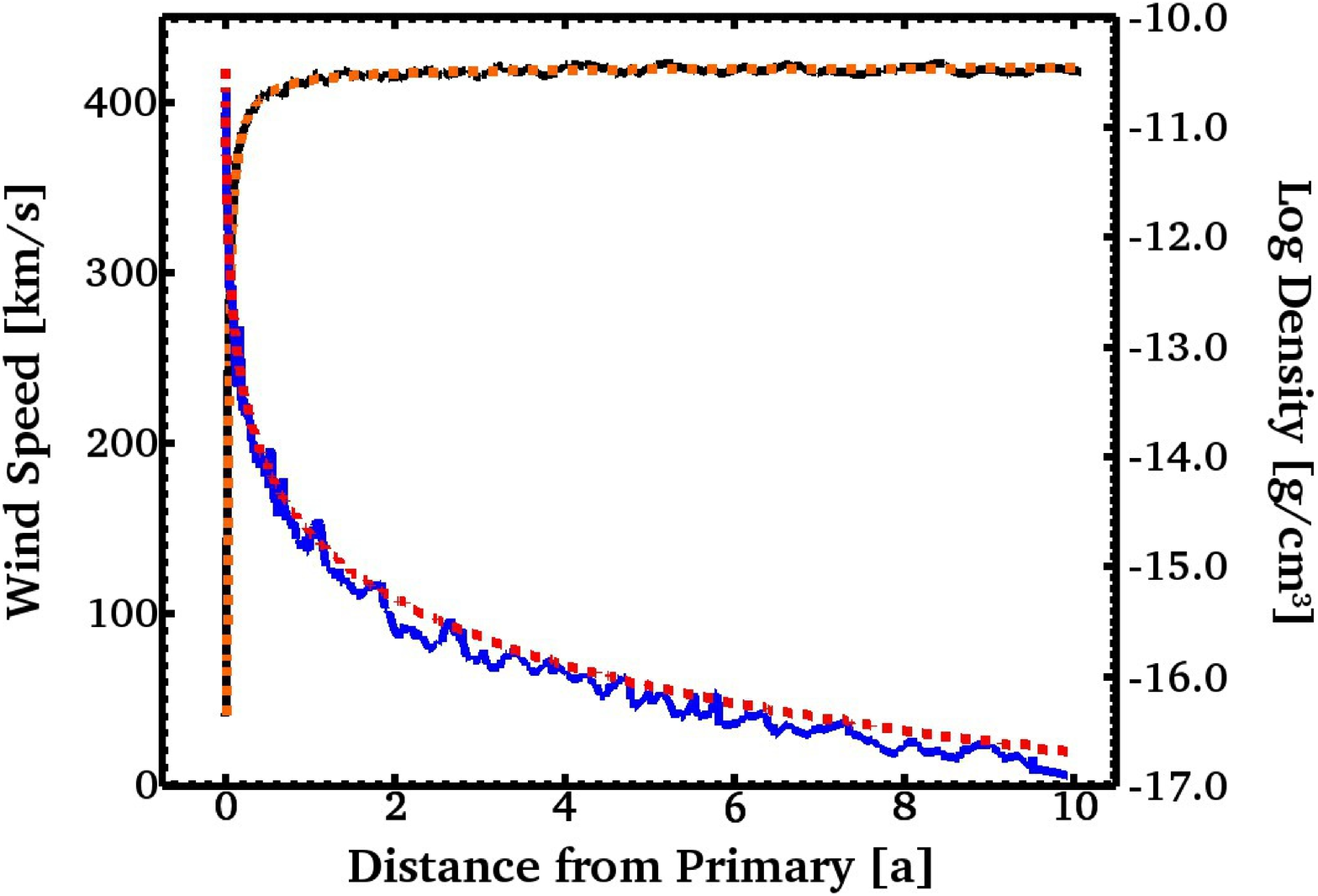}
\end{center}
\caption{Line plots of the total wind speed (orange and black) and log density (red and blue) as a function of distance from \eca\ (in units of $a = 15.45$~au) taken from the \mdota $= 8.5 \times 10^{-4} \ M_{\odot}$~yr$^{-1}$ 3D SPH simulation at apastron (solid lines) compared to the simple analytic expressions for $v(r)$ and $\rho(r)$ for an isolated star (dashed lines).}
\label{figA1}
\end{figure}

The stellar winds are initiated at twice the sound speed in order to ensure information characteristics that point from the star into the computational domain. The SPH particles from each star $i$ are injected at a radius $r_{\mathrm{inj},i} = R_{i}/[1 - (2c_{s}/v_{\infty,i})^{1/\beta_{i}}]$, where the sound speed $c_{s} = \sqrt{T\gamma k_{B}/\mu}$ and $T$, $\gamma$, $k_{B}$, and $\mu$ are the photospheric temperature, ratio of specific heats, Boltzmann's constant, and mean molecular weight, respectively.

\begin{figure}
\begin{center}
\includegraphics[width=8.0cm]{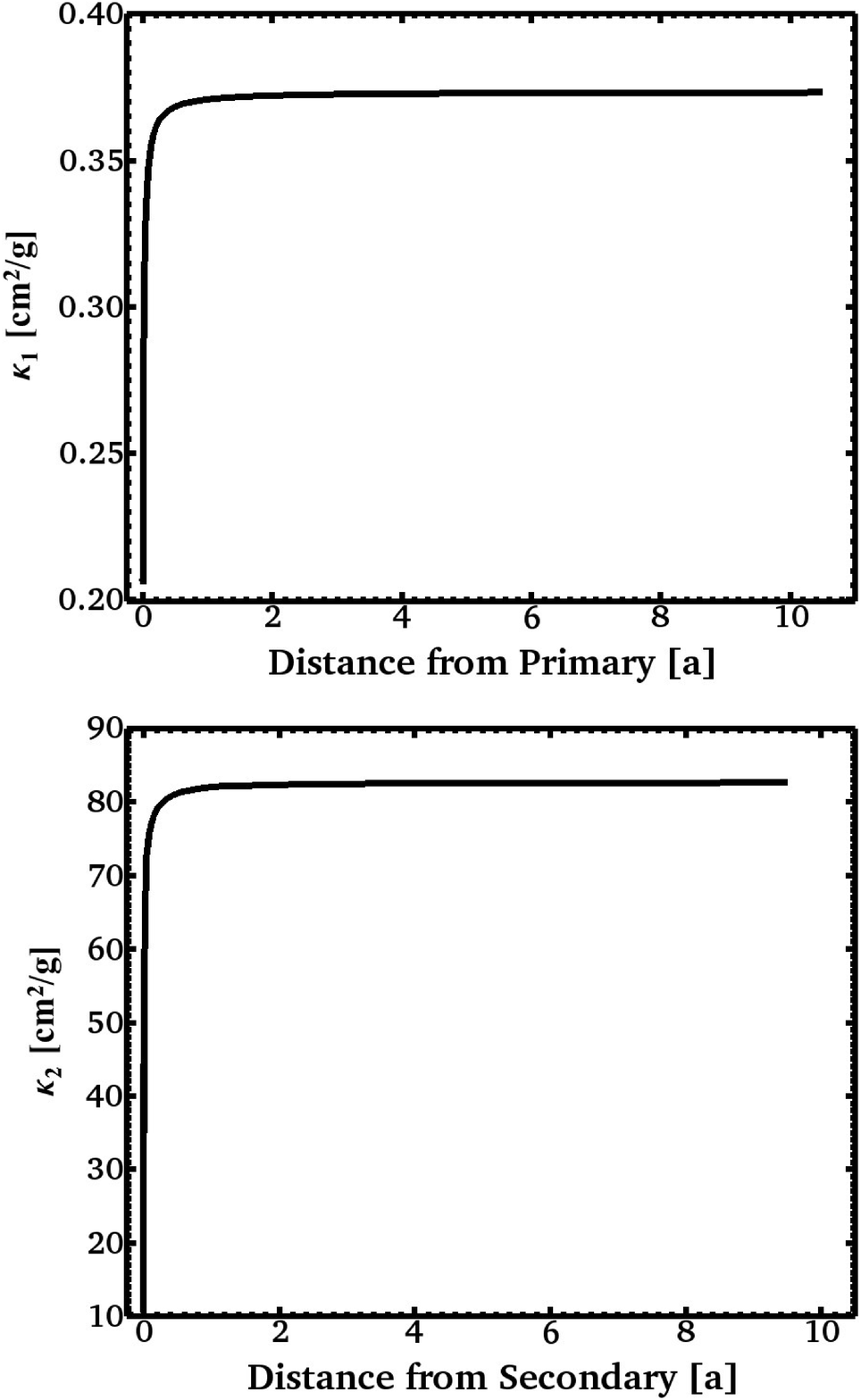}
\end{center}
\caption{Line plots of the opacity $\kappa_{i}$ (Eq.~\protect\ref{eqA5}) as a function of distance from the star (in units of $a = 15.45$~au) for \eca\ (top) and \ecb\ (bottom) for the \mdota~$= 8.5 \times 10^{-4} \ M_{\odot}$~yr$^{-1}$ simulation.}
\label{fig_kappas}
\end{figure}

Figure~\ref{figA1} shows that the anti-gravity method produces the correct wind-velocity and density structures. Plots of $v(r)$ and $\rho(r)$ through the back side of \eca's wind, taken at apastron from the \mdota$= 8.5 \times 10^{-4} \ M_{\odot}$~yr$^{-1}$ 3D SPH simulation, are compared to the analytic expressions $v(r)~=~v_{\infty}(1-\rstar/r)^{\beta}$ and $\rho(r) = \mdot / [4 \pi r^{2} v(r)]$ for the wind structure of an isolated star, assuming the values for \eca\ from Table~\ref{tab1}. The SPH simulation results are in very good agreement with the analytic scalings, which helps verify our approach and results. Figure~\ref{fig_kappas} shows line plots of the opacity $\kappa_{i}$ used to drive the individual stellar winds for the desired mass-loss rates and wind terminal speeds. Note that $\kappa_{2}$ is significantly larger than $\kappa_{1}$. More extensive testing and analysis of the anti-gravity method can be found in \citet{russell13}.

\subsection{An analytic solution for the flow along the line of centers}\label{appa2}

Since the radiative force from both stars is present throughout the computational domain, the velocity-altering effects of radiative inhibition (RI) and stellar gravity are included in the simulations \citep{stevens94,russell13}. To assist in the interpretation and understanding of the 3D SPH simulations, we derive here a sample analytic 1D solution for the effect of the anti-gravity radiative forces on the flow along the line of centers between the stars. Because the SPH simulations use $\beta = 1$ velocity laws to parameterize the stellar winds, we focus on the $\beta = 1$ case.

As \ecb\ has a much lower luminosity than \eca, its wind will be the one most strongly affected by the radiative force of its companion. We therefore investigate how \ecb's wind velocity along the line of centers is affected by the radiation and gravity of \eca. The equation of motion along the line of centers in terms of the radius $r$ measured from \ecb, for a given binary separation $D$, is

\begin{equation}
v_{2} \frac{dv_{2}}{dr} = \frac{G M_{1}}{(D - r)^{2}}\left(1 - \frac{\bar{\kappa} L_{1}}{4 \pi G M_{1} c} \right) - \frac{G M_{2}}{r^{2}} \left(1 - \frac{\bar{\kappa} L_{2}}{4 \pi G M_{2}~c} \right) \ , \label{eqA10}
\end{equation}\\
where subscripts $1$ and $2$ refer to \eca\ and \ecb, respectively. Because we are interested in the effect of \eca\ in slowing \ecb's wind before the WWC, we can set $\bar{\kappa} = \kappa_{2}$. Multiplying both sides of Equation~(\ref{eqA10}) by $r^{2}/(G M_{2})$ and using the definition of the Eddington parameter $\Gamma_{i} = \kappa_{i} L_{i}/(4\pi G M_{i} c)$, we get

\begin{equation}
\frac{v_{2} r^2}{G M_{2}} \frac{dv_{2}}{dr} = (\Gamma_{2} - 1) - \frac{r^2}{(D - r)^2} \left(\frac{M_{1}}{M_{2}} \right) \left[\frac{\kappa_{2} \Gamma_{1}}{\kappa_{1}} - 1 \right] \ . \label{eqA11}
\end{equation}\\
Defining $x \equiv 1 - R_{2}/r$, $w \equiv (v_{2}^{2}R_{2})/(2 G M_{2})$, $q \equiv L_{1}/L_{2}$, $m \equiv M_{1}/M_{2}$, and using the fact that $\kappa_{i} = (4\pi G M_{i} c \Gamma_{i} )/L_{i}$, Equation~(\ref{eqA11}) becomes

\begin{equation}
\frac{dw}{dx} = (\Gamma_{2} - 1) - \frac{q \Gamma_{2} - m}{[D(1-x)/R_{2} - 1]^2} \ . \label{eqA12}
\end{equation}\\
Equation~(\ref{eqA4}) gives the spatial variation of $\Gamma$ with $x$. Therefore, for $\beta = 1$ and $w_{\infty} \equiv (v_{\infty, 2}/v_{\mathrm{esc},2})^{2}$, $\Gamma_{2}(x) = 1 + 2w_{\infty} x$ and

\begin{equation}
\frac{dw}{dx} = 2w_{\infty}x - \frac{q (1 + 2w_{\infty}x) - m}{[D(1-x)/R_{2} - 1]^2} \ . \label{eqA13}
\end{equation}\

Equation~(\ref{eqA13}) can be integrated straightforwardly to give the variation of the scaled wind kinetic energy $w$ as a function of the scaled inverse radius $x$. Letting $d \equiv D/R_{2}$, we find

\begin{eqnarray}
w(x) & = & w_{\infty}x^{2} - \frac{2qw_{\infty}\ln[(1-x)d - 1]}{d^2} \nonumber\\ \nonumber\\
& & + \frac{qd - md - 2qw_{\infty} +2qdw_{\infty}}{d^{2} [1 + (x-1)d]} \ . \label{eqA14}
\end{eqnarray}\\
With $w/w_{\infty} = (v_{2}/v_{\infty,2})^{2}$, the radial variation of the velocity of \ecb's wind along the line of centers can be found from

\begin{eqnarray}
\frac{v_{2}^{2}}{v_{\infty, 2}^{2}} & = & \left(1 - \frac{R_{2}}{r}\right)^{2} - \frac{2qR_{2}^{2}}{D^{2}}\ln \left[\frac{D}{r} - 1 \right] + \frac{2qR_{2}(1 - \frac{R_{2}}{D})}{D(1 - D/r)} \nonumber\\ \nonumber\\
& & + \frac{2 G M_{2}(q - m)}{v_{\infty, 2}^{2} D (1 - D/r)} \ . \label{eqA15}
\end{eqnarray}\\
The corresponding expression for the radial variation of \eca's wind velocity $v_{1}$ along the line of centers is similarly

\begin{eqnarray}
\frac{v_{1}^{2}}{v_{\infty, 1}^{2}} & = & \left(1 - \frac{R_{1}}{r'}\right)^{2} - \frac{2q' R_{1}^{2}}{D^{2}}\ln \left[\frac{D}{r'} - 1 \right] + \frac{2q' R_{1}(1 - \frac{R_{1}}{D})}{D(1 - D/r')} \nonumber\\ \nonumber\\
& & + \frac{2 G M_{1}(q' - m')}{v_{\infty, 1}^{2} D (1 - D/r')} \ . \label{eqA15b}
\end{eqnarray}\\
where $q' = 1/q$, $m' = 1/m$, and $r'$ is measured from \eca.

In addition to the effects of RI, orbital motion will increase the wind speeds as the system moves from apastron to periastron. The stellar velocity along the line of centers for each star $i$ with respect to the system center of mass (COM) is $v_{r,i} = \sqrt{v_{\star}^{2} - v_{t,i}^{2}}$, where

\begin{eqnarray}
&v_{t,1} = h M_{2}/(M_{\mathrm{T}} D) \nonumber\\
&v_{t,2} = h M_{1}/(M_{\mathrm{T}} D) \ \label{eqA15c}
\end{eqnarray}\\
are the transverse components of the stellar velocities with respect to the COM, and $M_{\mathrm{T}} = M_{1} + M_{2}$, $h^{2} = G a M_{\mathrm{T}} (1 - e^2)$, $G$ is the gravitational constant, $a$ is the semimajor axis length, $e$ is the orbital eccentricity, and $v_{\star}$ is the stellar orbital speed given by

\begin{equation}
v_{\star}^{2} = G M_{\mathrm{T}} \left(\frac{2}{D} - \frac{1}{a} \right) \ . \label{eqA15d}
\end{equation}\

\begin{figure}
\begin{center}
\includegraphics[width=8.4cm]{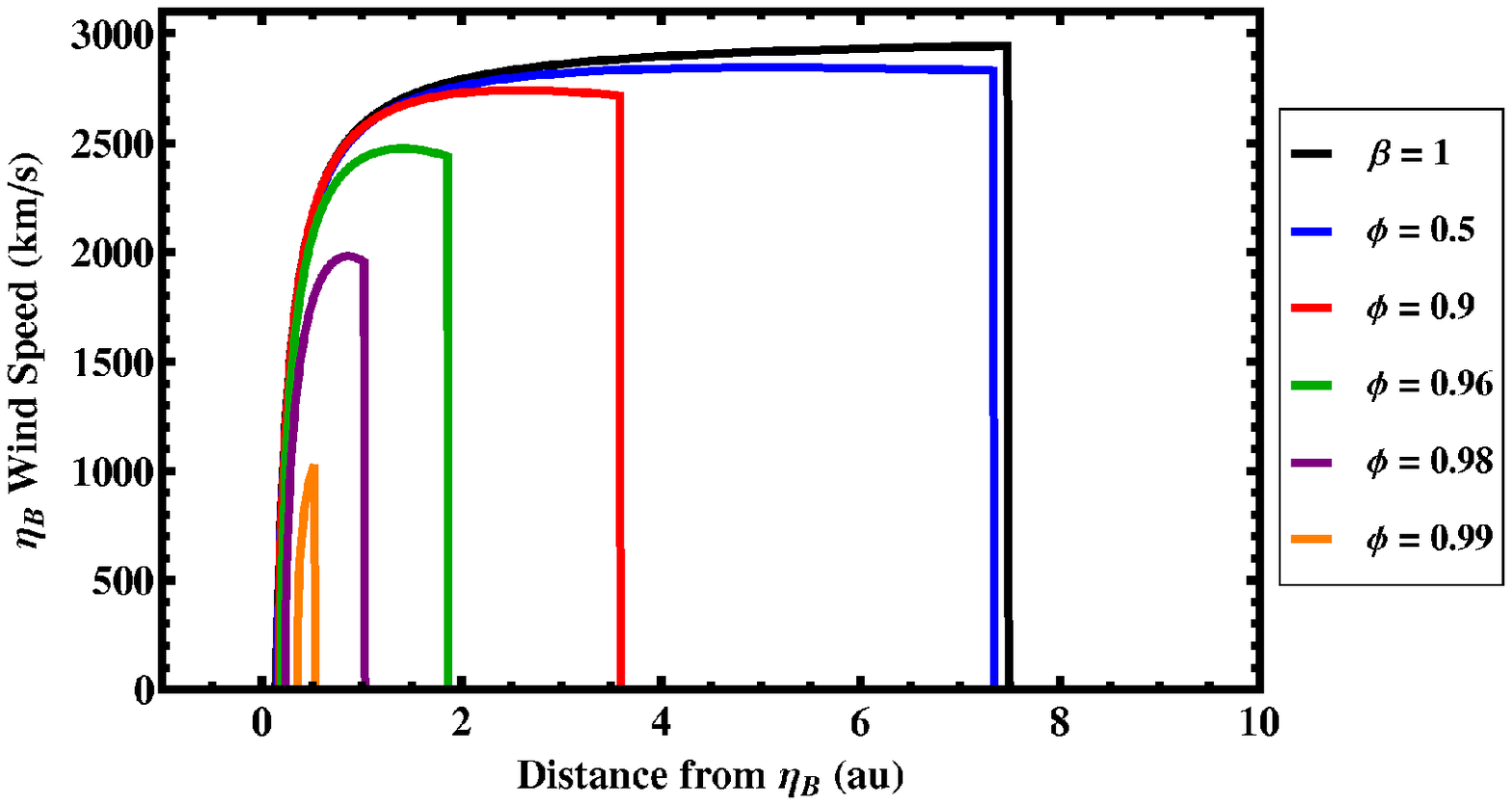}
\end{center}
\caption{Total \ecb\ wind speed $v_{2, \mathrm{tot}} = v_{2} + v_{r,2}$ along the line of centers as a function of distance from \ecb\ assuming the parameters in Table~1 and \mdota $= 8.5 \times 10^{-4} \ M_{\odot}$~yr$^{-1}$, $q = 22.28$, for binary separations $D$ corresponding to $\phi = 0.5$ (apastron), 0.9, 0.96, 0.98, and 0.99. For reference, the solid black line is the standard $\beta = 1$ law without RI effects. Each curve is truncated at the radius of ram pressure balance $r_{b}$ given by Equation~(\ref{eqA16}), indicated by the vertical lines. There is no stable balance at periastron ($\phi = 1$), implying that \ecb\ cannot drive a wind toward \eca.}
\label{figA2}
\end{figure}

Figure~\ref{figA2} plots the total \ecb\ wind speed along the line of centers $v_{2, \mathrm{tot}} = v_{2} + v_{r,2}$ versus $r$ for several $D$ corresponding to various phases leading up to periastron. Each curve is truncated at the radius of wind-wind ram pressure balance $r_{b}$ measured from \ecb\ (indicated by the vertical lines), which is found by solving the ram pressure balance condition

\begin{equation}
\frac{D}{r_{b}} = 1 + \sqrt{\frac{\dot{M}_{1} [v_{1}(r' = D - r_{b}) + v_{r,1}]}{\dot{M}_{2} [v_{2}(r = r_{b}) + v_{r,2}]}} \ , \label{eqA16}
\end{equation}\\
where $v_{2}(r = r_{b})$ and $v_{1}(r' = D - r_{b})$ are given by Equations~(\ref{eqA15}) and (\ref{eqA15b}). The standard $\beta = 1$ velocity-law profile for \ecb\ is included for comparison.

\begin{figure}
\begin{center}
\includegraphics[width=8.45cm]{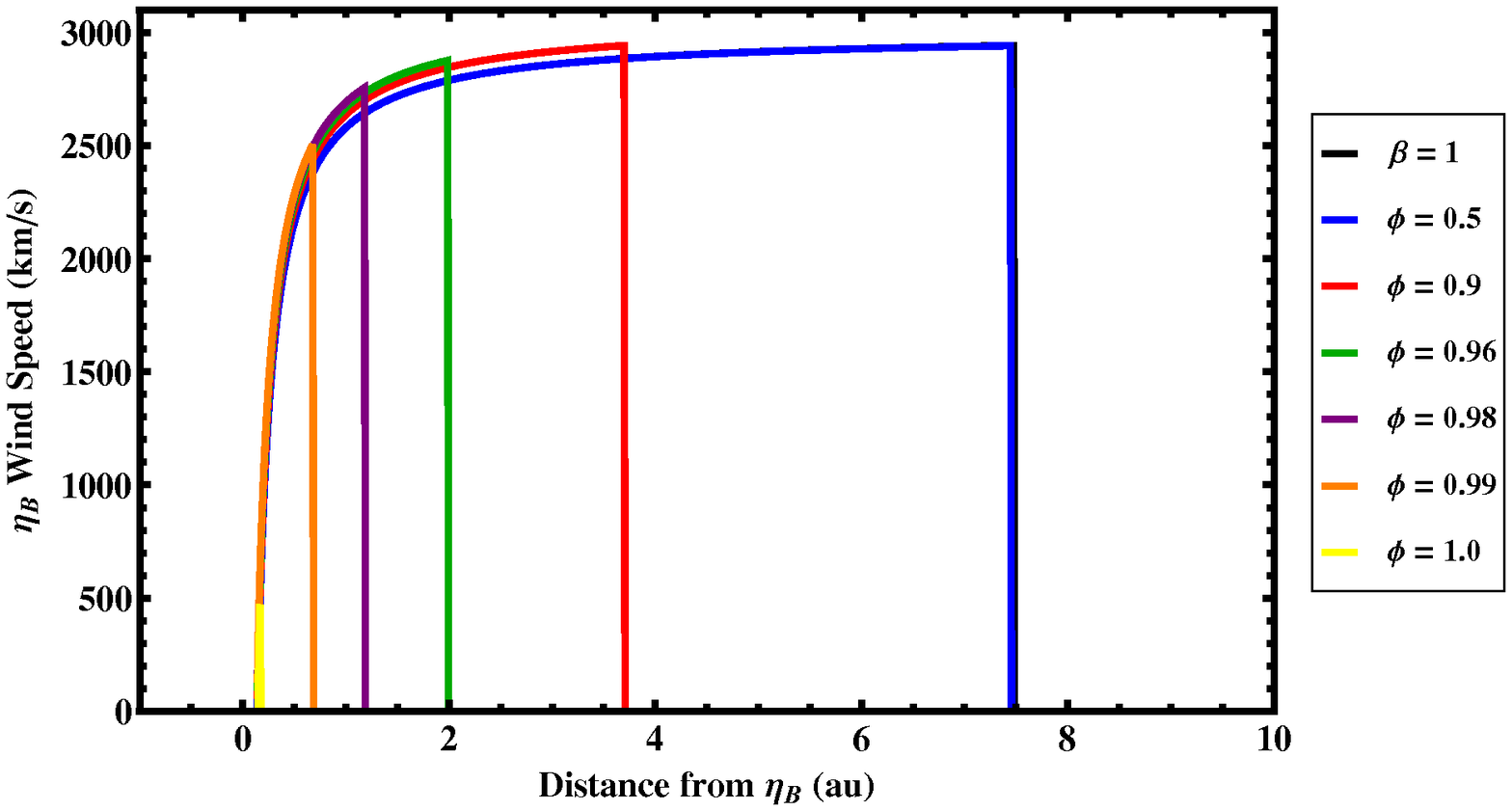}
\end{center}
\caption{Same as Figure~\protect\ref{figA2}, but using $\bar{\kappa} = \kappa_{1}$ for the influence of \eca's radiation field on \ecb's wind and including periastron ($\phi = 1.0$). The solid blue line for $\phi = 0.5$ overlaps almost exactly with the solid black line that corresponds to the standard $\beta = 1$ law without RI effects. Unlike Figure~\ref{figA2}, there is a stable ram balance at periastron.}
\label{fig_k1}
\end{figure}

Note how when the effects of RI are included, \ecb's wind is initially accelerated from the stellar surface, but then eventually decelerates as it approaches \eca. The offset position from \eca\ of the deceleration depends on the luminosity ratio $q$. This deceleration weakens the colliding wind shocks in the 3D SPH simulations and enhances the dominance of \eca's wind over that of \ecb, thus moving the WWC apex closer to \ecb\ and narrowing the opening angle of the WWC zone around \ecb. Because it has a strong dependence on the binary separation $D$, these velocity-altering effects are time-dependent and most pronounced near periastron.

We find there is no stable balance solution at $\phi = 1$, i.e. according to Equation~(\ref{eqA15}), \ecb\ cannot drive a wind along the line of centers toward \eca\ at periastron. This implies that in the absence of other effects, \eca's wind would collide with the surface of \ecb. However, because we set $\bar{\kappa} = \kappa_{2}$ in Equation~(\ref{eqA10}), there is a strong coupling between \eca's radiation field and \ecb's wind. To investigate the effect of a weaker coupling analogous to that used by \citetalias{parkin11}, we rederived Equation~(\ref{eqA15}) assuming $\bar{\kappa} = \kappa_{1}$ for the coupling between \eca's radiation and \ecb's wind. Figure~\ref{fig_k1} plots the resulting total \ecb\ wind speed along the line of centers. One notices immediately that the weaker coupling decreases greatly the importance of RI and leads to much higher pre-shock \ecb\ wind speeds. The higher pre-shock wind speeds reduce the importance of radiative cooling in the post-shock gas and there is no longer a `collapse' of the WWC zone at periastron, but rather a stable ram balance. This is the same result obtained by \citetalias{parkin11}, indicating that a strong coupling between \eca's radiation field and \ecb's wind is necessary for RI effects to be important.

\subsection{Possible changes to $\mathbf{\dot{M}}$}\label{appa3}

Because we fix the mass-loss rates in our anti-gravity approach, possible changes to the stellar mass-loss due to radiative inhibition are not included. Such changes though are not expected to be significant in \ec, with the exception of at periastron for \ecb.

As demonstrated in \citet{russell13}, the mass-loss rate factor $\dot{m}$ that describes the change in star 1's mass-loss rate due to the radiation and gravity of star 2 is,

\begin{eqnarray}
\dot{m} & = & \left[1 - \frac{L_{2}}{L_{1}} \left(\frac{R_{1}}{R_{2}} \right)^{2} \left(1 - \left[1 - \left(\frac{R_{2}}{D - R_{1}} \right)^{2} \right]^{1+\alpha} \right) \right]^{\frac{1}{\alpha}} \nonumber\\ \nonumber\\
& & \times \left[1 - \frac{M_{2}/M_{1}}{(D/R_{1} -1)^{2}} \right]^{1 - \frac{1}{\alpha}} \ , \label{eqA17}
\end{eqnarray}\\
Using the values in Table~\ref{tab1} and $\alpha = 2/3$, we find that the radiation and gravity of \ecb\ are insufficient to modify \eca's mass-loss rate at any phase, and so $\dot{m} = 1$ for the entire 5.54-year orbit and there is no difference in \mdota\ compared to the expected single-star values.

For most of \ec's cycle ($0.5 \leq \phi \leq 0.98$), $D$ is large enough that \eca\ does not modify \ecb's mass-loss and $\dot{m} \approx 1$. At $\phi = 0.99$, $\dot{m} \approx 0.87$ and there is a slight reduction in \ecb's \mdot. Only between $\phi = 0.996$ ($\dot{m} \approx 0.66$) and periastron ($\dot{m} \approx 0.52$) is there a significant reduction in the \mdot\ of \ecb. Since this occurs only for an extremely brief period at periastron, it should not greatly affect our results or conclusions. We further note that the above results assume $\mdota=8.5~\times~10^{-4}~M_{\odot}$~yr$^{-1}$ and $q = 22.28$ (Table~\ref{tab1}). Because our lower \mdota\ simulations use slightly larger \ecb\ luminosities, the reduction in \ecb's \mdot\ will be less, with $\dot{m} \approx 0.66$ and $0.79$ at periastron for $\mdota=4.8~\times~10^{-4}~M_{\odot}$~yr$^{-1}$ and $2.4~\times~10^{-4}~M_{\odot}$~yr$^{-1}$, respectively.

\section{The Structure of Eta Car's Binary Colliding Winds in the $\mathbf{xz}$ and $\mathbf{yz}$ Planes Perpendicular to the Orbital Plane}\label{appendb}

\subsection{Results from the small-domain simulations}\label{appb1}

\begin{figure*}
\includegraphics[width=15.5cm]{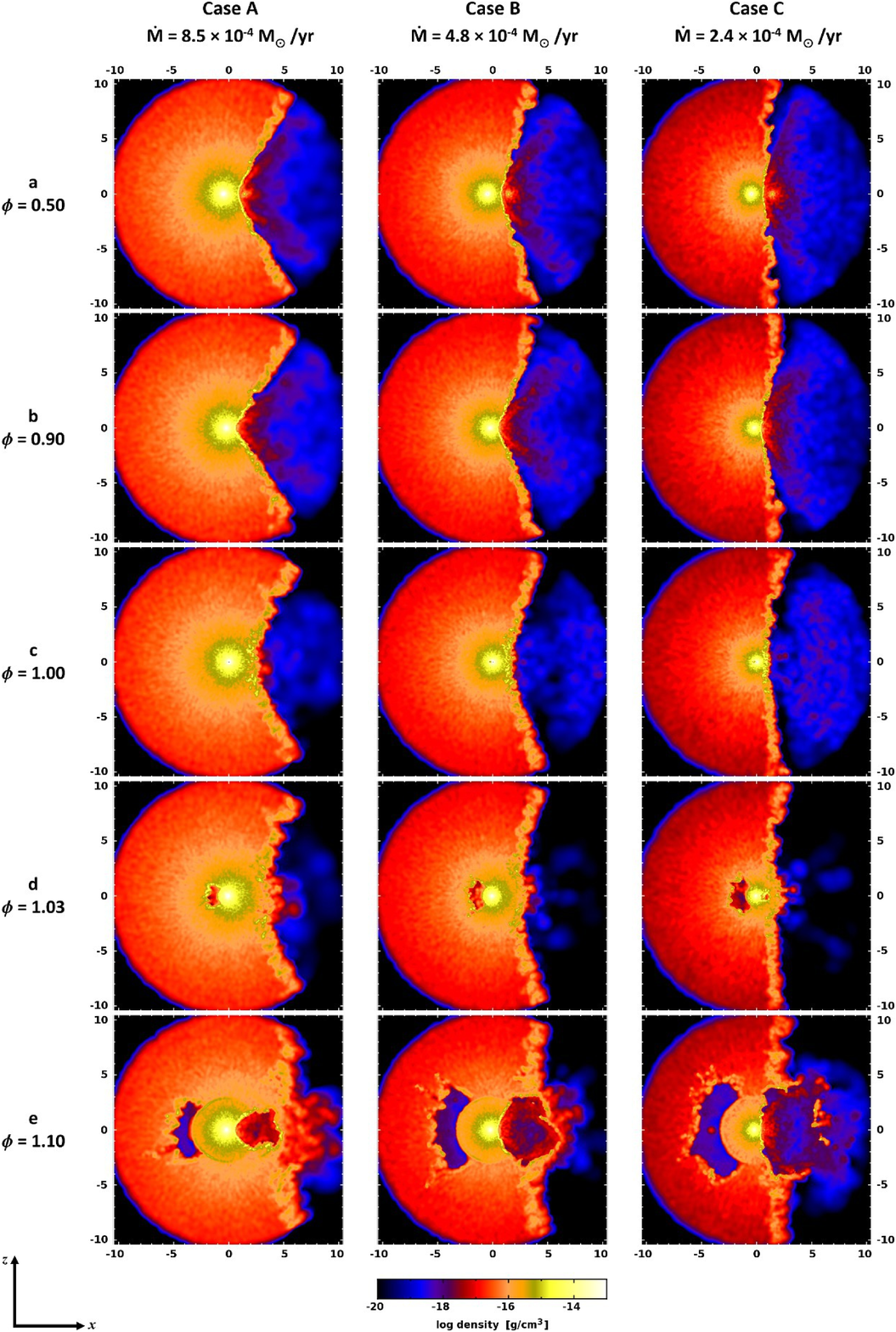}
\caption{Same as Figure~\protect\ref{fig1}, but for slices in the $xz$ plane containing the binary semimajor ($x$) and orbital angular momentum ($z$) axes.}
\label{figB1}
\end{figure*}

\begin{figure*}
\includegraphics[width=15.5cm]{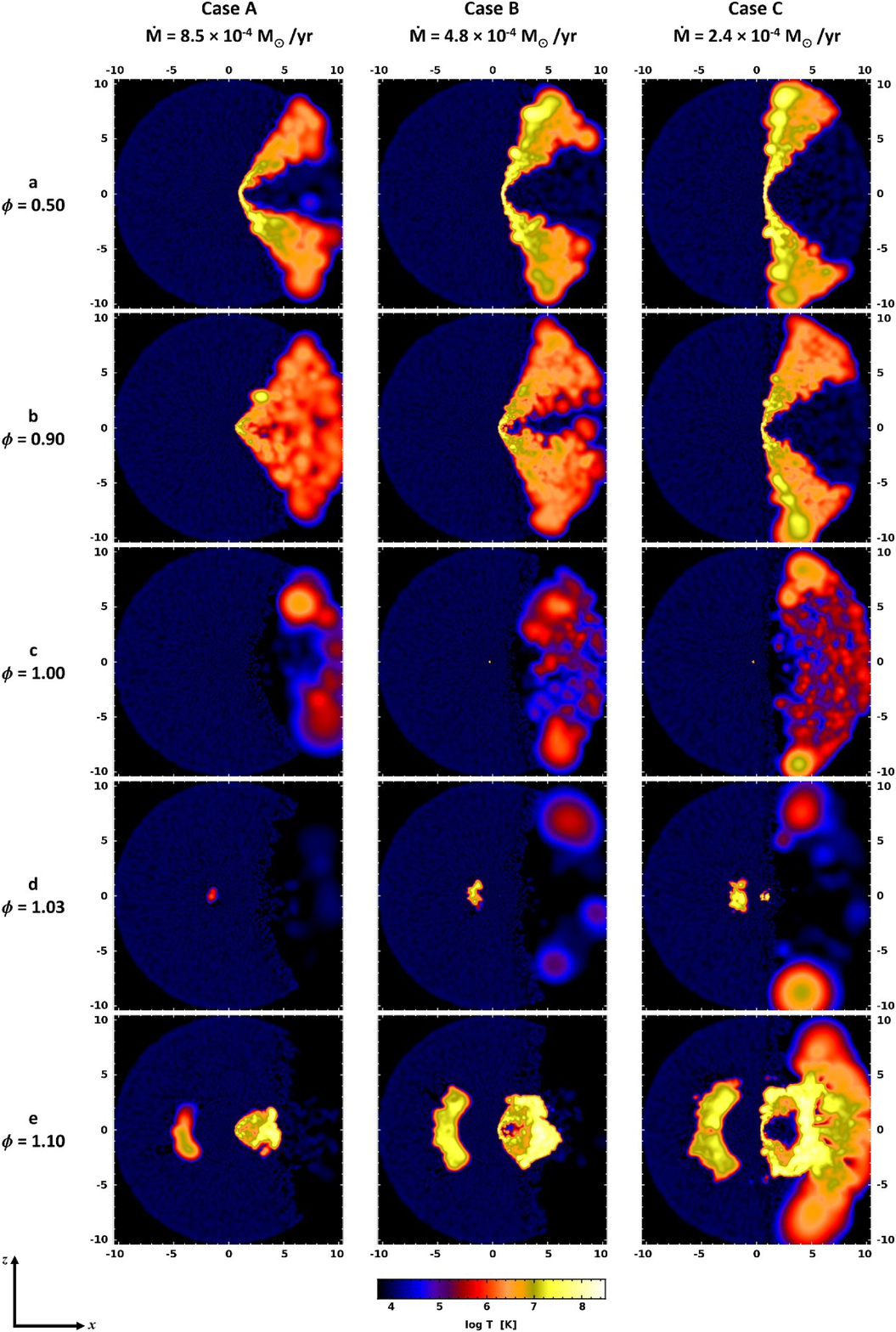}
\caption{Same as Figure~\protect\ref{figB1}, but with color showing log temperature.}
\label{figB2}
\end{figure*}

\begin{figure*}
\includegraphics[width=15.5cm]{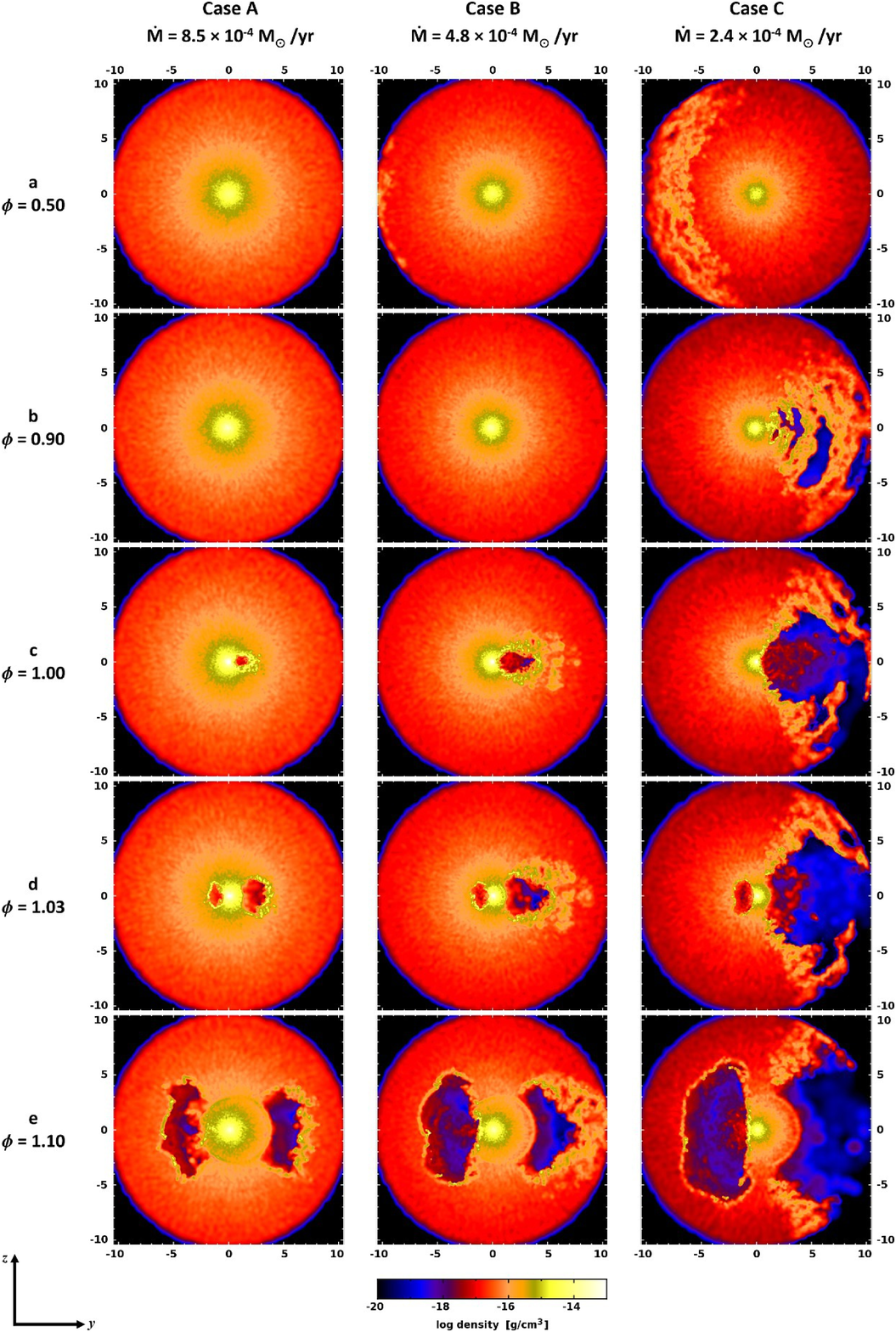}
\caption{Same as Figure~\protect\ref{fig1}, but for slices in the $yz$ plane containing the binary semiminor ($y$) and orbital angular momentum ($z$) axes.}
\label{figB3}
\end{figure*}

\begin{figure*}
\includegraphics[width=15.5cm]{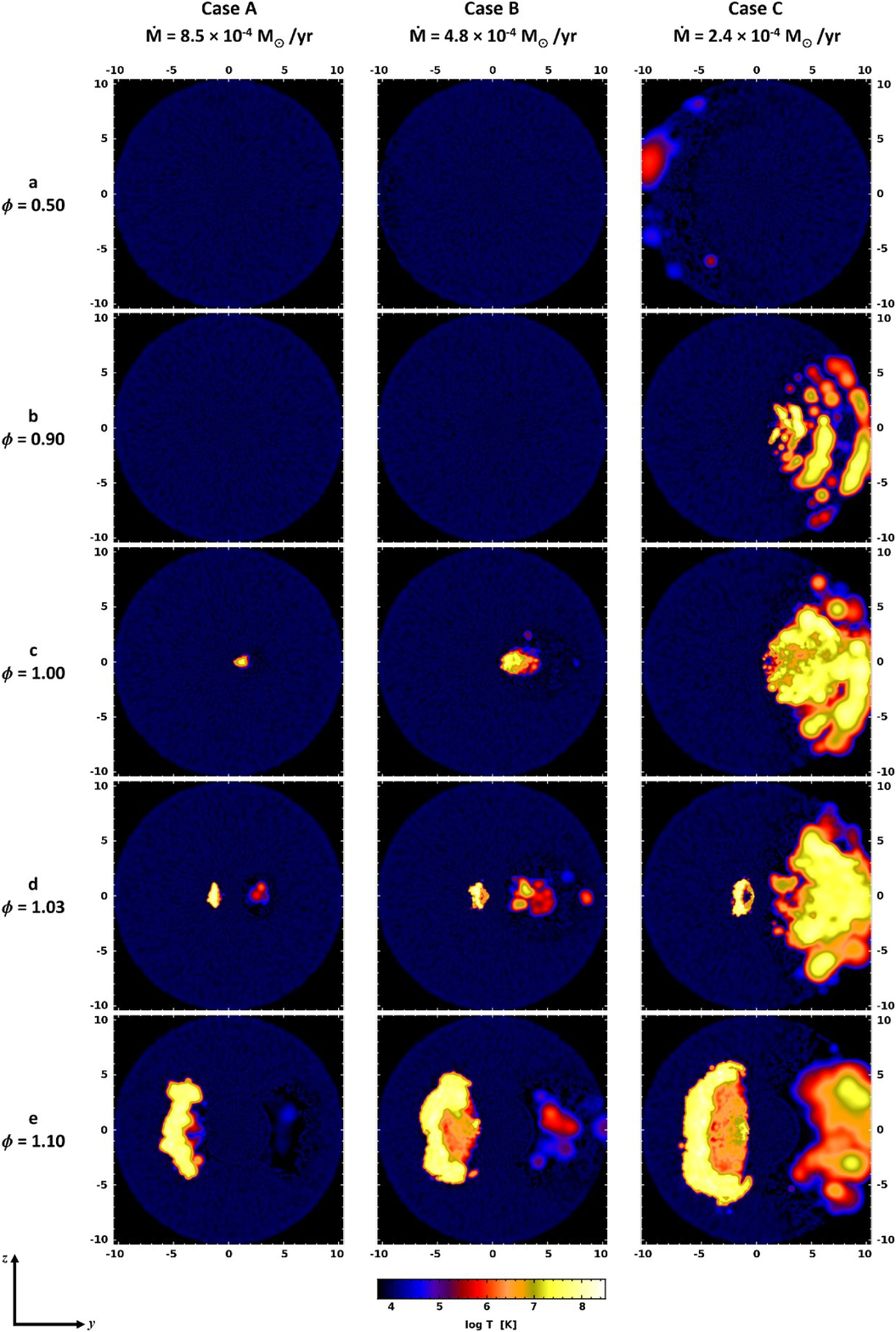}
\caption{Same as Figure~\protect\ref{figB3}, but with color showing log temperature.}
\label{figB4}
\end{figure*}

Figures~\ref{figB1} through \ref{figB4} show density and temperature slices in the $xz$ and $yz$ planes perpendicular to the orbital plane at the same five orbital phases of Figures~\ref{fig1} and \ref{fig2} for the small-domain 3D SPH simulations, allowing one to more fully appreciate the complex 3D structure of the winds and their collision. Row a of Figures~\ref{figB1} and \ref{figB2} further demonstrates the axisymmetry of the shock cone at phases around apastron, the increase in cavity opening angle with decreasing \mdota, and the increase in distance between the WWC apex and \eca\ as \mdota\ is increased. Decreases in the thickness, density, and stability of the cold, post-shock primary wind region with decreasing \mdota\ are also apparent.

At $\phi = 0.9$, the WWC zone starts to become distorted and \ecb\ begins to move to the back side of \eca. Slices centered in the $xz$ plane sample the leading edge of the trailing arm of the WWC zone above and below the orbital plane. Figure~\ref{figB2} shows the changing temperature structure of the hot post-shock \ecb\ wind as the trailing arm sweeps through the $xz$ plane. The $xz$ slice at $\phi = 0.9$ samples more of the leading edge of the trailing arm the higher the value of \mdota. This is because higher \mdota\ produce smaller shock-cone opening angles. The overall result is the appearance that the arms of the hot shock increase in width in the $xz$ plane as the system moves from $\phi = 0.5$ to $0.9$, filling in the cooler wind cavity.

At periastron (row c), \ecb\ is behind \eca, and there is a brief moment when the wind of \eca\ can flow unimpeded in the $+x$ (apastron) direction. During this time, the apex of the low density cavity on the $+x$ side of the system fills with dense \eca\ wind. The inner \eca\ wind region in the $+x$ direction also starts to look more spherical (see e.g. row d of Figure~\ref{figB1}). The thickness and density of this inner primary wind region increases with \mdota. The wind momentum balance is responsible for this since a higher value of \mdota\ moves the apex of the WWC zone farther from \eca, allowing the primary wind to fill a larger volume in the $+x$ direction during periastron passage. Slices in the $xz$ plane at periastron also sample mostly cold ($\sim 10^{4}$~K) material from both winds, although the \ecb\ wind cavity in the lower \mdota\ simulations contains warmer material due to the larger WWC opening angle and less oblique shocks.

Following periastron, \ecb\ emerges from \eca's dense wind, carving a new low-density wind cavity that is visible on the $-x$ side of the $xz$-plane slices (Figures~\ref{figB1} and \ref{figB2}, row d). Eventually \ecb\ moves back to the $+x$ side of the system and the WWC zone is reestablished (row e). A new wind cavity is slowly carved in \eca's wind on the $+x$ side while the spiral WWC region and low-density cavity created on the $-x$ side flow outward. Due to the more similar wind momenta, the wind cavity is larger and hotter in the lower \mdota\ simulations. The size and geometry of the cavities on the $+x$ and $-x$ sides of the system are also very distinct. The wind of \ecb\ is able to plough through the dense shell of primary wind on the $+x$ side of the system at earlier phases after periastron in the lower \mdota\ simulations (Figure~\ref{figB1}, row e).

For most of the orbit, slices of the density and temperature in the $yz$ plane are very similar for all three simulations since only the dense wind of \eca\ is sampled. The exception is around periastron, mainly for Case~C (Figures~\ref{figB3} and \ref{figB4}). There is a clear density and temperature difference between the $+y$ and $-y$ sides of the system in the $yz$-plane slice for Case~C at $\phi = 0.9$ (row b). This is because of Case~C's extremely large WWC-cavity opening angle. Examination of row~b of Figure~\ref{fig1} shows that at $\phi = 0.9$, only in Case~C has the leading arm of the WWC zone passed through the $yz$ plane. In Cases~A and B, the leading arm does not pass through the $yz$ plane until close to $\phi = 1$ (Figures~\ref{figB3} and \ref{figB4}, row c).

Again, following periastron passage, \ecb\ returns to the $+x$ side of the system, carving a new wind cavity along the way. We see in the density and temperature slices taken at $\phi = 1.03$ (Figures~\ref{figB3} and \ref{figB4}, row d) that this produces a difference in the size of the cavity on the $+y$ and $-y$ sides of the system. The magnitude of this difference increases with decreasing \mdota. The larger cavity on the $+y$ side of the system is the remnant of the WWC cavity from before periastron, while the smaller cavity on the $-y$ side is part of the new cavity created after periastron. A temperature difference also exists within the cavity, with the $+y$ side being much colder than the $-y$ side in the simulations for Cases~A and B. After enough time passes, the new cavity on the $-y$ side of the system expands to a size comparable to the cavity on the $+y$ side, although there is still a clear difference in the temperature and density structure of the two cavities (Figures~\ref{figB3} and \ref{figB4}, row e).

\subsection{Results from the large-domain simulations}\label{appb2}

\begin{figure*}
\includegraphics[width=15.5cm]{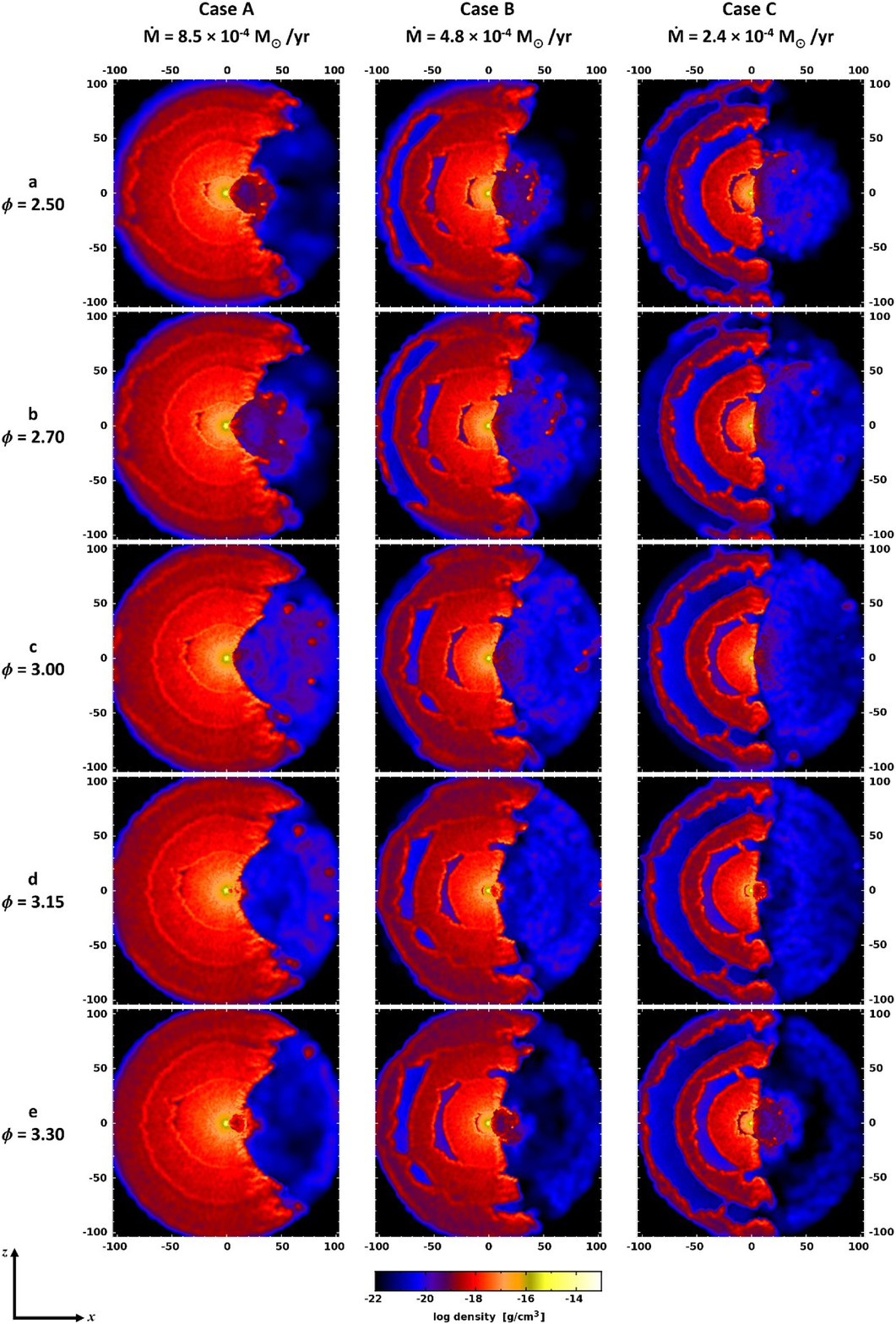}
\caption{Same as Figure~\protect\ref{fig8}, but for slices in the $xz$ plane containing the binary semimajor ($x$) and orbital angular momentum ($z$) axes.}
\label{figB5}
\end{figure*}

\begin{figure*}
\includegraphics[width=15.5cm]{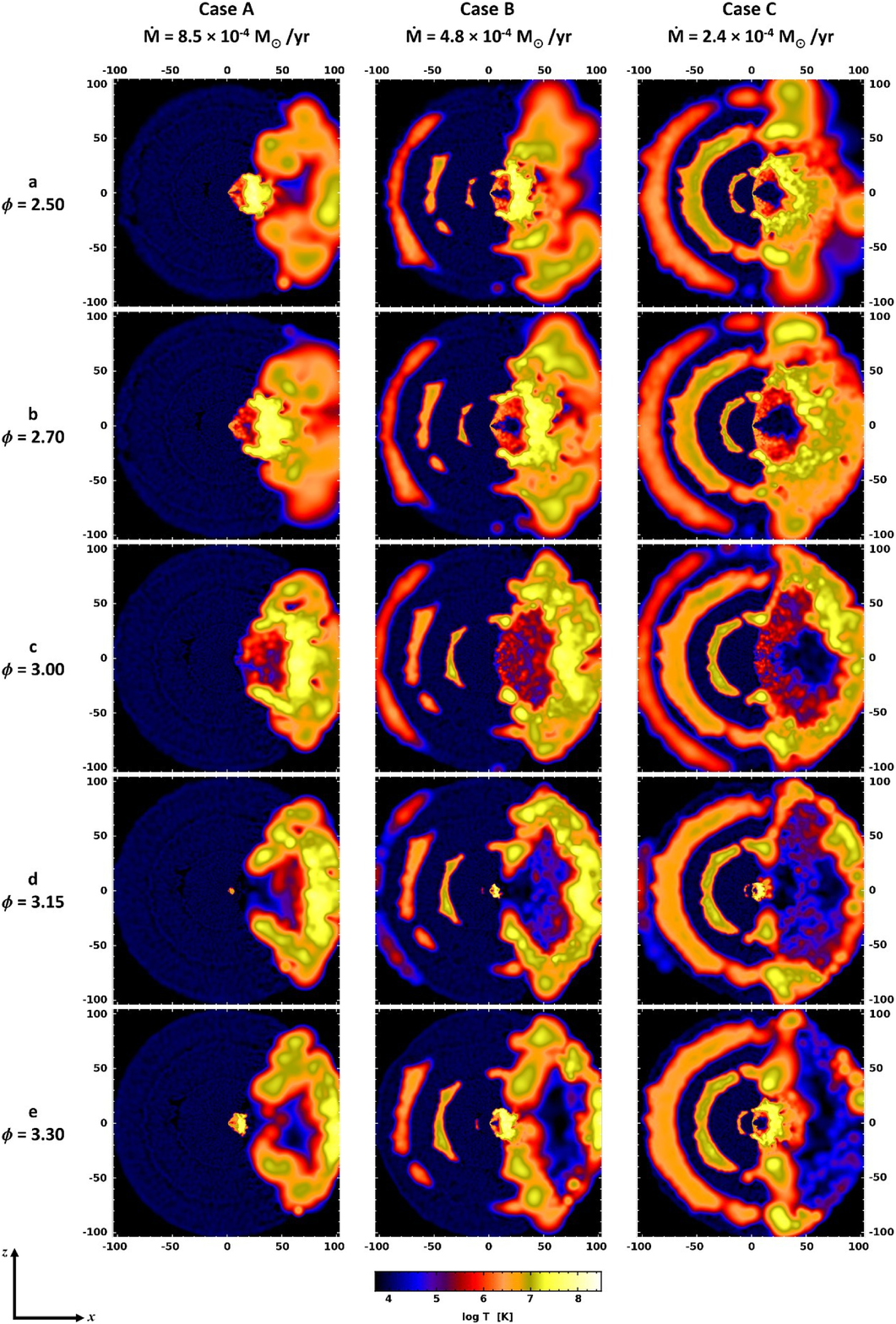}
\caption{Same as Figure~\protect\ref{figB5}, but with color showing log temperature.}
\label{figB6}
\end{figure*}

\begin{figure*}
\includegraphics[width=15.5cm]{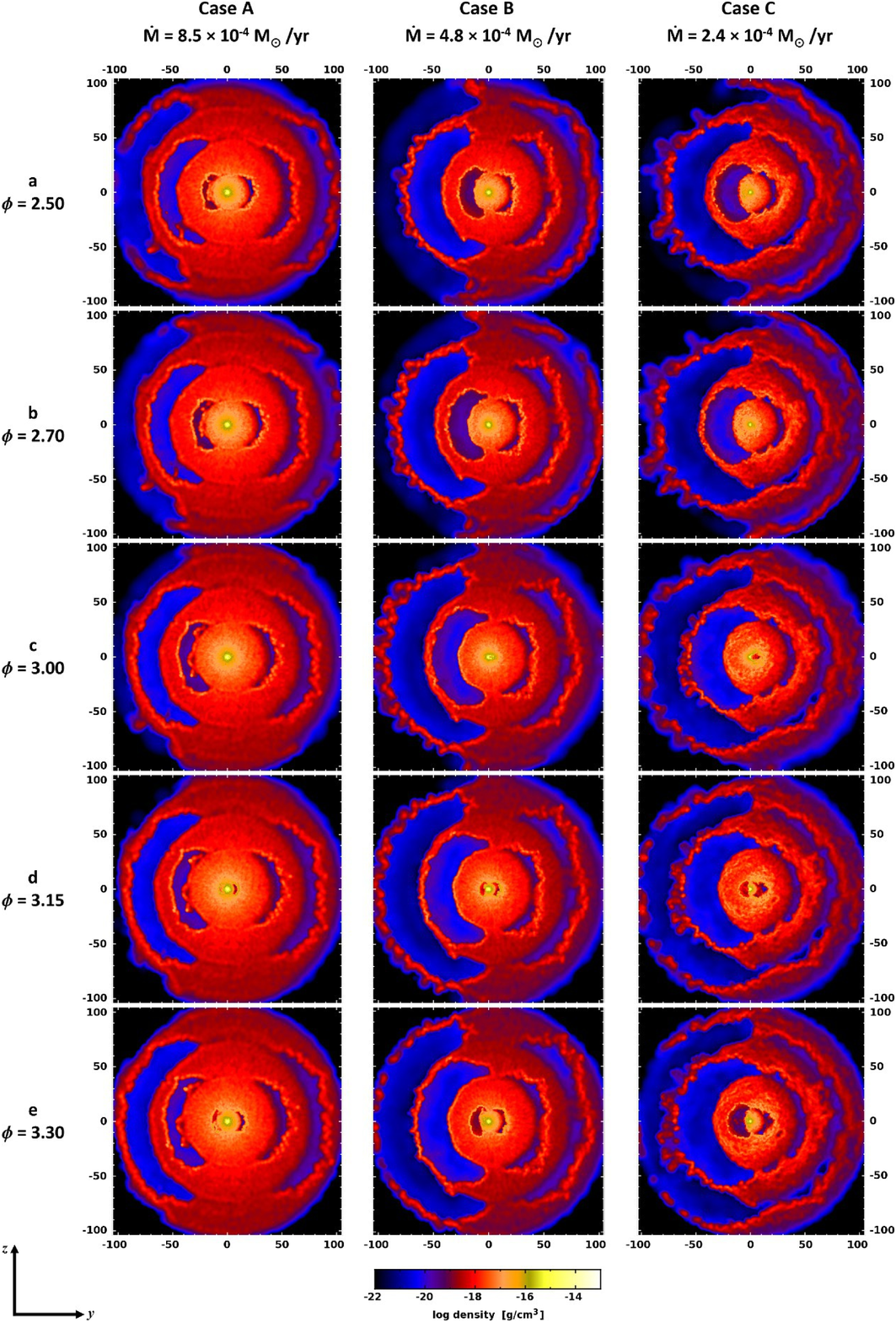}
\caption{Same as Figure~\protect\ref{fig8}, but for slices in the $yz$ plane containing the binary semiminor ($y$) and orbital angular momentum ($z$) axes.}
\label{figB7}
\end{figure*}

\begin{figure*}
\includegraphics[width=15.5cm]{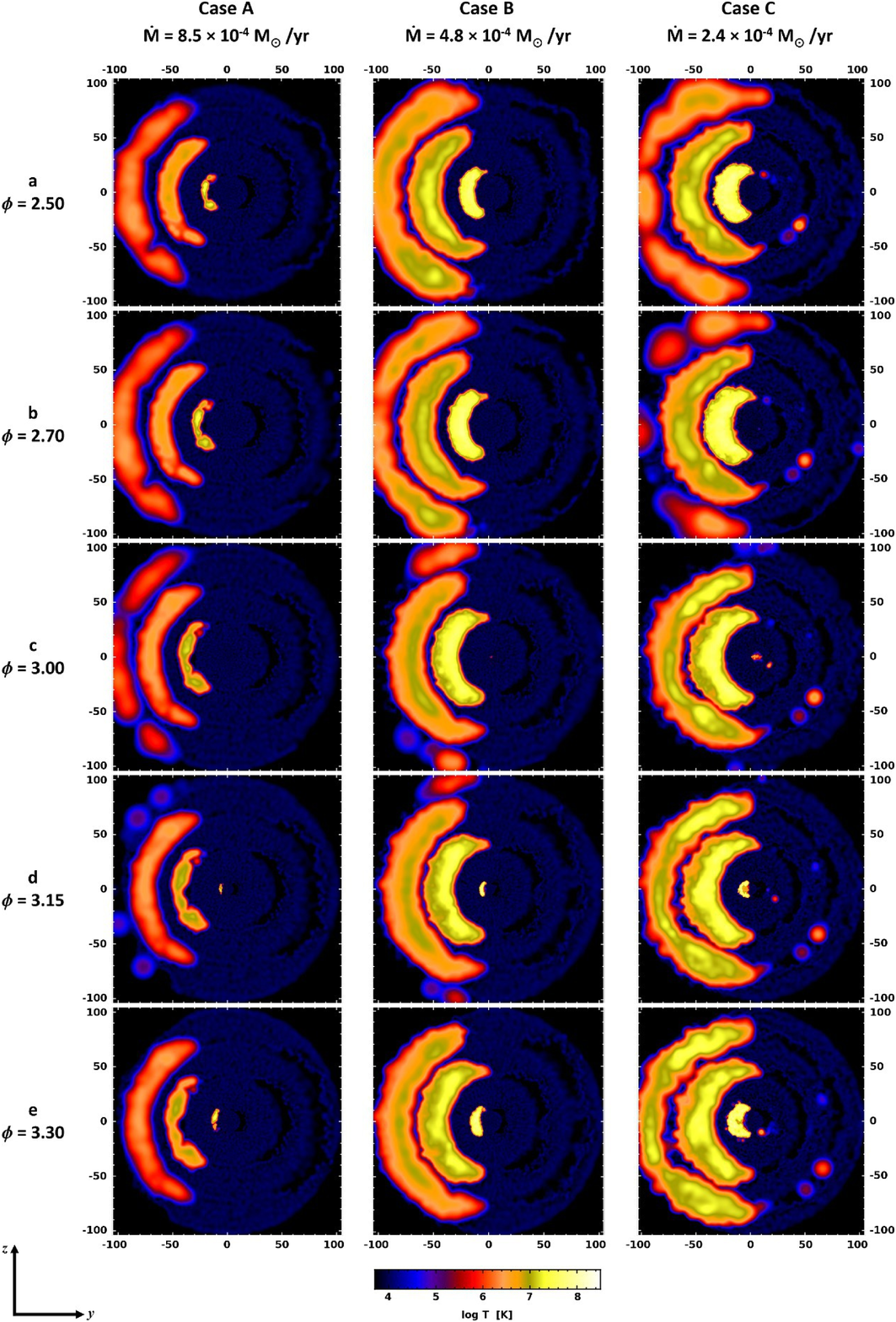}
\caption{Same as Figure~\protect\ref{figB7}, but with color showing log temperature.}
\label{figB8}
\end{figure*}

Figures~\ref{figB5} through \ref{figB8} show density and temperature slices in the $xz$ and $yz$ planes at the same five orbital phases of Figures~\ref{fig8} and \ref{fig9} for the large-domain 3D SPH simulations. There is a clear left-right asymmetry in the density and temperature in each panel of the figures. Figure~\ref{figB5} shows that the higher the value of \mdota, the smaller the wind cavities carved by \ecb\ on the $-x$ side of the system become in the $xz$ plane. They are almost not visible in the Case~A simulation, with the wind of \eca\ in the $-x$ direction looking continuous, but containing shells of enhanced density at periodic intervals. Figure~\ref{figB5} also shows that the gradual fragmentation of the shell of dense \eca\ material on the $+x$ side of the system produces a series of small, cold, dense blobs within the larger, low-density wind cavity created by \ecb. The higher the value of \mdota, the larger and denser these blobs are, and the longer they last before mixing with the surrounding \ecb\ wind. Figure~\ref{figB6} illustrates how in Case~A, the winds on the $-x$ side of the system in the $xz$ plane remain cold, while the wind cavity on the $+x$ side remains hot. In Cases~B and C, the low-density cavities on both sides of the system remain hot, with only the intermediate regions of dense \eca\ wind cold.

In Figure~\ref{figB7} we see that the extended narrow wind cavities on the $+y$ side of the system in the $yz$ plane are always smaller than the wind cavities carved on the $-y$ side. These smaller cavities are always cold, while the larger cavities are always hot (Figure~\ref{figB8}). The difference in size between these cavities grows with decreasing \mdota. In Case~A, the cavities are of comparable size, but in Case~C, the cavities on the $-y$ side are much larger. The temperature of the gas in the $-y$ cavities is also nearly an order of magnitude hotter in Cases~B and C than in Case~A. Finally, the shells of dense \eca\ material on the $-y$ side of the system are thicker and remain intact longer in the higher \mdota\ simulations.

\bsp

\label{lastpage}

\end{document}